\documentclass[a4paper,11pt]{article}
\pdfoutput=1  

\usepackage{jheppub}  
\usepackage{graphicx}
\usepackage{epstopdf}
\usepackage{bm}
\usepackage{epsfig}
\usepackage{graphics}
\usepackage{xspace}
\usepackage{hyperref}
\usepackage{slashed}
\usepackage{xcolor}
\usepackage{array}
\usepackage[small]{subfigure}
\usepackage[normalem]{ulem}

\newcommand{\nl}{n_{\ell}}
\newcommand{\nq}{n_Q}
\newcommand{\MSR}{\mathrm{MSR}}
\newcommand{\pole}{\mathrm{pole}}
\newcommand{\mbar}{\overline{m}}
\newcommand{\mpole}{m^{\pole}}
\newcommand{\msr}{m^{\rm MSR}}
\newcommand{\delbar}{\overline{\delta}}

\newcommand{\MSb}{\overline{\mathrm{MS}}}	
\newcommand{\Ord}{\mathcal{O}}
\newcommand{\LQCD}{\Lambda_\mathrm{QCD}}

\newcommand{\dd}{\mathrm{d}}

\newcommand{\betazero}{\beta_0^{(\nq)}}
\newcommand{\betaone}{\beta_1^{(\nq)}}
\newcommand{\betazeroell}{\beta_0^{(\nl)}}

\begin{document}
\title{On the Light Massive Flavor Dependence of the Large Order Asymptotic Behavior and the Ambiguity\\ of the Pole Mass}

\preprint{\begin{flushright} UWThPh-2017-13\\MIT-CTP 4915 \end{flushright}\vspace*{-1cm}}

\author[a,b]{Andr\'e H. Hoang}
\author[a]{Christopher Lepenik}
\author[a,c]{Moritz Preisser}

\affiliation[a]{University of Vienna, Faculty of Physics\\Boltzmanngasse 5, A-1090 Wien, Austria}
\affiliation[b]{Erwin Schr\"odinger International Institute for Mathematical Physics,\\
University of Vienna, Boltzmanngasse 9, A-1090 Wien, Austria}
\affiliation[c]{Center for Theoretical Physics, Massachusetts Institute of Technology,\\Cambridge, MA 02139, USA}

\emailAdd{andre.hoang@univie.ac.at}
\emailAdd{christopher.lepenik@univie.ac.at}
\emailAdd{moritz.preisser@univie.ac.at}

\abstract{
We provide a systematic renormalization group formalism for the mass effects in the
relation of the pole mass $m_Q^{\rm pole}$ and short-distance masses such as the $\MSb$ mass $\mbar_Q$ of a heavy quark $Q$, coming from virtual loop insertions of massive quarks lighter than $Q$. 
The formalism reflects the constraints from heavy quark symmetry and entails a combined matching and evolution procedure that allows to disentangle and successively integrate out the corrections coming from the lighter massive quarks and the momentum regions between them and to precisely control the large order asymptotic behavior. With the formalism we systematically sum logarithms of ratios of the lighter quark masses and $m_Q$, relate the QCD corrections for different external heavy quarks to each other, predict the ${\cal O}(\alpha_s^4)$ virtual quark mass corrections in the pole-$\MSb$ mass relation, calculate the pole mass differences for the top, bottom and charm quarks with a precision of around $20$~MeV and analyze the decoupling of the lighter massive quark flavors at large orders.
The summation of logarithms is most relevant for the top quark pole mass $m_t^{\rm pole}$, where the hierarchy to the bottom and charm quarks is large. We determine the ambiguity of the pole mass for top, bottom and charm quarks in different scenarios with massive or massless  bottom and charm quarks in a way consistent with heavy quark symmetry, and we find that it is $250$~MeV.
The ambiguity is larger than current projections for the precision of top quark mass measurements in the high-luminosity phase of the LHC. 
}

\maketitle
\section{Introduction}
\label{sec:intro}

The masses of the heavy charm, bottom and top quarks belong to the most important input parameters in precise theoretical predictions of the Standard Model and models of new physics. Due to the effects of quantum chromodynamics (QCD) and because quarks are states with color charge, however, the mass of a heavy quark $Q$ is not a physical observable and should, in general, be better thought of as a renormalized and scheme-dependent parameter of the theory. This concept is incorporated most cleanly in the so-called $\MSb$ mass $\mbar_Q(\mu)$, which is defined through the same renormalization prescription as the $\MSb$ QCD coupling $\alpha_s(\mu)$. It can be measured from experimental data very precisely, but does not have any kinematic meaning, and it can be thought of incorporating short-distance information on the mass from scales larger than $\mu$. On the other hand, the so-called pole mass $m_Q^{\rm pole}$ is defined as the single particle pole in correlation functions involving the massive quark $Q$ as an external on-shell particle, and it determines the kinematic mass of the quark $Q$ in the context of perturbation theory. It is therefore unavoidable that the pole mass scheme appears in one way or another in higher order QCD calculations involving external massive quarks. For perturbative predictions involving the production of top quarks at hadron colliders, the pole mass scheme is therefore the main top quark mass scheme used in the literature, and switching scheme is cumbersome since these computations are predominantly numerical where the pole scheme provides the most efficient approach for the computations. In Refs.~\cite{Tarrach:1980up, Gray:1990yh,Chetyrkin:1999ys,Chetyrkin:1999qi,Melnikov:2000qh,Marquard:2007uj,Marquard:2015qpa,Marquard:2016dcn} the relation between the $\MSb$ and the pole mass has been computed up to ${\cal O}(\alpha_s^4)$ in the approximation that all quarks lighter than $Q$ are massless. Assuming the values $\mbar_t\equiv\mbar_t(\mbar_t)=163$~GeV, $\mbar_b\equiv\mbar_b(\mbar_b)=4.2$~GeV and
$\mbar_c\equiv\mbar_c(\mbar_c)=1.3$~GeV we obtain\footnote{
We assume $\alpha^{(5)}(M_Z)=0.1180$ for $M_Z=91.187$~GeV for the $\MSb$ QCD coupling and account for 5-loop evolution~\cite{Baikov:2016tgj} and flavor matching at the scales $\mbar_{c,b,t}$~\cite{Kniehl:2006bg}, which gives
$\alpha_s^{(6)}(\mbar_t)=0.10847$,
$\alpha_s^{(5)}(\mbar_b)=0.22430$,
$\alpha_s^{(4)}(\mbar_c)=0.38208$.}
\begin{eqnarray}
\label{eqn:mtpolevalues}
m_t^{\rm pole} &= 163 \,+\, 7.5040  \,+\,1.6005 \,+\,0.4941 \,+\, (0.1944\pm 0.0004)\,\mbox{GeV}\,,  \\
\label{eqn:mbpolevalues}
m_b^{\rm pole} &= 4.2 \,+\, 0.3998  \,+\,0.1986 \,+\,0.1443 \,+\, (0.1349\pm 0.0002)\,\mbox{GeV}\,,  \\
\label{eqn:mcpolevalues}
m_c^{\rm pole} &= 1.3 \,+\, 0.2108  \,+\,0.1984 \,+\,0.2725 \,+\, (0.4843\pm 0.0005)\,\mbox{GeV}\,,
\end{eqnarray}
where the terms show the series in powers of the strong coupling $\alpha_s(\mbar_Q)$ in the 
scheme that includes $Q$ as a dynamical flavor. The fourth order coefficient displays the numerical uncertainties from~\cite{Marquard:2016dcn}, which are, however, much smaller than other types of uncertainties considered in this paper.
 
The pole mass renormalization scheme is infrared-safe and gauge-invariant~\cite{Tarrach:1980up,Kronfeld:1998di}, but suffers from large corrections in the QCD perturbation series. This is because the pole mass scheme involves subtractions of on-shell quark self energy corrections containing virtual gluon and massless quark fluctuations which are linearly sensitive to small momenta. The on-shell approximation of the self energy diagrams entails that this sensitivity increases strongly with the order. The effect this has for the form of the corrections can be seen in Eqs.~\eqref{eqn:mtpolevalues}--\eqref{eqn:mcpolevalues}, which in the asymptotic large order limit have the form
\begin{align} 
\label{eqn:polemsbarseriesLL}
m_Q^{\rm pole}-\mbar_Q(\mbar_Q) &  \sim \mu\, \sum_{n=0}^\infty \frac{16}{3}(2\beta_0^{(\nl)})^{n} \,n!\,
\bigg(\frac{\alpha_s^{(\nl)}(\mu)}{4\pi}\bigg)^{n+1}\,,
\end{align}
in the $\beta_0$/LL approximation, which means that the terms in the QCD $\beta$-function,
\begin{equation}\label{eqn:betafct}
\frac{\dd\alpha_s^{(\nl)}(\mu)}{\dd\log \mu}=\beta^{(\nl)}(\alpha_s(\mu))\,=\,-\,2\,\alpha_s^{(\nl)}(\mu)\sum_{n=0}^\infty\beta_n^{(\nl)}\bigg(\frac{\alpha_s^{(\nl)}(\mu)}{4\pi}\bigg)^{n+1}\,,
\end{equation}
beyond the leading logarithmic level (i.e.\ $\beta_{n>0}$) are neglected. Here $\nl$ is the number of massless quark flavors. 

The factorially diverging pattern of the perturbation series and the linear dependence on the renormalization scale $\mu$ of the strong coupling displayed in Eq.~\eqref{eqn:polemsbarseriesLL} are called {\it the ${\cal O}(\Lambda_{\rm QCD})$ renormalon} of the pole mass~\cite{Bigi:1994em,Beneke:1994sw}. The form of the series on the RHS of Eq.~\eqref{eqn:polemsbarseriesLL} implies that at asymptotic large orders, and up to terms suppressed by inverse powers of $n$, the series becomes independent of its intrinsic physical scale $m_Q$. This and the $n$-factorial growth is an artifact of the pole mass scheme itself and not related to any physical effect. Technically this issue entails that for computing differences of series containing ${\cal O}(\Lambda_{\rm QCD})$ renormalon ambiguities using fixed-order perturbation theory one must consistently expand in powers of the strong coupling at the same renormalization scale such that the renormalon can properly cancel. 

The ${\cal O}(\Lambda_{\rm QCD})$ renormalon problem of the pole mass has received substantial attention in the literature as it turned out to be  not just an issue of pedagogical interest, but one that is relevant phenomenologically~\cite{Beneke:1998ui}. This is because for $\mu=\mbar_Q$ the known coefficients of the series in Eqs.~\eqref{eqn:mtpolevalues}--\eqref{eqn:mcpolevalues} agree remarkably well with the corresponding large order asymptotic behavior already beyond the terms of ${\cal O}(\alpha_s)$ (so that the terms of the series are known quite precisely to all orders) and because even for orders where the QCD corrections still decrease with order they can be very large numerically and make phenomenological applications difficult.
The pole mass scheme has therefore been abandoned in high precision top, bottom and charm quark mass analyses
in favor of quark mass schemes such as $\MSb$ or low-scale short distance masses such as
the kinetic mass~\cite{Czarnecki:1997sz}, the potential-subtracted (PS)~mass~\cite{Beneke:1998rk}, the 1S mass~\cite{Hoang:1998ng,Hoang:1998hm,Hoang:1999ye}, the renormalon-subtracted (RS) mass~\cite{Pineda:2001zq}, the jet mass~\cite{Jain:2008gb,Fleming:2007qr} or the MSR mass~\cite{Hoang:2008yj,Hoang:2017suc}.
These mass schemes do not have an ${\cal O}(\LQCD)$ renormalon and are called short-distance masses. 
It is commonly agreed from many studies that it is possible to determine short-distance masses with theoretical uncertainties of a few $10$~MeV~\cite{Hoang:2000yr,Olive:2016xmw}, and we therefore neglect any principle ambiguity in their values in this paper.

Using the theory of asymptotic series one can show that the best possible approximation to the LHS of Eq.~\eqref{eqn:polemsbarseriesLL} is to truncate the series on the RHS at the minimal term at order $n_{\rm min}$ which is approximately $n_{\rm min}\approx 2\pi/(\beta_0^{(\nl)}\alpha_s^{(\nl)}(\mu))$. The size of the correction of the minimal term is approximately $\Delta(n_{\rm min})\approx(4\pi\alpha_s^{(\nl)}(\mu)/\beta_0^{(\nl)})^{1/2}\Lambda_{\rm QCD}^{(\nl)}$, and there is a region in the orders $n$ around $n_{\rm min}$ of width $\Delta n\approx(2\pi^2/(\beta_0^{(\nl)}\alpha_s^{(\nl)}(\mu)))^{1/2}$ in which all series terms have a size close to the minimal term. At orders above $n_{\rm min}+\Delta n/2$ the series diverges quickly and the series terms from these orders are useless even if they are known through an elaborate loop calculation. The uncertainty with which the pole mass can be determined {\it in principle} given the full information about the perturbative series is called the {\it pole mass ambiguity}. It is universal, independent on the choice of the renormalization scale $\mu$ and exists in equivalent size in any context without the possibility to be circumvented.
However, the $\mu$-dependence of $n_{\rm min}$, $\Delta(n_{\rm min})$ and $\Delta n$ indicates that the way how the renormalon problem appears in practical applications based on perturbative QCD can differ substantially depending on the physical scale of the quantity under consideration and the corresponding choice of the renormalization scale $\mu$. Using the method of Borel resummation the pole mass ambiguity can be estimated to be of order $\Lambda_{\rm QCD}^{(\nl)}$, where the superscript $(\nl)$ stands for the dependence of the hadronization scale on the number of massless quark flavors. The norm of the ambiguity,  which we call $N_{1/2}^{(\nl)}$ in this paper, and the resulting pattern of the large order asymptotic behavior of the series can be determined very precisely and have been studied in many analyses (see e.g.\ the recent work of Refs.~\cite{Ayala:2014yxa,Beneke:2016cbu,Komijani:2017vep,Hoang:2017suc}). However, when quoting a concrete numerical size of the ambiguity, criteria common for converging series cannot be applied, and it is instrumental to consider more global aspects of the series and the quantity it describes. An essential aspect of the low-energy quantum corrections in heavy quark masses is {\it heavy quark symmetry (HQS)}~\cite{Isgur:1989vq} on which we put particular focus in this work.

An issue that has received less attention in the literature so far is how the masses of the lighter massive quarks affect the large order asymptotic behavior of the pole-$\MSb$ mass relation, where we refer to the effects of quarks with masses that are larger than $\Lambda_{\rm QCD}$. 
These corrections come from insertions of virtual quark loops and are known up to ${\cal O}(\alpha_s^3)$~\cite{Gray:1990yh,Bekavac:2007tk} from explicit loop calculations. It is known that the masses of lighter massive quarks provide an infrared cutoff and effectively reduce the number $n_\ell$ of massless flavors governing the large order asymptotic behavior~\cite{Ball:1995ni}. Due to the $\nl$-dependence of the QCD $\beta$-function the finite bottom and charm quark masses lead to an increased infrared sensitivity of the top quark pole mass and a stronger divergence pattern of the series, as can be seen from Eq.~\eqref{eqn:polemsbarseriesLL}. The ambiguity therefore inflates following the $\nl$-dependent increase of $\Lambda_{\rm QCD}$.
In Refs.~\cite{Hoang:1999us,Hoang:2000fm} it was pointed out that the ${\cal O}(\alpha_s^2)$ and ${\cal O}(\alpha_s^3)$ virtual quark mass corrections are already dominated by the infrared behavior related to the ${\cal O}(\Lambda_{\rm QCD})$ renormalon. In Ref.~\cite{Ayala:2014yxa} it was further observed  that the ${\cal O}(\alpha_s^3)$ charm mass corrections in the bottom pole-$\MSb$ mass relation can be rendered small when the series is expressed in terms of $\alpha_s^{(\nl=3)}$ rather than $\alpha_s^{(\nl=4)}$, i.e.\ the charm quark effectively decouples. A systematic and precise understanding of the intrinsic structure of the lighter massive quark effects from the point of view of disentangling the different momentum modes and their interplay has, however, not been provided so far in the literature. The task is complicated since apart from being a problem in connection with the behavior of perturbation theory at large orders, it also represents a multi-scale problem with scales given by the quark masses as well as $\Lambda_{\rm QCD}$ and where, for the top quark, logarithms of mass ratios can be large. 

It is the main purpose of this paper to present a formalism that can do exactly that. It is based on the concept of the renormalization group (RG) and allows to successively integrate out momentum modes from the pole-$\MSb$ mass relation of a heavy quark $Q$ in order to disentangle the contributions coming from the lighter massive quarks and to systematically sum logarithms of the mass ratios. The approach allows to quantify and formulate precisely the effects the masses of the lighter massive quarks have on the pole-$\MSb$ mass relation and therefore on the pole mass itself and may find interesting applications in other contexts.
As the essential new feature the RG formalism entails {\it linear scaling with the renormalization scale}. The common logarithmic scaling, as known for the strong coupling, cannot capture the linear momentum dependence of QCD corrections to the heavy quark mass for scales below $m_Q$. 
The formalism is in particular useful since it fully accounts for all aspects of HQS. It can be used to concretely formulate and study in a transparent way two important properties of the heavy quark pole masses following from HQS:  (1) The pole mass ambiguity is independent of the mass of the heavy quark and (2) the ambiguities of all heavy quarks are equal up to power corrections of order $\Lambda_{\rm QCD}^2/m_Q$. 

The essential technical tool to set up the formalism is the MSR mass $m_Q^{\rm MSR}(R)$~\cite{Hoang:2008yj,Hoang:2017suc}.
Like the perturbative series for the pole-$\MSb$ mass relation, the pole-MSR mass relation is calculated from on-shell heavy quark self energy diagrams, but has also linear dependence on $R$. It is the basis of the RG formalism we propose,
allows to precisely capture the QCD corrections from the different quark mass scales and, in particular, to encode and study issue (1) coming from HQS. The renormalization group evolution in the scale $R$ is described by R-evolution~\cite{Hoang:2008yj,Hoang:2017suc}, which is free of the ${\cal O}(\Lambda_{\rm QCD})$ renormalon, and allows to sum large logarithms of ratios of the quark masses in the evolution between the quark mass scales. Using the concepts of the MSR mass and the R-evolution it is then possible to relate the pole-$\MSb$ masses of the top, bottom and charm quarks to each other. This allows to systematically encode and study issue (2) coming from HQS, and to interpret the small effects of HQS breaking as matching corrections in a renormalization group flow that connects the QCD correction of the top, bottom and charm quarks. 
The resulting formula can be used to specify the heavy quark pole mass ambiguity in the context of lighter massive quarks and to derive a generalized expression for the large order asymptotic behavior accounting accurately for the light massive flavor dependence. Concerning the accuracy of our description of the virtual quark loop mass effects in the large order asymptotic behavior we reach a precision of a few MeV, which applies equally for top, bottom and charm quarks.

The second main purpose of this paper is to use the RG formalism to specify concretely the ambiguity of the top quark pole mass and also the pole mass of the bottom and charm quarks assuming that  their $\MSb$ masses are given. We in particular address the question how the outcome depends on different scenarios for treating the bottom and charm quarks as massive or massless, and we explicitly take into account the consistency requirements of HQS. The aim is to provide a concrete numerical specification of the ambiguity of the top quark pole mass beyond the qualitative statement that the ambiguity is ``of order $\Lambda_{\rm QCD}^{(\nl)}$'' and to make a concrete statement up to which principle precision the top quark pole mass may still be used as a meaningful phenomenological parameter.
We stress that in this context we adopt the view that the pole masses have well-defined and unique meaning, so that the pole mass ambiguity acquires the meaning of an intrinsic numerical uncertainty. This differs from the view sometimes used in high-precision analyses, where the pole mass is employed as an intrinsic order-dependent parameter to effectively parameterize the use of a short-distance mass scheme. 

Apart from specifying the ambiguity of the pole masses we are also interested in studying the dependence of their value on the different scenarios for treating the bottom and charm quarks as massive or massless. 
The issue is of particular interest for the top quark pole mass which is still widely used for theoretical predictions and phenomenological studies in top quark physics.
The top quark pole mass is, due to its linear sensitivity to small momenta, also linearly sensitive to the masses of the lighter massive quarks. Since many short-distance observables used for top quark pole mass determinations are at most quadratically sensitive to small momenta, the dominant effects of the bottom and charm masses may well come from the top quark pole definition itself. A large dependence of the top quark pole mass value on whether the bottom and charm quarks are treated as massive or massless would therefore affect the ambiguity estimate if one considers the top quark pole mass as a globally defined mass scheme (valid for any scenario for the bottom and charm quark masses).  We can address this question precisely because the RG-formalism we use  
allows for very accurate numerical calculations of the lighter quark mass effects. Within the size of the ambiguity, we do not find any such dependence.
The outcome of our analysis is that the top quark pole mass ambiguity, and the ambiguity of the bottom and charm quark pole masses, is around $250$~MeV. 

Prior to this work the best estimate and the ambiguity of the top quark pole mass were studied in
Ref.~\cite{Beneke:2016cbu}. They analyzed the top quark pole-$\MSb$ mass series of Eq.~\eqref{eqn:mtpolevalues} for $\mu=\mbar_t$ and massless bottom and charm quarks and in an extended analysis also for massive bottom and charm quarks. 
They argued that the ambiguity of the top, bottom and charm quark pole masses amounts to $110$~MeV. We believe that their ambiguity estimate of $110$~MeV is too optimistic, and we explain this in detail from the requirements of HQS.
They also quantified the bottom and charm mass effects coming from beyond the known corrections at  ${\cal O}(\alpha_s^2)$ and ${\cal O}(\alpha_s^3)$ by using a heuristic prescription based on an order-dependent reduction of the flavor number. This does not represent a systematic calculation, but we find it to be an adequate approximation for the task of estimating the top quark pole mass renormalon ambiguity.

The paper is organized as follows:
In Sec.~\ref{sec:prelim} we review the explicitly calculated corrections up to ${\cal O}(\alpha_s^4)$ for the pole-$\MSb$ and the pole-MSR mass relations for the case that all quarks lighter than quark $Q$ are massless and we explain our notation for parameterizing the virtual quark mass corrections due to the light massive quarks. This notation is essential for our setup of the flavor number dependent RG evolution of the MSR mass, which we also review to the extend needed for our studies in the subsequent sections. We also review known basic issues about the large order asymptotic behavior and the renormalon ambiguity of the pole-$\MSb$ and the pole-MSR mass relations, including their dependence on the number of massless quarks.
In Sec.~\ref{sec:hardmodes} we explain details about the matching procedures that allow to integrate out the virtual corrections coming from the heavy quark $Q$ and the lighter massive quarks, and to relate the pole-MSR mass relation of quark $Q$ to the pole-$\MSb$ mass relation of the next lighter massive quark, which is based on heavy quark symmetry. These considerations and the numerical analysis of the latter matching corrections allow us to derive a prediction for the yet uncalculated ${\cal O}(\alpha_s^4)$ virtual quark mass corrections and to discuss the large order asymptotic form of the virtual quark mass corrections. As an application of the RG formalism devised in our work we compute the difference of the pole masses of the top, bottom and charm quarks. Since their differences are short-distance quantities we can compute them with a precision of around $20$~MeV. We also analyze the validity of the effective flavor decoupling at large orders in the context of the top quark pole mass.
In Sec.~\ref{sec:topmass} we finally discuss in detail the best possible estimate of the top quark pole mass and in particular its ambiguity in the context of three different scenarios for the bottom and charm quark masses. We discuss these three scenarios separately because the pole mass concept, strictly speaking, depends on the setup for the lighter quark masses, and we also discuss our results in the context of adopting the view that the top quark pole mass is a general concept. 
Finally, in Sec.~\ref{sec:conclusions} we conclude.
In  App.~\ref{app:Delta} we provide explicit results for the virtual quark mass corrections at ${\cal O}(\alpha_s^3)$ in our notation, using the results from Ref.~\cite{Bekavac:2007tk}, and we complete them concerning the
corrections coming from the insertion of two quark loops involving quarks with two arbitrary masses.

\section{Preliminaries and Notation}
\label{sec:prelim}

\subsection{\texorpdfstring{$\overline{\rm MS}$}{MS-bar} Mass}
\label{sec:MSbarmass}

The perturbative series of the difference between the $\MSb$ mass $\mbar_Q(\mu)$ at the scale $\mu=\mbar_Q(\mbar_Q)$ and the pole mass $\mpole_Q$ of a heavy quark $Q$ is the basic relation from which we start our analysis of the renormalon ambiguity of the pole mass. To be more specific we consider 
\begin{align}\label{msbardefine}
\mbar_Q \equiv \mbar_Q^{(\nq+1)}(\mbar_Q^{(\nq+1)})\,,
\end{align}
which is the $\MSb$ mass defined for $(\nq+1)$ active dynamical flavors, where 
\begin{align}
\nq\, \equiv \,\mbox{number of flavors lighter than quark $Q$}\,.
\end{align}
{\it In this work we use these two definitions for all massive quarks, and depending on the context we also use the lower case letter $q$ for massive quarks.} We also define
\begin{align}
\nl\, \equiv \,\mbox{number of flavors lighter than $\Lambda_{\rm QCD}$}\,,
\end{align}
which we strictly treat in the massless approximation. 

Assuming that $q_1,\dots,q_n$ are the massive quarks lighter than $Q$  {\it in the order of decreasing mass} (i.e.\ $m_Q > m_{q_1} > \ldots > m_{q_n}> \LQCD$ with $n < n_Q$ and $\nl=\nq-n$), the pole-$\MSb$ mass relation for the heavy quark $Q$ can be written in the form
\begin{align}\label{eqn:mpoleMSbar}
 \mpole_Q &= \mbar_Q + \mbar_Q\,\sum_{n=1}^\infty\,a_n(\nq+1,0)\,\left(\frac{\alpha_s^{(\nq+1)}(\mbar_Q)}{4\pi}\right)^n \\[2mm]
 &+\mbar_Q\left[\delbar_Q^{(Q,q_1,\dots,q_n)}(1,r_{q_1Q},\dots,r_{q_nQ}) + \delbar_Q^{(q_1,\dots,q_n)}(r_{q_1Q},\dots,r_{q_nQ}) + \dots + \delbar_Q^{(q_n)}(r_{q_nQ})\right]\nonumber\,,
\end{align}
with
\begin{align}\label{eqn:coeffanmsbar}
 a_1(\nq,n_h) &= {\textstyle \frac{16}{3}}\,,\\
 a_2(\nq,n_h) &= 213.437 + 1.65707\, n_h - 16.6619\, \nq\,,\nonumber\\
 a_3(\nq,n_h) &= 12075. + 118.986\, n_h + 4.10115\, n_h^2 - 1707.35\, \nq + 1.42358\, n_h\, \nq + 41.7722\, \nq^2\,,\nonumber\\
 a_4(\nq,n_h) &= (911588.\pm 417.) + (1781.61\pm 30.72)\,n_h - (60.1637\pm 0.6912)\,n_h^2 \nonumber\\ &\quad- (231.201\pm 0.102)\,n_h\,\nq - (190683.\pm10.)\,\nq + 9.25995\,n_h^2\,\nq \nonumber\\&\quad + 6.35819\,n_h^3 + 4.40363\,n_h\,\nq^2   + 11105.\,\nq^2 - 
 173.604\,\nq^3\,,\nonumber
\end{align}
where $\alpha_s^{(\nq+1)}$ is the strong coupling that evolves with $(\nq+1)$ active dynamical flavors, see Eq.~\eqref{eqn:betafct}.

The coefficients $a_n(\nq,n_h)$ encode the QCD corrections to $\mpole_Q - \mbar_Q$ for the case that the $\nq$ quarks lighter than $Q$ are assumed to be massless, and $n_h=1$ is just an identifier for the corrections coming from virtual loops of the quark $Q$. The coefficients $a_{1,2,3}$ are known analytically from Refs.~\cite{Tarrach:1980up, Gray:1990yh,Chetyrkin:1999ys,Chetyrkin:1999qi,Melnikov:2000qh,Marquard:2007uj}, and $a_4$ was determined numerically in Refs.~\cite{Marquard:2015qpa,Marquard:2016dcn}, where the quoted numerical uncertainties have been taken from Ref.~\cite{Marquard:2016dcn}.
In Ref.~\cite{Kataev:2015gvt} an approach was suggested to further reduce the uncertainties of the $\nq$-dependent terms. The numerical uncertainties of the coefficient $a_4$ are, however, tiny and irrelevant for the analysis carried out in this work. We quote them just for completeness throughout this work.

The terms $\delbar_Q^{(q,q^\prime,\dots)}(r_{qQ},r_{q^\prime Q},\dots)$ contain the mass corrections coming from the quark $Q$ on-shell self-energy Feynman diagrams with insertions of virtual massive quark loops. We remind the reader that the quarks with mass below the hadronization scale are taken as massless and do not contribute. The superscript $(q,q^\prime,\dots)$ indicates that {\it each diagram contains at least one insertion of the massive quark $q$} and in addition all possible insertions of the (lighter) massive quarks $q^\prime,\dots$ as well as of massless quark and gluonic loops. From each diagram the corresponding diagram with all the quark loops in the massless limit is subtracted in the scheme compatible with the flavor number scheme for the strong coupling $\alpha_s$. The fraction
\begin{align}\label{rdefine}
r_{qq^\prime} \,\equiv \,\frac{\mbar_q}{\mbar_{q^\prime}} \,,
\end{align}
stands for the ratio of $\MSb$ masses for massive quarks $q$ and $q^\prime$ as defined in Eq.~\eqref{msbardefine}. In the pole-$\MSb$ mass relation for the heavy quark $Q$ only mass ratios with respect to the heavy quark mass $\mbar_Q$ arise. 
By construction, the sum of all virtual quark mass corrections contained in the functions $\delbar_Q^{(q,q^\prime,\dots)}(r_{qQ},r_{q^\prime Q},\dots)$ are RG-invariant and do not contain effects from quarks heavier than the external quark $Q$. The effects on the mass of the quark $Q$ related to quarks heavier than $Q$ are accounted for in the renormalization group evolution of the $\MSb$ mass $\mbar_Q(\mu)$ for scales $\mu>m_Q$ and are not considered here.
The virtual quark mass corrections satisfy the following two relations to all orders of perturbation theory
\begin{align}
 \delbar_Q^{(q_1,q_2,\dots,q_n)}(0,0,\dots,0) &= 0 \, ,\label{eqn:Delta1}\\
 \delbar_Q^{(Q,q_1,\dots,q_n)}(1,0,\dots,0) &= \sum_{n=2}^\infty\,\left[\,a_n(\nq,1) - a_n(\nq+1,0)\,\right]\left(\frac{\alpha_s^{(\nq+1)}(\mbar_Q)}{4\pi}\right)^n \label{eqn:Delta2} \,.
\end{align}
Due to Eq.~\eqref{eqn:Delta2} the pole-$\MSb$ mass relation of Eq.~\eqref{eqn:mpoleMSbar} can be rewritten in the alternative form
\begin{align}\label{eqn:mpoleMSbarv2}
 \mpole_Q &= \mbar_Q + \mbar_Q\,\sum_{n=1}^\infty\,a_n(\nq,1)\,\left(\frac{\alpha_s^{(\nq+1)}(\mbar_Q)}{4\pi}\right)^n \nonumber\\
 &\qquad+\mbar_Q\left[\delbar_Q^{(Q,q_1,\dots,q_n)}(1,r_{q_1Q},\dots,r_{q_nQ}) - \delbar_Q^{(Q,q_1,\dots,q_n)}(1,0,\dots,0) \right.\\[2mm]
 &\hspace{2.5cm}\left.+\; \delbar_Q^{(q_1,\dots,q_n)}(r_{q_1Q},\dots,r_{q_nQ}) + \dots + \delbar_Q^{(q_n)}(r_{q_nQ})\right]\nonumber\,.
\end{align}
In the limit that all quarks lighter than $Q$ are massless, all $\delbar$ terms cancel or vanish in Eq.~\eqref{eqn:mpoleMSbarv2}, and only  
the first line involving the $a_n$ coefficients remains. 

The perturbative expansion of the virtual quark mass corrections in the pole-$\MSb$ mass relation of Eq.~\eqref{eqn:mpoleMSbar} and \eqref{eqn:mpoleMSbarv2} can be written in the form
\begin{align}\label{eqn:Deltadef}
 \delbar_Q^{(q,q^\prime,\dots)}(r_{qQ},r_{q^\prime Q},\dots) &=
 \delta_2(r_{qQ})\left(\frac{\alpha_s^{(\nq+1)}(\mbar_Q)}{4\pi}\right)^2\, \nonumber\\ &\qquad+\,   \sum_{n=3}^\infty\,\delta_{Q,n}^{(q,q^\prime,\dots)}(r_{qQ},r_{q^\prime Q},\dots)\left(\frac{\alpha_s^{(\nq+1)}(\mbar_Q)}{4\pi}\right)^n\,,
\end{align}
which together with Eq.~\eqref{eqn:Delta2} implies that
\begin{align}
& \delta_2(1)  = a_2(\nq,1) - a_2(\nq+1,0)= 18.3189 \,, \nonumber\\
& \delta_{Q,n}^{(Q,q,q^\prime,\dots)}(1,0,0,\dots)  = a_n(\nq,1) - a_n(\nq+1,0)\,.
\label{eqn:Delta2an}
\end{align}
The ${\cal O}(\alpha_s^2)$ correction comes from the on-shell self energy diagram of quark $Q$ with the insertion of a loop of the massive quark $q$. The result was determined analytically in Ref.~\cite{Gray:1990yh}. At ${\cal O}(\alpha_s^3)$, in Ref.~\cite{Bekavac:2007tk}, the virtual quark mass corrections were determined in a semi-analytic form for arbitrary quark masses for insertions of loops of the quark $Q$ and one other massive quark $q$. The expressions for these virtual quark mass corrections are for convenience collected in App.~\ref{app:Delta} after adapting the results of Ref.~\cite{Bekavac:2007tk} to our notation. We also provide the ${\cal O}(\alpha_s^3)$ result for insertions of loops with two arbitrary massive quarks, which were not given in Ref.~\cite{Bekavac:2007tk}. The ${\cal O}(\alpha_s^4)$ virtual quark mass corrections have not been determined through an explicit loop calculation.

\begin{figure}
	\center
	\includegraphics[scale=.3]{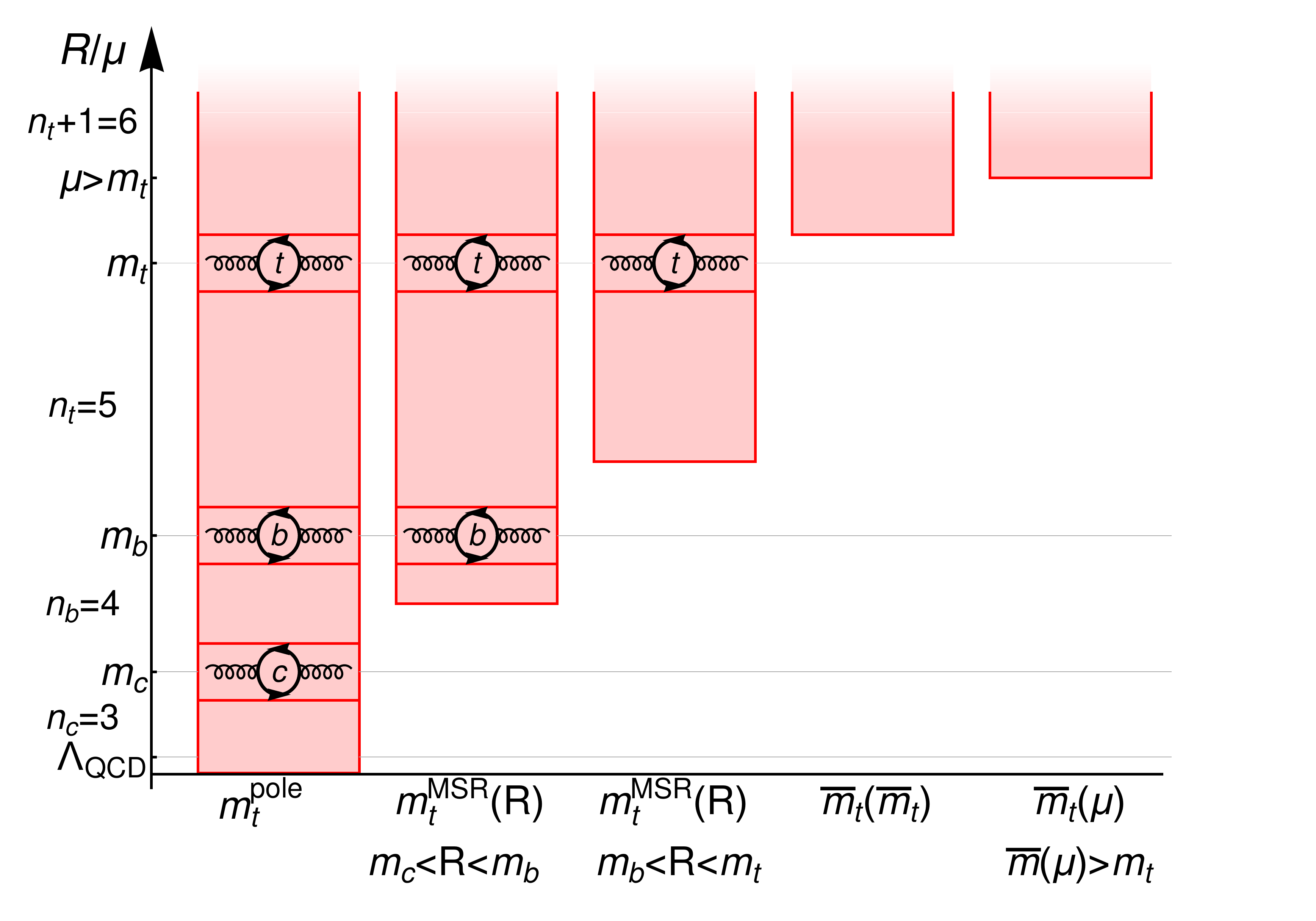}
	\caption{Graphical illustration of the physical contributions contained in the pole, MSR and $\MSb$ mass schemes coming from the different momentum scales for the case of the top quark. The quark loops stand for the contributions of the virtual massive quark loops contained in the masses. \label{fig:massschemes}
	}
\end{figure}

One can interpret the $\MSb$ mass $\mbar_Q= \mbar_Q^{(\nq+1)}(\mbar_Q^{(\nq+1)})$ as the pole mass {\it minus} all self-energy corrections coming from scales {\it at and below  $\mbar_Q$}. So $\mbar_Q$ only contains mass contributions from momentum fluctuations from above $\mbar_Q$, which illustrates that it is a short-distance mass that is strictly insensitive to issues related to low momentum fluctuations at the hadronization scale $\Lambda_{\rm QCD}$. See Fig.~\ref{fig:massschemes} for illustration.

\subsection{MSR Mass and R-Evolution}
\label{sec:MSRmass}

In order to integrate out high momentum contributions and formulate the renormalization group flow of momentum contributions in the heavy quark masses we use the MSR mass $\msr_Q(R)$ introduced in Ref.~\cite{Hoang:2017suc}\footnote{In Ref.~\cite{Hoang:2017suc,Hoang:2008yj} the natural and the practical MSR masses were introduced. In this paper we employ the natural MSR mass and call it just the MSR mass for convenience.}, extending its definition to account for the mass effects of the lighter massive quarks.

The MSR mass for the heavy quark $Q$ is derived from on-shell self-energy diagrams just like the pole-$\MSb$ mass relation of Eq.~\eqref{eqn:mpoleMSbar}, but it does not include any diagrams involving virtual loops of the heavy quark $Q$, i.e.\ the contributions from heavy quark $Q$ virtual loops are integrated out. Like the $\MSb$ mass, the MSR mass is a short-distance mass, and since the corrections from the heavy quark $Q$ are short-distance effects, its relation to the pole mass fully contains the pole mass ${\cal O}(\Lambda_{\rm QCD})$ renormalon (just as the pole-$\MSb$ mass relation of Eqs.~\eqref{eqn:mpoleMSbar} and \eqref{eqn:mpoleMSbarv2}). Furthermore the MSR mass depends on the arbitrary scale $R\lesssim m_Q$ to describe contributions in the mass from the momenta below the scale $m_Q$, and therefore represents the natural extension of the concept of the $\MSb$ mass for scales below $m_Q$.

Assuming that $q_1,\dots,q_n$ are the massive quarks lighter than $Q$ in the order of decreasing mass (i.e.\ $m_Q > m_{q_1} > \ldots > m_{q_n}> \LQCD$ with $n < n_Q$ and $\nl=\nq-n$), the MSR mass $\msr_Q(R)$ is \textit{defined} by the relation
\begin{align}\label{eqn:mpoleMSR}
 \mpole_Q &= \msr_Q(R) + R\,\sum_{n=1}^\infty\,a_n(\nq,0)\left(\frac{\alpha^{(\nq)}_s(R)}{4\pi}\right)^n \nonumber\\
 &\qquad+ \mbar_Q\,\left[\delta_Q^{(q_1,\dots,q_n)}(r_{q_1Q},\dots,r_{q_nQ}) + \dots + \delta_Q^{(q_n)}(r_{q_nQ})\right]\,,
\end{align}
where the coefficients $a_n$ are given in Eqs.~\eqref{eqn:coeffanmsbar} and the perturbative expansion is in powers of the strong coupling in the $n_Q$-flavor scheme since the quark $Q$ is integrated out. The $R$-dependence of the strong coupling entails that the scale $R$ has to be chosen sufficiently larger than $\Lambda_{\rm QCD}$ to stay away from the Landau pole. The definition generalizes the one already provided in Ref.~\cite{Hoang:2017suc}, which only considered $\nq$ massless quarks.

The notation used for the virtual quark mass corrections involving the functions $\delta_{Q}^{(q,q^\prime,\dots)}(r_{qQ},r_{q^\prime Q},\dots)$ is the same as the one for the $\MSb$ mass described above, and their sum is by construction RG-invariant. Their perturbative expansion has the 
form
\begin{align}\label{eqn:Deltadef2}
 \delta_Q^{(q,q^\prime,\dots)}(r_{qQ},r_{q^\prime Q},\dots) &=
 \delta_2(r_{qQ})\left(\frac{\alpha_s^{(\nq)}(\mbar_Q)}{4\pi}\right)^2 \nonumber\\ &\qquad+ \sum_{n=3}^\infty\,\delta_{Q,n}^{(q,q^\prime,\dots)}(r_{qQ},r_{q^\prime Q},\dots)\left(\frac{\alpha_s^{(\nq)}(\mbar_Q)}{4\pi}\right)^n\,,
\end{align}
where the coefficient functions $\delta_{Q,n}^{(q,q^\prime,\dots)}(r_{qQ},r_{q^\prime Q},\dots)$ are identical to the ones appearing in Eq.~\eqref{eqn:Deltadef}. 

In our definition of the MSR mass, the virtual quark mass corrections are independent of $R$.
This entails that the renormalization group evolution of the MSR mass in $R$ does not depend on the masses of the $n_Q$ lighter quarks. So  $\msr_Q(R)$ is defined in close analogy to the $\mu$-dependent $\MSb$ strong coupling and the $\MSb$ masses, whose renormalization group evolution only depends on the number of active dynamical quarks (which is typically the number of quarks lighter than $\mu$) and where mass effects are implemented by threshold corrections when $\mu$ crosses a flavor threshold.
Moreover, because the ${\cal O}(\LQCD)$ renormalon ambiguity of the series proportional to $R$ is independent of $R$ and because the corrections from the virtual loops of the heavy quark $Q$ are short-distance effects, the series of the pole-MSR mass relation in Eq.~\eqref{eqn:mpoleMSR} suffers from the same ${\cal O}(\LQCD)$ renormalon ambiguity as the pole-$\MSb$ mass relation of Eqs.~\eqref{eqn:mpoleMSbar} and \eqref{eqn:mpoleMSbarv2}. It can therefore also be used to study and quantify the ${\cal O}(\LQCD)$ renormalon of the pole mass $m_Q^{\rm pole}$.

As explained below Eq.~\eqref{eqn:polemsbarseriesLL}, in order to expand the difference of MSR masses at two scales $R$ and $R^\prime$
in the fixed-order expansion in powers of $\alpha_s^{(\nq)}$ it is necessary to do that at a common renormalization scale $\mu$ so that the renormalon in the $R$-dependent corrections of Eq.~\eqref{eqn:mpoleMSR} cancels order by order. This unavoidably leads to large logarithms if the scale separation is large, similarly to when considering the fixed-order expansion of the difference of the strong coupling at widely separated scales. To sum the logarithms in the difference of MSR masses we use its RG-evolution equation in $R$, which reads
\begin{equation}\label{eqn:revolvdef}
 R\frac{\dd}{\dd R}m_Q^{\MSR}(R)=-\,R\,\gamma^{R,(n_Q)}(\alpha^{(\nq)}_s(R))
 =-\,R\sum_{n=0}^\infty\gamma_n^{R,(n_Q)}\bigg(\frac{\alpha^{(\nq)}_s(R)}{4\pi}\bigg)^{\!\!n+1}\;,
\end{equation}
where the coefficients are known up to four loops and given by \cite{Hoang:2008yj,Hoang:2017suc}
\begin{align}\label{eqn:gammadef}
\gamma_0^{R,(n_Q)} & = {\textstyle \frac{16}{3}}\,,\\
\gamma_1^{R,(n_Q)} & = 96.1039 - 9.55076\, \nq\,,\nonumber\\
\gamma_2^{R,(n_Q)} & = 1595.75 - 269.953\, \nq - 2.65945\, \nq^2\,,\nonumber\\
\gamma_3^{R,(n_Q)} & = (12319.\pm417.) - (9103.\pm10.)\, \nq + 610.264\, \nq^2 - 6.515\, \nq^3\,.\nonumber
\end{align} 
The difference of MSR masses at two scales $R^\prime$ and $R$ can then be computed from solving the evolution equation
\begin{equation}\label{eqn:rrge}
\Delta m^{(n_Q)}(R,R^\prime) = \msr_Q(R^\prime) - \msr_Q(R) = \sum_{n=0}^\infty \,\gamma_n^{R,(n_Q)} \int_{R^\prime}^R \dd R\,\left(\frac{\alpha^{(\nq)}_s(R)}{4\pi}\right)^{n+1}\,,
\end{equation}
which accounts for the RG-evolution in the presence of $\nq$ active dynamical quark flavors.

\begin{figure}
	\center
	\includegraphics[scale=0.275]{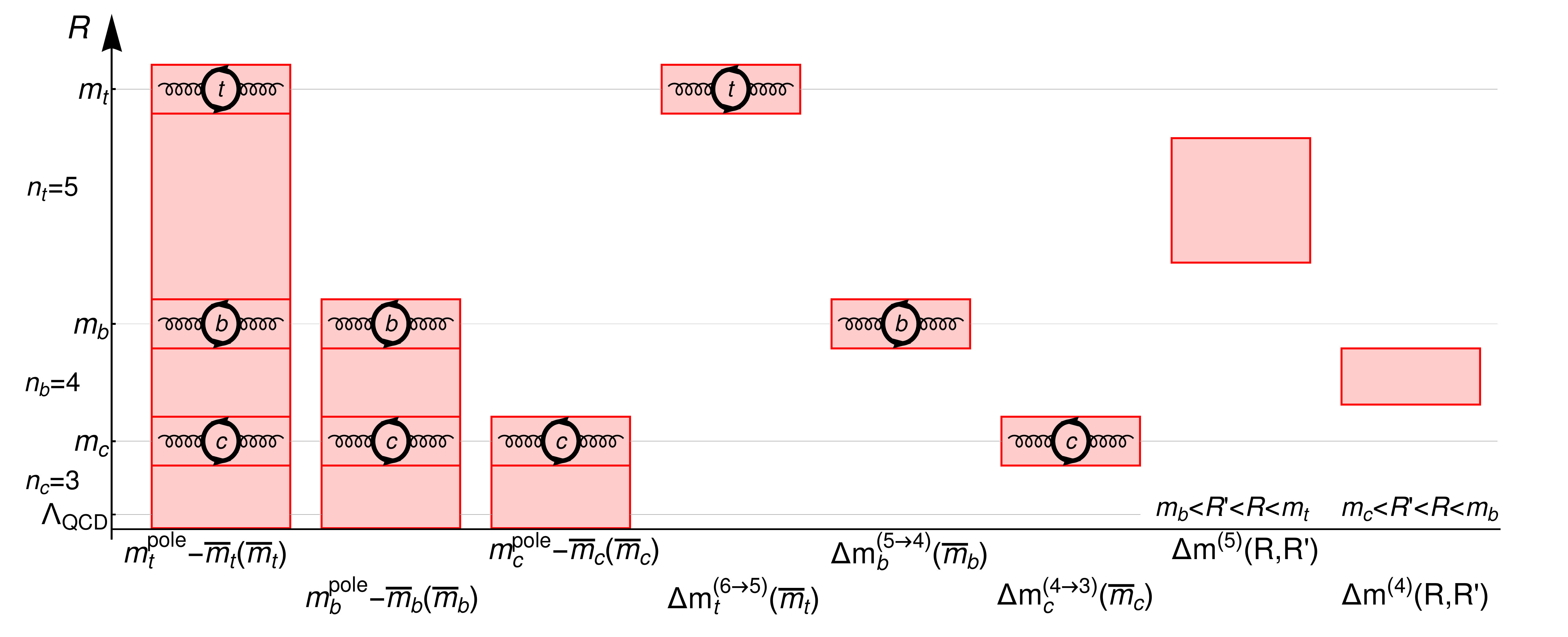}
	\caption{Graphical illustration for pole-$\MSb$ mass differences, the MSR-$\MSb$ mass matching corrections and MSR mass differences for different $R$ scales. They constitute the major contributions in the RG analysis of the heavy quark pole masses. \label{fig:matchrun}
	}
\end{figure}

The RG-equation of the MSR mass has a linear as well as logarithmic dependence on $R$ and thus differs from the usual logarithmic RG-equations for $\alpha_s$ and the $\MSb$ mass. Since its linear dependence on $R$ allows to systematically probe linear sensitivity to small momenta it can be used to systematically study the ${\cal O}(\LQCD)$ renormalon behavior of perturbative series~\cite{Hoang:2008yj,Hoang:2017suc}. Since this is impossible for usual logarithmic RG-evolution equations, Eq.~\eqref{eqn:revolvdef} was called the {\it R-evolution equation} in Refs.~\cite{Hoang:2008yj,Hoang:2017suc}.
Continuing on the thoughts made at the end of Sec.~\ref{sec:MSbarmass} we note that one can interpret the MSR mass $\msr_Q(R)$ as the pole mass {\it minus} all self-energy contributions coming from scales {\it below} $R$ and all virtual quark mass corrections from quarks lighter than $Q$, see Fig.~\ref{fig:massschemes}. This also illustrates that the MSR mass $\msr_Q(R)$ is a short-distance mass. The negative overall sign on the RHS of Eq.~\eqref{eqn:revolvdef} expresses that self-energy contributions are added to the MSR mass when $R$ is evolved to smaller scales, and that $\Delta m^{(n_Q)}(R,R^\prime)$ for $R>R^\prime$ is positive and represents the self-energy contributions to the mass in the presence of $n_Q$ active dynamical flavors coming from the scales between $R^\prime$ and $R$. This is illustrated in Fig.~\ref{fig:matchrun}.

In the context of the analyses in this work the essential property is that the ${\cal O}(\LQCD)$ renormalon ambiguity in the series on the RHS of Eq.~\eqref{eqn:mpoleMSR} is $R$-independent. This entails that the R-evolution equation is free of the ${\cal O}(\LQCD)$ renormalon, and solving the R-evolution equation in Eq.~\eqref{eqn:rrge} allows to relate MSR masses at different scales in a way that is renormalon free and, in addition, systematically sums logarithms $\ln(R/R^\prime)$ to all orders in a way free of the ${\cal O}(\LQCD)$ renormalon. So the R-evolution equation resolves the problem of the large logarithms that arise when computing MSR mass differences in the fixed-order expansion. The integral of Eq.~\eqref{eqn:rrge} can be readily computed numerically, and an analytic solution has been discussed in detail in~\cite{Hoang:2017suc}. The analytic solution also allows to derive the large-order asymptotic form of the perturbative coefficients $a_n$. To implement renormalization scale variation in Eq.~\eqref{eqn:rrge} one expands $\alpha_s^{(\nq)}(R)$ as a series in $\alpha_s^{(\nq)}(\lambda R)$, and by varying $\lambda$ in some interval around unity. We note that in our analysis we consider the top, bottom and charm mass scales, and using the R-evolution equation is instrumental for our discussion of the top quark pole mass.

In Tab.~\ref{tab:DeltaMRRprime} we show numerical results for various MSR mass differences $\Delta m^{(n_Q)}$ relevant in our examinations below for $\nq=3,4,5$. We display the results obtained from using the R-evolution equation at ${\cal O}(\alpha_s^n)$ for $n=1,2,3,4$. The uncertainties are from $\lambda$ variations in the interval $[0.5,2]$ for the cases where scales above the charm mass scale $1.3$~GeV are considered, and in the interval $[0.6,2.5]$ for cases which involve the charm mass scale. We see an excellent convergence and stability of the results and a significant reduction of scale variation with the order, illustrating that the mass differences $\Delta m^{(n_Q)}(R,R^\prime)$ are free of an ${\cal O}(\Lambda_{\rm QCD})$ renormalon ambiguity. For our analyses below we use the most precise ${\cal O}(\alpha_s^4)$ results shown in the respective lowest lines.

\begin{table}
\renewcommand{\arraystretch}{1.2}
\newcolumntype{A}{>{\centering} m{1.5cm} }
\newcolumntype{B}{>{\centering\arraybackslash} m{3.95cm} }
\centering
\begin{tabular}{|A|B|B|B|}
	\hline
	${\cal O}(\alpha_s^n)$ & $\Delta m^{(5)}(163,20)$ & $\Delta m^{(5)}(163,4.2)$ & $\Delta m^{(5)}(163,1.3)$  \\\hline
	$n=1$ & $7.358 \pm 0.811$ & $8.536 \pm 1.008$ & $8.864 \pm 1.047$  \\
	$n=2$ & $8.007 \pm 0.168$ & $9.336 \pm 0.225$ & $9.728 \pm 0.311$  \\
	$n=3$ & $8.031 \pm 0.024$ & $9.368 \pm 0.035$ & $9.764 \pm 0.066$  \\
	$n=4$ & $8.006 \pm 0.009$ & $9.331 \pm 0.016$ & $9.716 \pm 0.023$  \\\hline
	${\cal O}(\alpha_s^n)$ & $\Delta m^{(4)}(163,4.2)$ & $\Delta m^{(4)}(20,4.2)$ & $\Delta m^{(4)}(4.2,1.3)$ \\\hline
	$n=1$ & $8.181 \pm 1.026$ & $1.153 \pm 0.211$ &   $0.337 \pm 0.098$ \\
	$n=2$ & $9.064 \pm 0.270$ & $1.326 \pm 0.073$ &  $0.419 \pm 0.063$  \\
	$n=3$ & $9.139 \pm 0.054$ & $1.346 \pm 0.018$ &  $0.434 \pm 0.026$ \\
	$n=4$ & $9.114 \pm 0.014$ & $1.337 \pm 0.007$ &  $0.423 \pm 0.017$   \\\hline
	${\cal O}(\alpha_s^n)$  & $\Delta m^{(3)}(163,1.3)$ & $\Delta m^{(3)}(20,1.3)$ & $\Delta m^{(3)}(4.2,1.3)$  \\\hline
	$n=1$  & $8.009 \pm 1.044$ & $1.419 \pm 0.296$ & $0.328 \pm 0.106$  \\
	$n=2$  & $9.008 \pm 0.404$ & $1.691 \pm 0.166$ & $0.418 \pm 0.078$  \\
	$n=3$  & $9.130 \pm 0.126$ & $1.741 \pm 0.067$ & $0.440 \pm 0.037$  \\
	$n=4$  & $9.111 \pm 0.032$ & $1.729 \pm 0.023$ & $0.434 \pm 0.020$  \\\hline
\end{tabular}
\caption{\label{tab:DeltaMRRprime} MSR mass differences $\Delta m^{(n_Q)}(R,R^\prime)$ computed from R-evolution, for $\nq=3,4,5$ active dynamical flavors for scale differences involving top, bottom and charm masses and the scale $20$~GeV. The central values are obtained for $\lambda=1$ and the uncertainties are symmetrized $\lambda$ variations in the interval $[0.5,2]$. For entries involving the scale $\mbar_c$ the interval $[0.6,2.5]$ is used for $\lambda$ variations. The numbers for $\Delta m^{(n_Q)}(R,R^\prime)$ are given in units of GeV.}
\end{table}

\subsection{Asymptotic High Order Behavior and Borel  Transform for Massless Lighter Quarks}
\label{sec:asym}

In this section we review a number of known results relevant for the analyses in the subsequent parts of the paper. The results are already known since Refs.~\cite{Bigi:1994em,Beneke:1994sw,Beneke:1998ui}. We adapt them according to our notation and present updated numerical results accounting for the recent perturbative calculations of the pole-$\MSb$ mass relation and the QCD $\beta$-function.

The Borel transform of an $\alpha_s$ power series
\begin{equation}
 f(\alpha_s(R)) =  
 R\,\sum_{n=0}^\infty\,a_{n+1}\left(\frac{\alpha_s(R)}{4\pi}\right)^{n+1}\,,
\end{equation}
is defined as
\begin{equation}\label{eqn:borel1}
 B[f](u) = R\,\sum_{n=0}^\infty\,a_{n+1} \,\frac{u^n}{n!\, \beta_0^{n+1}} \,,
\end{equation}
where $\beta_0$ is the one-loop $\beta$-function coefficient in the flavor number scheme of $\alpha_s$. For the approximation that all quarks lighter than the heavy quark $Q$ are massless (i.e.\ $\nl=\nq$) the Borel transform of the series for the pole-MSR mass reads
\begin{align}\label{eqn:borel2}
 B&\left[\mpole_Q-\msr_Q(R)\right](u) \,= \nonumber\\
 &\qquad N_{1/2}^{(\nl)} \;R\,\frac{4\pi}{\betazeroell}\,\sum_{k=0}^\infty\,g_k^{(\nl)}\frac{\Gamma(1+\hat{b}_1^{(\nl)}-k)}{\Gamma(1+\hat{b}_1^{(\nl)})}\,(1-2\,u)^{-1-\hat{b}_1^{(\nl)}+k} + \dots \,,
\end{align}
where the non-analytic (and singular) terms multiplied by the normalization factor $N_{1/2}^{(\nl)}$ single out the ${\cal O}(\LQCD)$ renormalon behavior of the pole-MSR mass series and the ellipses stand for contributions not affected by an ${\cal O}(\LQCD)$ renormalon. Their form is unambiguously determined by the coefficients $\beta_n^{(\nl)}$ of the QCD $\beta$-function in Eq.~\eqref{eqn:betafct}, and the sum over $k$ parametrizes the subleading effects due to the higher order coefficients of the QCD $\beta$-function. 
The coefficients $g^{(\nl)}_k$ can be determined from the recursion formulae~\cite{Hoang:2017suc}
 \begin{align}
\hat b_{n+1} &= 2\sum_{i\,=\,0}^n\, \frac{\hat b_{n-i}\,\beta_{i+1}}{(-2\beta_0)^{i+2}}\,,\nonumber \\	
g_{n+1} &= \frac{1}{1+n}\sum_{i=0}^n\,(-1)^i\,\hat b_{i+2}\,g_{n-i}
\end{align}
with $\hat b_{0}=g_0=1$, where we dropped the superscript $(\nl)$ for simplicity. Currently, coefficients $g^{(\nl)}_k$ are known up to $k=3$.
The factor $N_{1/2}^{(\nl)}$ precisely quantifies the overall normalization of the ${\cal O}(\LQCD)$ renormalon behavior and can be determined quite precisely from the coefficients $a_n(\nl,0)$ known from explicit computations. Accounting for the coefficients up to ${\cal O}(\alpha_s^4)$ the normalization was determined with very small errors for the relevant flavor numbers $\nl = 3,4,5$ in Refs.~\cite{Ayala:2014yxa,Beneke:2016cbu,Hoang:2017suc}, all of which are in agreement. We use the results from Ref.~\cite{Hoang:2017suc}:
\begin{align}\label{eqn:N12msr}
 N_{1/2}^{(\nl=3)} = 0.526\pm0.012 \,,\nonumber\\
 N_{1/2}^{(\nl=4)} = 0.492\pm0.016 \,,\\
 N_{1/2}^{(\nl=5)} = 0.446\pm0.024 \,.\nonumber
\end{align}
The uncertainties are not essential for the outcome of our analysis and quoted for completeness. Their small size reflects that the large-order asymptotic behavior of the series is known very precisely. 

The inverse Borel transform
\begin{equation}
 \int_0^\infty\mathrm du\, B[f](u)\;\mathrm e^{-\frac{4\pi u}{\beta_0\alpha_s(R)}}\, ,
\end{equation}
has the same $\alpha_s$ power series as the original series $f(\alpha_s(R))$ and provides the exact result if it can be calculated unambiguously from the Borel transform $B[f](u)$. However, for the case of Eq.~\eqref{eqn:borel2}, due to the singularity at $u=1/2$ and the cut along the positive real axis for $u>1/2$, the integral cannot be computed without further prescription and an ambiguity remains. Using an $i\epsilon$ prescription $(1-2u)^\alpha\to(1-2u-i\epsilon)^\alpha$ to shift the cut to the lower complex half plane, the resulting imaginary part of the integral is
\begin{align}\label{eqn:borelambiguity}
 \Delta m_\mathrm{Borel}^{(\nl)}&\equiv \left|\mathrm{Im}\int_0^\infty\mathrm du\,\exp\left(-\frac{4\pi u}{\beta_0^{(\nl)}\alpha_s^{(\nl)}(R)}\right) \right . \nonumber\\ & \hspace{2.3cm}\left . \times \left[N_{1/2}^{(\nl)}\, R\, \frac{4\pi}{\beta_0^{(\nl)}}\sum_{k=0}^\infty g_k^{(\nl)}\frac{\Gamma(1+\hat b_1^{(\nl)}-k)}{\Gamma(1+\hat b_1^{(\nl)})}\,(1-2u)^{-1-\hat b_1^{(\nl)}+k}\right] \right|\nonumber\\
 &=N_{1/2}^{(\nl)}\,\frac{2\pi^2}{\beta_0^{(\nl)}\Gamma(1+\hat b_1^{(\nl)})}\Lambda_\mathrm{QCD}^{(\nl)}\,,
\end{align} 
and represents a quantification of the ambiguity of the pole mass, where $\Lambda_\mathrm{QCD}^{(\nl)}$ is 
given by the expression ($t_R=-2\pi/\beta_0^{(\nl)}\alpha_s^{(\nl)}(R)$)
\begin{equation}\label{eqn:lambda}
\Lambda_\mathrm{QCD}^{(\nl)}=R\,\exp\left(t_R+\hat b_1^{(\nl)}\log(-t_R)-\sum_{k=2}^\infty\frac{\hat b_k^{(\nl)}}{(k-1)t_R^{k-1}}\right).
\end{equation}
In this work we use this expression as the definition of $\Lambda_\mathrm{QCD}$ for $\nl$ massless flavors. The RHS is $R$-independent, and truncating at $k=4$ provides the results
\begin{align}
 \Lambda_{\rm QCD}^{(\nl=3)}&=253\text{ MeV}\,,\nonumber \\
 \label{eqn:lambdanum}
 \Lambda_{\rm QCD}^{(\nl=4)}&=225\text{ MeV}\,,\\
 \Lambda_{\rm QCD}^{(\nl=5)}&=166\text{ MeV}\,,\nonumber
\end{align}
with uncertainties below $0.5$~MeV. $\Lambda_\mathrm{QCD}^{(\nl)}$ increases for smaller flavor numbers $\nl$ since the scale-dependence of $\alpha_s$, and thus also the infrared sensitivity of QCD quantities, increases with $\nl$. The expressions for 
$\Delta m_\mathrm{Borel}^{(\nl)}$ for the size of the imaginary part of the inverse Borel transform in Eq.~\eqref{eqn:borelambiguity}
provide a parametric estimate for the ambiguity of the pole mass.
Using Eqs.~\eqref{eqn:N12msr} and \eqref{eqn:lambdanum} they give
$ \Delta m_\mathrm{Borel}^{(3,4,5)}=(329\pm 8,295\pm 10,213\pm 11)$~MeV
which are around a factor $1.3$ larger than the corresponding values for $\Lambda_{\rm QCD}^{(\nl)}$.

From the expression for the Borel transform given in Eq.~\eqref{eqn:borel2} one can derive the large order asymptotic form of the perturbative coefficients $a_n$ of the pole-MSR mass series (which describe the case that all quarks lighter than $Q$ are massless, i.e.\ $\nq=\nl$):
\begin{equation}\label{eqn:asy1}
 a_n^\mathrm{asy}(\nl,n_h)= a_n^\mathrm{asy}(\nl,0)=4\pi N_{1/2}^{(\nl)}(2\beta_0^{(\nl)})^{n-1}\sum_{k=0}^\infty g_k^{(\nl)}\frac{\Gamma(n+\hat b_1^{(\nl)}-k)}{\Gamma(1+\hat b_1^{(\nl)})}\,,
\end{equation}
where the value of $n_h$ is insignificant because the virtual effects of quark $Q$ do not affect the large order asymptotic behavior.
The sum in $k$ is convergent, and
truncating at $k=3$ one can use the results for $n>4$ as an approximation for the yet uncalculated series coefficients. The results up to $n=12$ for $\nl=3, 4, 5$ using the values for the $N_{1/2}^{(\nl)}$ from Eq.~\eqref{eqn:N12msr} are displayed in Tab.~\ref{tab:aasy}.

With the normalization factors $N_{1/2}^{(\nl)}$, which are known to a precision of a few percent and which also entails the same precision for $\Delta m_\mathrm{Borel}^{(\nl)}$ and the asymptotic coefficients $a_n^{\rm asy}$, the series for the pole-MSR and also for the pole-$\MSb$ mass relation are essentially known to all orders for the case of $\nl=\nq$. The task to determine the ambiguity of the pole mass involves to specify how this precisely known pattern limits the principle capability to determine the pole mass numerically, see the discussion in Sec.~\ref{sec:method}. In other words, the ambiguity of the pole mass is known to be proportional to $\Delta m_\mathrm{Borel}^{(\nl)}$ or $\Lambda_{\rm QCD}^{(\nl)}$, but the factor of proportionality has to be determined from an additional dedicated analysis.

\begin{table}
\renewcommand{\arraystretch}{1.2}
\newcolumntype{A}{>{\centering} m{1cm} }
\newcolumntype{B}{>{\centering\arraybackslash} m{4.1cm} }
\centering
\begin{tabular}{|A|B|B|B|}
\hline
 $n$ & $a^\mathrm{asy}_n(\nl=3,0)$ & $a^\mathrm{asy}_n(\nl=4,0)$ & $a^\mathrm{asy}_n(\nl=5,0)$\\\hline
 $5$ & $(3.394\pm 0.077)\times 10^{7\phantom{0}}$ & $(2.249\pm 0.075)\times 10^{7\phantom{0}}$ & $(1.379\pm 0.074)\times 10^{7\phantom{0}}$\\
 $6$ & $(3.309 \pm 0.075)\times 10^{9\phantom{0}}$ & $(2.019 \pm 0.067)\times 10^{9\phantom{0}}$ & $(1.128 \pm 0.060)\times 10^{9\phantom{0}}$\\
 $7$ & $(3.819 \pm 0.087)\times 10^{11}$ & $(2.147 \pm 0.071)\times 10^{11}$ & $(1.095 \pm 0.059)\times 10^{11}$\\
 $8$ & $(5.093 \pm 0.115)\times 10^{13}$ & $(2.641 \pm 0.088)\times 10^{13}$ & $(1.231 \pm 0.066)\times 10^{13}$\\
 $9$ & $(7.706 \pm 0.175)\times 10^{15}$ & $(3.687 \pm 0.123)\times 10^{15}$ & $(1.572 \pm 0.084)\times 10^{15}$\\
 $10$ & $(1.305 \pm 0.030)\times 10^{18}$ & $(5.762 \pm 0.192)\times 10^{17}$ & $(2.250 \pm 0.120)\times 10^{17}$\\
 $11$ & $(2.443 \pm 0.055)\times 10^{20}$ & $(9.964 \pm 0.332)\times 10^{19}$ & $(3.563 \pm 0.191)\times 10^{19}$\\
 $12$ & $(5.014 \pm 0.114)\times 10^{22}$ & $(1.889 \pm 0.063)\times 10^{22}$ & $(6.190 \pm 0.331)\times 10^{21}$\\\hline
		\end{tabular}
 \caption{Coefficients of the pole-MSR mass series for $a_{n>4}(\nl,0)$ for $\nl=3,4,5$ estimated from the asymptotic formula of Eq.~\eqref{eqn:asy1} and with uncertainties from Eq.~\eqref{eqn:N12msr}.}\label{tab:aasy}
\end{table}

\section{Integrating Out Hard Modes from the Heavy Quark Pole Mass}
\label{sec:hardmodes}

\subsection{MSR-\texorpdfstring{$\MSb$}{MS-bar} Mass Matching}
\label{sec:Qintout}

Using the MSR mass we can successively separate off, i.e.\ integrate out, hard momentum contributions from the pole-$\MSb$ mass difference, $\mpole_Q - \mbar_Q$.
We start with the matching relation between the MSR and the $\MSb$ masses at the common scale $\mu=R=\mbar_Q$, which can be obtained by eliminating the pole mass from Eqs.~\eqref{eqn:mpoleMSbarv2} and \eqref{eqn:mpoleMSR}. The matching relation accounts for the virtual top quark loop contributions and can be written in the form
\begin{equation}\label{eqn:MSRMSbmatch}
 \msr_Q(\mbar_Q) - \mbar_Q = \Delta m_Q^{(\nq+1\rightarrow\nq)}(\mbar_Q) + \delta m_{Q,q_1,\dots,q_n}^{(\nq+1\rightarrow\nq)}(\mbar_Q) \,.
\end{equation}
The term $\Delta m_Q^{(\nq+1\rightarrow\nq)}(\mbar_Q)$ contains the virtual top quark loop contributions in the approximation that all $n_Q$ quarks lighter than quark $Q$ are massless and has the form \cite{Hoang:2017suc}
\begin{align}\label{eqn:MSRMSbmatch2}
 \Delta& m_Q^{(\nq+1\rightarrow\nq)}(\mbar_Q) = \\ &\qquad \mbar_Q\left\{1.65707\,\left(\frac{\alpha_s^{(\nq+1)}(\mbar_Q)}{4\pi}\right)^2 + [110.05 + 1.424\,\nq]\,\left(\frac{\alpha_s^{(\nq+1)}(\mbar_Q)}{4\pi}\right)^3 \right. \nonumber\\  & \left. \hspace{1.6cm} + \left[352.\pm31. - (111.59\pm0.10)\,\nq + 4.40\,\nq^2\right]\left(\frac{\alpha_s^{(\nq+1)}(\mbar_Q)}{4\pi}\right)^4 + \dots \right\} \,, \nonumber
\end{align}
where we expressed the series in powers of the strong coupling in the $(\nq+1)$ flavor scheme. The series only contains the hard corrections coming from the virtual heavy quark $Q$ and therefore does not have any ${\cal O}(\Lambda_{\rm QCD})$ ambiguity, see Fig.~\ref{fig:matchrun} for illustration.

In Tab.~\ref{tab:MSRMSbmatch} the numerical values for $\Delta m_Q^{(\nq+1\rightarrow\nq)}(\mbar_Q)$ are shown at ${\cal O}(\alpha_s^{2,3,4})$ for the top, bottom, and charm quarks for ($\mbar_t,\mbar_b,\mbar_c$) = ($163,4.2,1.3$)~GeV. Also shown is the variation due to changes in the renormalization scale in the range $0.5\,\mbar_Q\leq\mu\leq2\,\mbar_Q$, for the top and bottom quark and $0.65\,\mbar_c\leq\mu\leq2.5\,\mbar_c$ for the charm quark. The ${\cal O}(\alpha_s^3)$ corrections are quite sizable compared to the ${\cal O}(\alpha_s^2)$ contributions, but the ${\cal O}(\alpha_s^4)$ corrections are small indicating that the ${\cal O}(\alpha_s^4)$ result and the uncertainty estimate based on the scale variations can be considered reliable. Overall, the matching corrections amount to $32,4$ and $5$~MeV for the top, bottom and charm quarks, respectively with an uncertainty at the level of $1$ to $2$~MeV. The numerical uncertainties of the ${\cal O}(\alpha_s^4)$ coefficients displayed in Eq.~\eqref{eqn:MSRMSbmatch2} are smaller than $0.1$~MeV for all cases and therefore irrelevant for practical purposes.

\begin{table}
\renewcommand{\arraystretch}{1.2}
\newcolumntype{A}{>{\centering} m{1cm} }
\newcolumntype{B}{>{\centering\arraybackslash} m{4.1cm} }
\centering
	\begin{tabular}{|A|B|B|B|}
		\hline
		${\cal O}(\alpha_s^n)$ & $\Delta m_t^{(6\rightarrow 5)} (\mbar_t)$ & $\Delta m_b^{(5\rightarrow 4)} (\mbar_b)$ & $\Delta m_c^{(4\rightarrow 3)} (\mbar_c)$ \\\hline
		$2$ & $ 0.021\pm 0.004$ & $ 0.003 \pm 0.001$ & $0.002 \pm 0.002$ \\
		$3$ & $ 0.033\pm 0.003$ & $ 0.006 \pm 0.002$ & $0.008 \pm 0.005$ \\
		$4$ & $ 0.032\pm 0.001$ & $ 0.004 \pm 0.001$ & $0.005 \pm 0.002$ \\\hline
	\end{tabular}
 \caption{The MSR-$\MSb$ mass matching corrections for the top, bottom and charm quarks for ($\mbar_t,\mbar_b,\mbar_c$) = ($163,4.2,1.3$)~GeV, given in units of GeV. The uncertainties are obtained from variations of the renormalization scale in the range  $0.5\,\mbar_Q\leq\mu\leq2\,\mbar_Q$ for the top and bottom quark and $0.65\,\mbar_c\leq\mu\leq2.5\,\mbar_c$ for the charm quark. The central value is the respective mean of the largest and smallest values obtained in the scale variation.
 	\label{tab:MSRMSbmatch}}
\end{table}

The term $\delta m_{Q,q_1,\dots,q_n}^{(\nq+1\rightarrow\nq)}(\mbar_Q)$ represents the virtual top quark loop contributions arising from the finite masses of the lighter massive quarks $q_1,\dots,q_n$. Since at ${\cal O}(\alpha_s^2)$ only the loop of quark $Q$ can be inserted, the series for $\delta m_{Q,q_1,\dots,q_n}^{(\nq+1\rightarrow\nq)}(\mbar_Q)$ starts at ${\cal O}(\alpha_s^3)$, where only self energy diagrams with one insertion of a loop of quark $Q$ and one insertion of a loop of one of the lighter massive quarks $q_1,\dots,q_n$ can contribute. At ${\cal O}(\alpha_s^3)$ $\delta m_{Q,q_1,\dots,q_n}^{(\nq+1\rightarrow\nq)}(\mbar_Q)$ has the form
\begin{align}\label{eqn:MSRMSbmatch3}
 \delta &m_{Q,q_1,\dots,q_n}^{(\nq+1\rightarrow\nq)}(\mbar_Q) = \mbar_Q\,\Bigg\{\bigg[\delta_{Q,3}^{(Q,q_1,\dots,q_n)}(1,r_{q_1 Q},\dots,r_{q_n Q}) \nonumber\\ &\hspace{5.5cm} -\, \delta_{Q,3}^{(Q,q_1,\dots,q_n)}(1,0,\dots,0)\bigg]\left(\frac{\alpha_s^{(\nq+1)}(\mbar_Q)}{4\pi}\right)^3 + \dots \Bigg\} \nonumber\\
 &\qquad= \mbar_Q\,\left\{\sum_{i=1}^n\bigg[
 14.2222 \,r_{q_i Q}^2 
 - 18.7157 \,r_{q_i Q}^3 + 
\bigg(7.3689 
-11.1477 \ln(r_{q_i Q})\bigg)\, r_{q_i Q}^4 \right .\nonumber\\ & \hspace{3cm} + \left . \dots \bigg]\left(\frac{\alpha_s^{(\nq+1)}(\mbar_Q)}{4\pi}\right)^3 + \dots \right\} \,,
\end{align}
where $r_{q Q} = \mbar_q/\mbar_Q$, and for simplicity we suppress the masses of the quarks $q_1,\dots,q_n$ in the argument of $\delta m_{Q,q_1,\dots,q_n}^{(\nq+1\rightarrow\nq)}$. Starting at ${\cal O}(\alpha_s^4)$ the finite quark mass corrections in $\delta m_{Q,q_1,\dots,q_n}^{(\nq+1\rightarrow\nq)}(\mbar_Q)$ become also dependent on the flavor threshold corrections relating $\alpha_s^{(\nq)}(\mbar_Q)$ and $\alpha_s^{(\nq+1)}(\mbar_Q)$. In Eq.~\eqref{eqn:MSRMSbmatch3} we have also displayed the first terms of the expansions in the mass ratios $r_{q_i Q}$. They start quadratically in the $r_{q_i Q}$ indicating that the corrections are governed by the scale $\mbar_Q$ just like the matching term $\Delta m_Q^{(\nq+1\rightarrow\nq)}(\mbar_Q)$ and do not have any linear sensitivity to small momenta and the lighter quark masses, in particular. This feature is realized at any order of perturbation theory.

Because the finite mass corrections $\delta m_{Q,q_1,\dots,q_n}^{\nq+1\rightarrow\nq}(\mbar_Q)$ start at ${\cal O}(\alpha_s^3)$ and are quadratic in the mass ratios $r_{q_i Q}$ they are extremely small and never exceed $0.01$~MeV for the top quark (due to the finite bottom or charm masses) and the bottom quark (due to the finite charm mass). We can expect that this is also exhibited at higher orders, so that $\delta m_{Q,q_1,\dots,q_n}^{(\nq+1\rightarrow\nq)}(\mbar_Q)$ can be neglected for all practical purposes and will not be considered and discussed any further in this work.

\subsection{Top-Bottom and Bottom-Charm Mass Matching}
\label{sec:MSRMSbmatch}

Comparing the pole-$\MSR$ mass relation \eqref{eqn:mpoleMSR} for the heavy quark $Q$ to the pole-$\MSb$ mass relation \eqref{eqn:mpoleMSbar} for the next lighter massive quark $q$, one immediately notices that for $R=\mbar_q$ the corrections 
are identical in the approximation that in the \textit{virtual} quark loops all $\nq$ lighter quarks (i.e.\ including the quark $q$) are treated as massless. This identity is a consequence of \textit{heavy quark symmetry} which states that the low-energy QCD corrections to the heavy quark masses coming from massless partons are flavor-independent.

For the top MSR and the bottom $\MSb$ masses (i.e.\ for $Q=t$ and $q=b$) the resulting matching relation reads
\begin{equation}\label{eqn:tbmatch}
 \left[\mpole_t-\msr_t(\mbar_b)\right] - \left[\mpole_b-\mbar_b\right] = \delta m_{b,c}^{(t\rightarrow b)}(\mbar_b,\mbar_c) \,,
\end{equation}
where $\delta m_{b,c}^{(t\rightarrow b)}(\mbar_b,\mbar_c)$ encodes the heavy quark symmetry breaking corrections coming from the finite virtual charm and bottom quark masses. Their form can be extracted directly from Eqs.~\eqref{eqn:mpoleMSbar} and \eqref{eqn:mpoleMSR} and written in the form ($r_{q q^\prime}=\mbar_q/\mbar_{q^\prime}$)
\begin{equation}\label{eqn:tbmatch2}
 \delta m_{b,c}^{(t\rightarrow b)}(\mbar_b,\mbar_c) = \mbar_t \left[ \delta_t^{(b,c)}(r_{bt},r_{ct}) + \delta_t^{(c)}(r_{ct})\right] - \mbar_b \left[ \overline{\delta}_b^{(b,c)}(1,r_{cb}) + \bar{\delta}_b^{(c)}(r_{cb})\right] \,,
\end{equation}
where the first term on the RHS (multiplied by $\mbar_t$) represents the virtual bottom and charm mass effects from the top quark self energy and the second term (multiplied by $\mbar_b$) represents the virtual bottom and charm mass effects from the bottom quark self energy. Their explicit form up to ${\cal O}(\alpha_s^3)$ reads
\begin{align}\label{eqn:tbmatcht}
 \mbar_t & \left[\delta_t^{(b,c)}(r_{bt}, r_{ct}) + \, \delta_t^{(c)}(r_{ct})\right] = \mbar_t \left[\,\delta_2(r_{bt}) + \delta_2(r_{ct}) \,\right] \left(\frac{\alpha_s^{(5)}(\mu)}{4\pi}\right)^2 \\\nonumber
  &+ \mbar_t \left[\,\delta_{t,3}^{(b,c)}(r_{bt},r_{ct}) +  \delta_{t,3}^{(c)}(r_{ct}) + 4 \beta_0^{(5)} \ln\left(\frac{\mu}{\mbar_t}\right)\left[\,\delta_2(r_{bt}) + \delta_2(r_{ct})\right] \right] \left(\frac{\alpha_s^{(5)}(\mu)}{4\pi}\right)^3 + \dots \,,
\end{align}
and
\begin{align}\label{eqn:tbmatchb}
 \mbar_b & \left[\overline{\delta}_b^{(b,c)}(1, r_{cb}) + \, \overline{\delta}_b^{(c)}(r_{cb})\right] = \mbar_b \left[\,\delta_2(1) + \delta_2(r_{cb}) \,\right] \left(\frac{\alpha_s^{(5)}(\mu)}{4\pi}\right)^2 \\\nonumber 
 &+ \mbar_b \left[\,\delta_{b,3}^{(b,c)}(1,r_{cb}) +  \delta_{b,3}^{(c)}(r_{cb}) + 4 \beta_0^{(5)} \ln\left(\frac{\mu}{\mbar_b}\right)\left[\,\delta_2(1) + \delta_2(r_{cb})\right] \right] \left(\frac{\alpha_s^{(5)}(\mu)}{4\pi}\right)^3 + \dots \,.
\end{align}
It is important that the quark mass corrections in \eqref{eqn:tbmatch2} are expressed coherently in powers of $\alpha_s$ at the common scale $\mu$ because the individual $\delta_n$ terms carry contributions that modify the infrared sensitivity and therefore each contain ${\cal O}(\LQCD)$ renormalon ambiguities. In Eq.~\eqref{eqn:tbmatch} these renormalon ambiguities mutually cancel.
We also note that $\delta m_{b,c}^{(t\rightarrow b)}(\mbar_b,\mbar_c)$ also depends on the top quark mass $\mbar_t$. We have suppressed $\mbar_t$ in the argument since $\delta m_{b,c}^{(t\rightarrow b)}(\mbar_b,\mbar_c)$ encodes symmetry breaking corrections due to the finite bottom and charm quark masses.

For the bottom MSR and the charm $\MSb$ masses the corresponding matching relation reads
\begin{equation}\label{eqn:bcmatch}
 \left[\mpole_b - \msr_b(\mbar_c)\right] - \left[\mpole_c - \mbar_c\right] = \delta m_c^{(b\rightarrow c)}(\mbar_c) \;,
\end{equation}
with
\begin{equation}\label{eqn:bcmatch2}
 \delta m_c^{(b\rightarrow c)} (\mbar_c) = \mbar_b \delta_b^{(c)}(r_{cb}) - \mbar_c \overline{\delta}_c^{(c)}(1) \,,
\end{equation}
where the first term on the RHS (multiplied by $\mbar_b$) represents the virtual charm mass effects from the bottom quark self energy and the second term (multiplied by $\mbar_c$) represent the virtual charm mass effects from the charm quark self energy. Their explicit form up to ${\cal O}(\alpha_s^3)$ reads
\begin{align}\label{eqn:bcmatchb}
 \mbar_b\,\delta_b^{(c)}(r_{cb}) &= \mbar_b \,\delta_2(r_{cb}) \left( \frac{\alpha_s^{(4)}(\mu)}{4\pi}\right)^2 \nonumber\\ &\qquad + \mbar_b \left[ \delta_{b,3}^{(c)}(r_{cb}) + 4\beta_0^{(4)} \delta_2(r_{cb})\ln\left(\frac{\mu}{\mbar_b}\right)\right]\left( \frac{\alpha_s^{(4)}(\mu)}{4\pi}\right)^3 + \dots \,,
\end{align}
and
\begin{align}\label{eqn:bcmatchc}
 \mbar_c\,\overline{\delta}_c^{(c)}(1) &= \mbar_c \,\delta_2(1) \left( \frac{\alpha_s^{(4)}(\mu)}{4\pi}\right)^2 \nonumber\\ &\qquad+ \mbar_c \left[ \delta_{c,3}^{(c)}(1) + 4\beta_0^{(4)} \delta_2(1)\ln\left(\frac{\mu}{\mbar_c}\right)\right]\left( \frac{\alpha_s^{(4)}(\mu)}{4\pi}\right)^3 + \dots \,,
\end{align}
where again we expanded both terms consistently for a common renormalization scale $\mu$ in the strong coupling.

\begin{figure}
\center
 \subfigure[]{\includegraphics[width=0.48\textwidth]{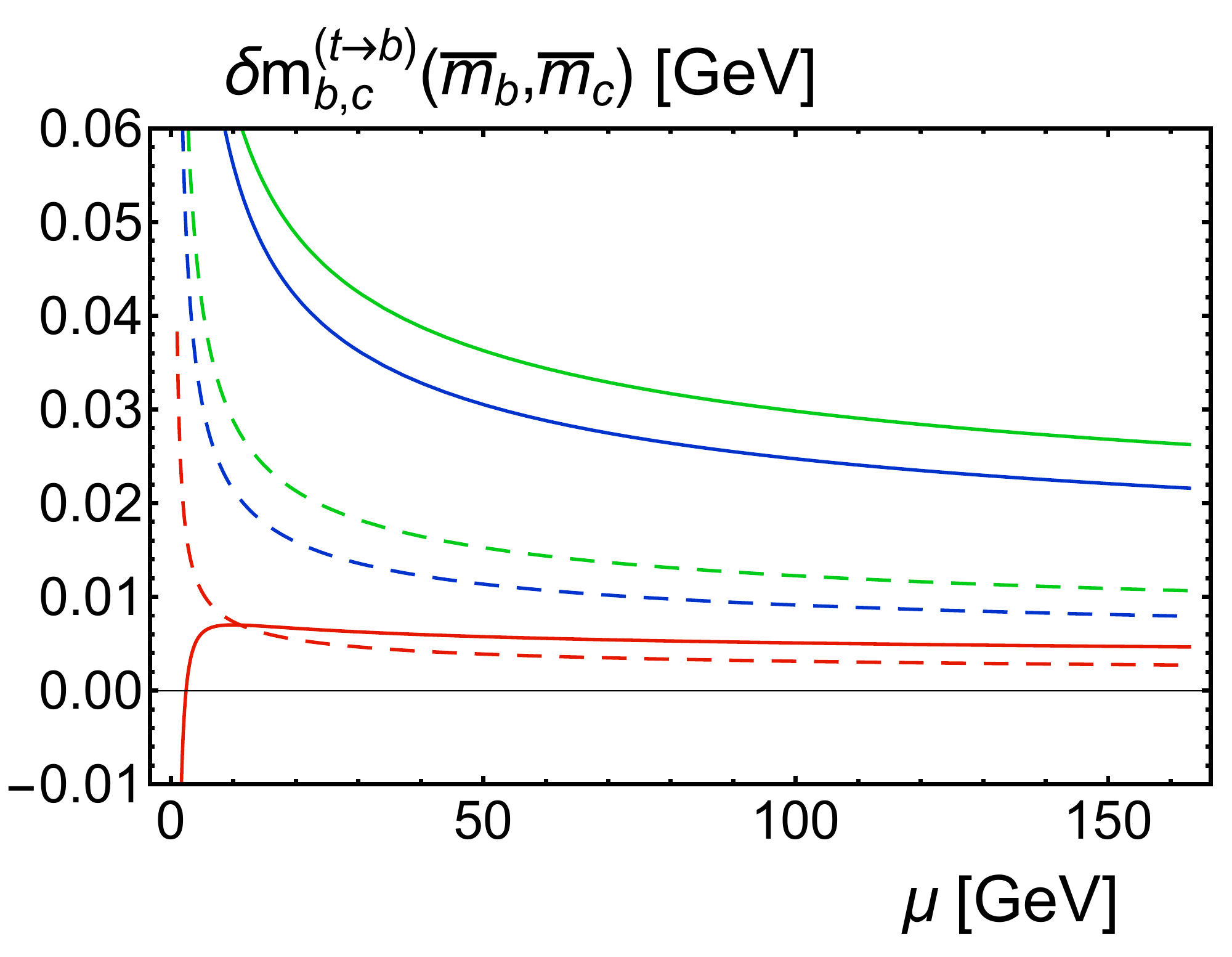}
 \label{fig:HQbreakinga}}  
 \subfigure[]{\includegraphics[width=0.48\textwidth]{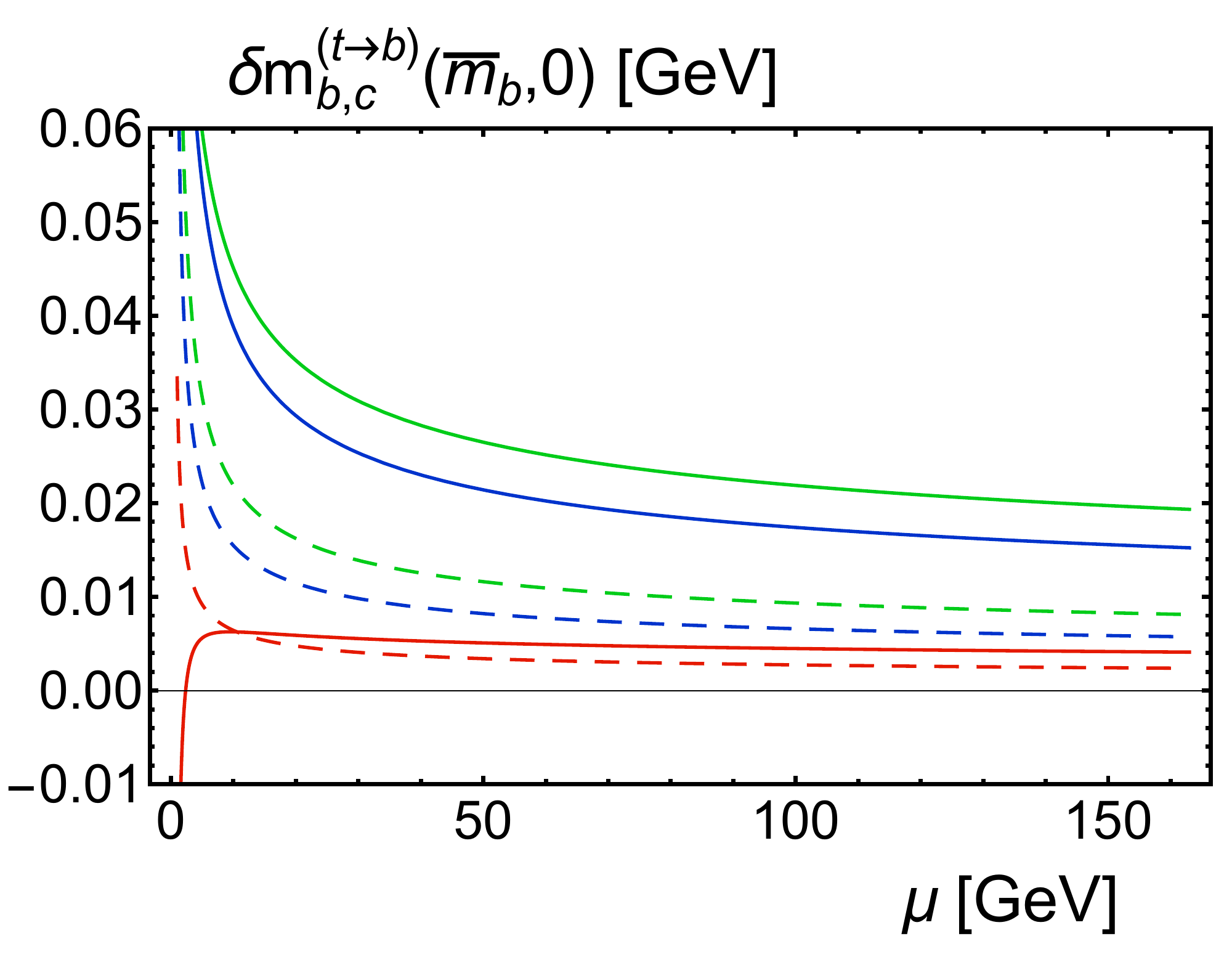}
 \label{fig:HQbreakingb}}
 \subfigure[]{\includegraphics[width=0.48\textwidth]{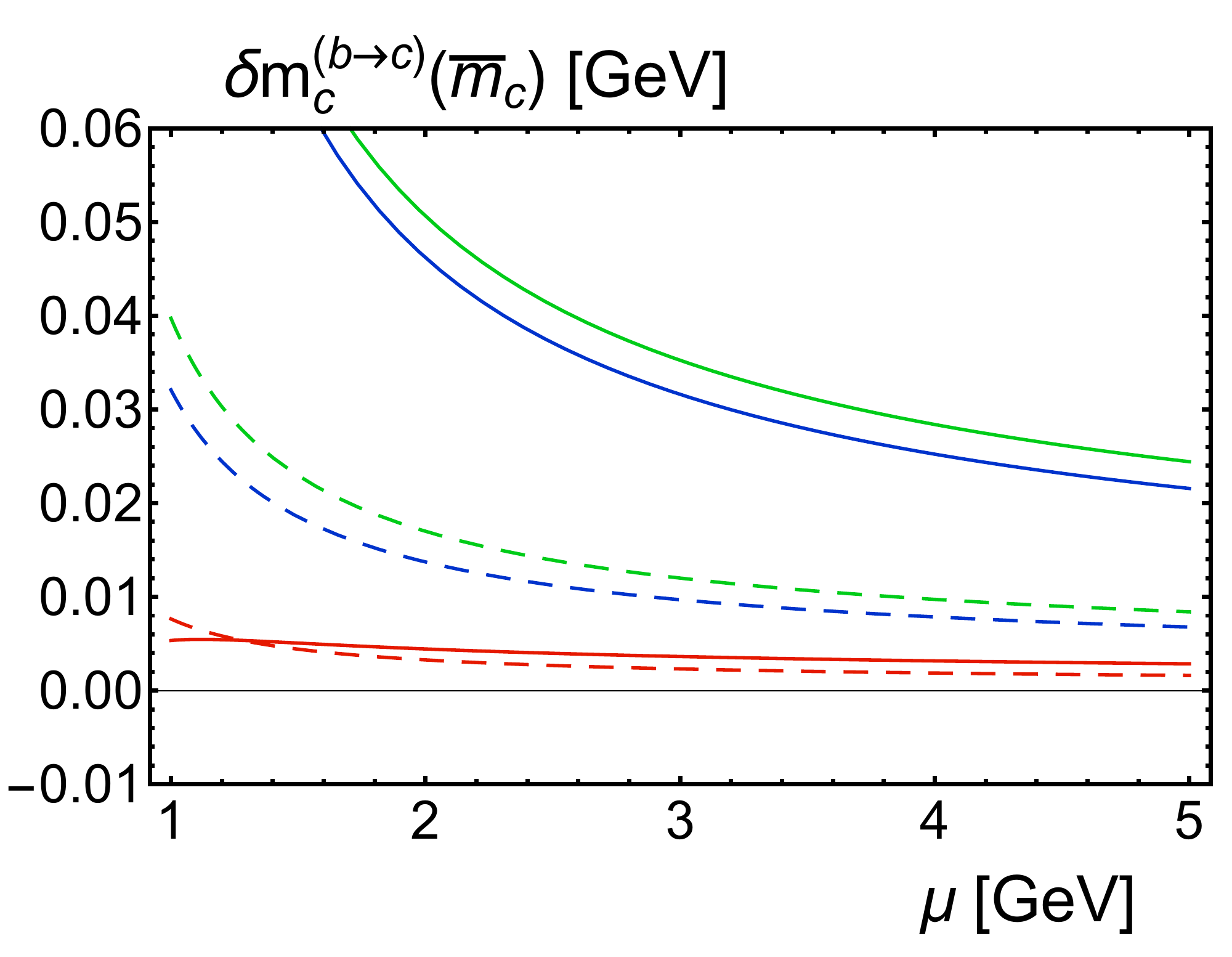}
 \label{fig:HQbreakingc}}
 
 \caption{\label{fig:HQbreaking} (a) Top-MSR bottom-$\MSb$ mass matching correction $\delta m_{b,c}^{(t\rightarrow b)}(\mbar_b,\mbar_c)$ at ${\cal O}(\alpha_s^2)$ (red dashed curve) and ${\cal O}(\alpha_s^3)$ (red solid curve) over the renormalization scale $\mu$. The virtual bottom and charm mass effects to the top quark self energy of Eq.~\eqref{eqn:tbmatcht} (green curves) and the virtual bottom and charm mass effects to the bottom quark self energy of Eq.~\eqref{eqn:tbmatchb} (blue curves) at ${\cal O}(\alpha_s^2)$ (dashed) and ${\cal O}(\alpha_s^3)$ (solid). For the masses of the top, bottom and charm quarks the values ($\mbar_t,\mbar_b,\mbar_c$) = ($163,4.2,1.3$)~GeV are used. 
 (b)~Same quantities as in panel (a) for $\mbar_c=0$. 
 (c) The bottom-MSR charm-$\MSb$ mass matching correction $\delta m_{c}^{(b\rightarrow c)}(\mbar_c)$ at ${\cal O}(\alpha_s^2)$ (red dashed curve) and ${\cal O}(\alpha_s^3)$ (red solid curve) over the renormalization scale $\mu$. The virtual charm mass effects to the bottom quark self energy of Eq.~\eqref{eqn:bcmatchb} (green curves) and the virtual charm mass effects to the charm quark self energy of Eq.~\eqref{eqn:bcmatchc} (blue curves) are shown at ${\cal O}(\alpha_s^2)$ (dashed) and ${\cal O}(\alpha_s^3)$(solid).
 }
\end{figure}

\begin{table}
\renewcommand{\arraystretch}{1.2}
\newcolumntype{A}{>{\centering} m{1cm} }
\newcolumntype{B}{>{\centering\arraybackslash} m{4.1cm} }
\centering
\begin{tabular}{|A|B|B|B|}
\hline
 ${\cal O}(\alpha_s^n)$ & $\delta m_{b,c}^{(t\rightarrow b)} (\mbar_b,\mbar_c)$ & $\delta m_{b,c}^{(t\rightarrow b)} (\mbar_b,0)$ & $\delta m_{c}^{(b\rightarrow c)} (\mbar_c)$ \\\hline
 $2$ & $ 0.007\pm 0.004$ & $ 0.006 \pm 0.004$ & $0.004 \pm 0.002$ \\
 $3$ & $ 0.006\pm 0.001$ & $ 0.005 \pm 0.001$ & $0.004 \pm 0.001$ \\\hline
\end{tabular}
 \caption{The top-bottom MSR-$\MSb$ mass matching corrections, given in units of GeV, for finite bottom and charm masses (second column), for finite bottom quark mass and massless charm quark (third column), and the bottom-charm  MSR-$\MSb$ mass matching correction (fourth column). 
 For the finite masses of the top, bottom and charm quarks the values ($\mbar_t,\mbar_b,\mbar_c$) = ($163,4.2,1.3$)~GeV are used. The uncertainties are obtained from variations of the renormalization scale in the range $\mbar_b\leq\mu\leq\mbar_t$ for $\delta m_{b,c}^{(t\rightarrow b)}$ and in the range $\mbar_c\leq\mu\leq\mbar_b$ for $\delta m_{c}^{(b\rightarrow c)}$. The central values are the respective mean of the largest and smallest values obtained in the scale variation.
  \label{tab:HQbreaking}}
\end{table}

In Fig.~\ref{fig:HQbreakinga} the top-MSR bottom-$\MSb$ mass matching correction $\delta m_{b,c}^{(t\rightarrow b)}(\mbar_b,\mbar_c)$ of Eq.~\eqref{eqn:tbmatch} is displayed as a function of the renormalization scale $\mu$ at ${\cal O}(\alpha_s^2)$ (red dashed line) and ${\cal O}(\alpha_s^3)$ (red solid line) for ($\mbar_t,\mbar_b,\mbar_c$) = ($163,4.2,1.3$)~GeV. The matching correction at ${\cal O}(\alpha_s^3)$ amounts to $6$~MeV and has a scale variation of only $1$~MeV for $\mbar_b\leq\mu\leq\mbar_t$. Compared to the ${\cal O}(\alpha_s^2)$ result we see a strong reduction of the scale-dependence at ${\cal O}(\alpha_s^3)$. The final numerical results at ${\cal O}(\alpha_s^2)$ and ${\cal O}(\alpha_s^3)$ are shown in the second column of Tab.~\ref{tab:HQbreaking} where
the uncertainties are obtained from variations of the renormalization scale in the range $\mbar_b\leq\mu\leq\mbar_t$ and the central values are the respective mean of the largest and smallest values obtained in the scale variation. The corresponding results for a vanishing charm quark mass are shown in Fig.~\ref{fig:HQbreakingb} and the third column of Tab.~\ref{tab:HQbreaking}. We see that the charm mass effects in the top-MSR bottom-$\MSb$ mass matching correction $\delta m_{b,c}^{(t\rightarrow b)} (\mbar_b,\mbar_c)$ are only around $1$~MeV, and the stability for $\mbar_c\rightarrow 0$ shows that the matching correction is governed by scales of order $\mbar_b$ and higher, which reconfirms the range $\mbar_b\leq\mu\leq\mbar_t$ for the variation of the renormalization scale.

In Fig.~\ref{fig:HQbreakingc} the bottom-MSR charm-$\MSb$ mass matching correction $\delta m_{c}^{(b\rightarrow c)}(\mbar_c)$ of Eq.~\eqref{eqn:bcmatch} is displayed as a function of the renormalization scale $\mu$ for $\mbar_b=4.2$~GeV and $\mbar_c=1.3$~GeV at ${\cal O}(\alpha_s^2)$ and ${\cal O}(\alpha_s^3)$ using the same color coding and curve styles as for Figs.~\ref{fig:HQbreakinga} and \ref{fig:HQbreakingb}. In the fourth column of Tab.~\ref{tab:HQbreaking} the final numerical results at ${\cal O}(\alpha_s^2)$ and ${\cal O}(\alpha_s^3)$ are shown using $\mbar_c\leq\mu\leq\mbar_b$ for the renormalization scale variation. The stability and convergence is again excellent, and at ${\cal O}(\alpha_s^3)$ the matching correction amounts to $4$~MeV with an uncertainty of $1$~MeV.

Given that the heavy quark symmetry breaking matching corrections $\delta m_{b,c}^{(t\rightarrow b)}(\mbar_b,\mbar_c)$ and $\delta m_{c}^{(b\rightarrow c)}(\mbar_c)$ amount to only $4$ to $6$~MeV, we note that they may be simply neglected in practical applications where they yield contributions that are much smaller than other sources of uncertainties. In fact, this also applies to our subsequent studies of the top, bottom and charm quark pole masses. However, we include them here for completeness. Due to their small size, we have not explicitly included the heavy quark symmetry breaking matching corrections in the graphical illustration of Fig.~\ref{fig:matchrun}.

\subsection{Light Virtual Quark Mass Corrections at \texorpdfstring{${\cal O}(\alpha_s^4)$}{4-Loop Order} and Beyond}
\label{sec:lightvirtual}

The excellent perturbative convergence of the top-MSR bottom-$\MSb$ mass matching correction $\delta m_{b,c}^{(t\rightarrow b)} (\mbar_b,\mbar_c)$ and of the bottom-MSR charm-$\MSb$ mass matching correction $\delta m_{c}^{(b\rightarrow c)} (\mbar_c)$ discussed in the previous section illustrates that they both are short-distance quantities and free of an ${\cal O}(\LQCD)$ renormalon ambiguity. This is also expected theoretically due to heavy quark symmetry. However, the facts that the overall size of the matching corrections only amounts to a few MeV, and that the ${\cal O}(\alpha_s^3)$ corrections are only around $1$~MeV allows us to draw interesting conceptual implications for the large order asymptotic behavior of the virtual quark mass corrections in the mass relations of Eqs.~\eqref{eqn:mpoleMSbar}, \eqref{eqn:mpoleMSbarv2} and \eqref{eqn:mpoleMSR}. We discuss these implications in the following. As a consequence we can {\it predict} the yet uncalculated virtual quark mass corrections at ${\cal O}(\alpha_s^4)$ to within a few percent without an additional loop calculation and draw important conclusions on their properties for the orders beyond.

To be concrete, we consider the matching correction $\delta m_{q}^{(Q\rightarrow q)} (\mbar_q)$ between the MSR mass of heavy quark $Q$ and the $\MSb$ mass of the next lighter massive quark $q$ assuming the massless approximation for all quarks lighter than quark $q$ i.e.\ $\nq = n_q+1 = \nl+1$ and $n_\ell=n_q$ being the number of massless quarks. This situation applies to the matching relation for the top-MSR and the bottom $\MSb$ masses for a massless charm quark or to the matching relation between the bottom-MSR and the charm-$\MSb$ masses.

In Fig.~\ref{fig:HQbreakinga} we have displayed separately the virtual bottom and charm mass effects to the top quark self energy of Eq.~\eqref{eqn:tbmatcht} (green curves) and the virtual bottom and charm mass effects to the bottom quark self energy of Eq.~\eqref{eqn:tbmatchb} (blue lines) at ${\cal O}(\alpha_s^2)$ (dashed) and ${\cal O}(\alpha_s^3)$ (solid). In Fig.~\ref{fig:HQbreakingb} the charm quark is treated as massless in the same quantities. In Fig.~\ref{fig:HQbreakingc} the virtual charm mass effects to the bottom quark self energy of Eq.~\eqref{eqn:bcmatchb} and the virtual charm mass effects to the charm quark self energy of Eq.~\eqref{eqn:bcmatchc} are shown at ${\cal O}(\alpha_s^2)$ and ${\cal O}(\alpha_s^3)$ with the analogous line styles and colors. We see that both types of contributions each are quite large and furthermore do not at all converge. The ${\cal O}(\alpha_s^3)$ corrections are even bigger than the ${\cal O}(\alpha_s^2)$ corrections, which indicates that the corresponding asymptotic large order behavior already dominates the ${\cal O}(\alpha_s^2)$ and ${\cal O}(\alpha_s^3)$ corrections.

The origin of this behavior has been already mentioned and is understood: The mass of the virtual quark $q$ acts as an infrared cutoff and therefore modifies the infrared sensitivity of the self energy diagrams (of quark $Q$ and of quark $q$) with respect to the case where the virtual loops of quark $q$ are evaluated in the massless approximation. As a consequence these corrections individually carry an ${\cal O}(\LQCD)$ renormalon ambiguity. Moreover, at large orders in perturbation theory the sensitivity of the self energy diagrams to infrared momenta increases due to high powers of logarithms from gluonic and massless quark loops. As a consequence, at large orders, the finite mass effects of the virtual loops of quark $q$ in the self energy diagrams of quark $Q$ and the self energy diagrams of quark $q$ become equivalent due to heavy quark symmetry. The strong cancellation in the sum of both types of corrections in $\delta m_{q}^{(Q\rightarrow q)}(\mbar_q)$ ($\sim75\%$ at ${\cal O}(\alpha_s^2)$ and $\gtrsim90\%$ at ${\cal O}(\alpha_s^3)$ for the cases displayed in Fig.~\ref{fig:HQbreaking}) thus confirms that the known ${\cal O}(\alpha_s^2)$ and ${\cal O}(\alpha_s^3)$ self energy corrections coming from virtual quark masses are already dominated by their large order asymptotic behavior.

From the observations that the series for $\delta m_{b,c}^{(t\rightarrow b)}(\mbar_b,\mbar_c)$ and $\delta m_{c}^{(b\rightarrow c)}(\mbar_c)$ converge very well and that their ${\cal O}(\alpha_s^3)$ corrections amount to only about $1$~MeV, we can therefore expect that the two types of corrections that enter $\delta m_{b,c}^{(t\rightarrow b)}(\mbar_b,\mbar_c)$ as well as $\delta m_{c}^{(b\rightarrow c)}(\mbar_c)$ agree to even better than $1$~MeV  at $ {\cal O}(\alpha_s^4)$ and beyond. This allows us to make an approximate \textit{prediction} for the yet uncalculated ${\cal O}(\alpha_s^4)$ finite mass corrections from virtual loops of quark $q$ in the pole-$\MSb$ mass relations of quark $Q$ of Eqs.~\eqref{eqn:mpoleMSbar} and \eqref{eqn:mpoleMSbarv2} by setting the ${\cal O}(\alpha_s^4)$ correction in $\delta m_{q}^{(Q\rightarrow q)}(\mbar_q)$ to zero:
\begin{align}\label{eqn:delta4predict}
 &\delta_{Q,4}^{(q)}(r_{qQ}) \\
 &\approx r_{qQ}\Bigg[\,\delta_{q,4}^{(q)}(1) + \left(6\,\betazero\delta_{q,3}^{(q)}(1) + 4\,\betaone\delta_2(1)\right) \ln\left(\frac{\mu}{\mbar_q}\right) + 12\,\delta_2(1)\left(\betazero\ln\left(\frac{\mu}{\mbar_q}\right)\right)^2\Bigg] \, \nonumber\\
 &\hspace{1.3cm} - \left( 6\,\betazero \delta_{Q,3}^{(q)}(r_{qQ}) + 4\, \betaone\delta_2(r_{qQ})\right)\ln\left(\frac{\mu}{\mbar_Q}\right) - 12\,\delta_2(r_{qQ})\left(\betazero\ln\left(
 \frac{\mu}{\mbar_Q}\right)\right)^2\nonumber \,.
\end{align}
The prediction has a residual $\mu$-dependence, which would vanish in the formal limit that the virtual quark $q$ mass corrections are entirely dominated by their large order asymptotic behavior. Therefore the dependence on the scale $\mu$ can be used as an uncertainty estimate of our approximation.

\begin{figure}
\center
 \subfigure[]{\includegraphics[width=.48\textwidth]{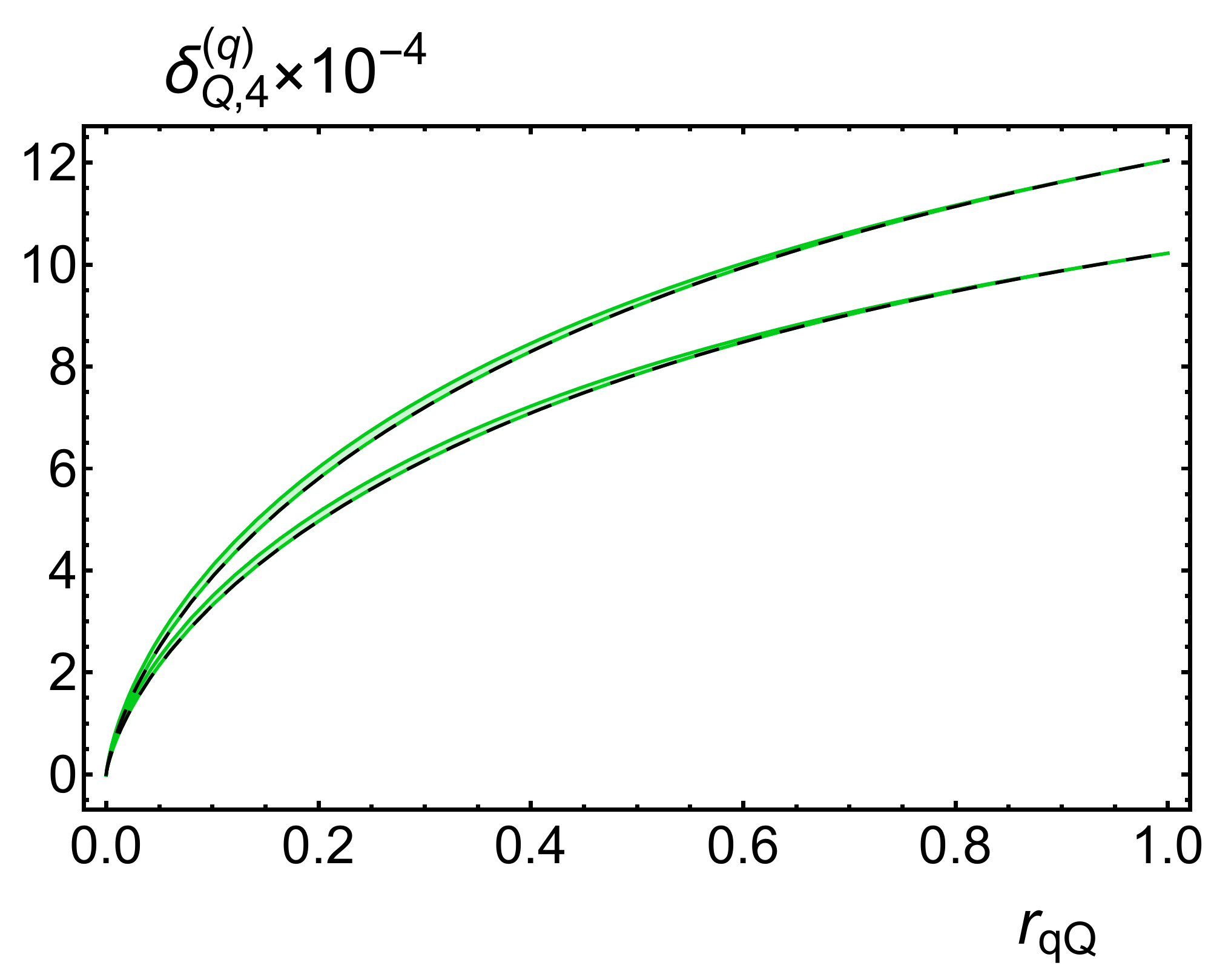}
 \label{fig:deltapredicta}}  
 \subfigure[]{\includegraphics[width=.48\textwidth]{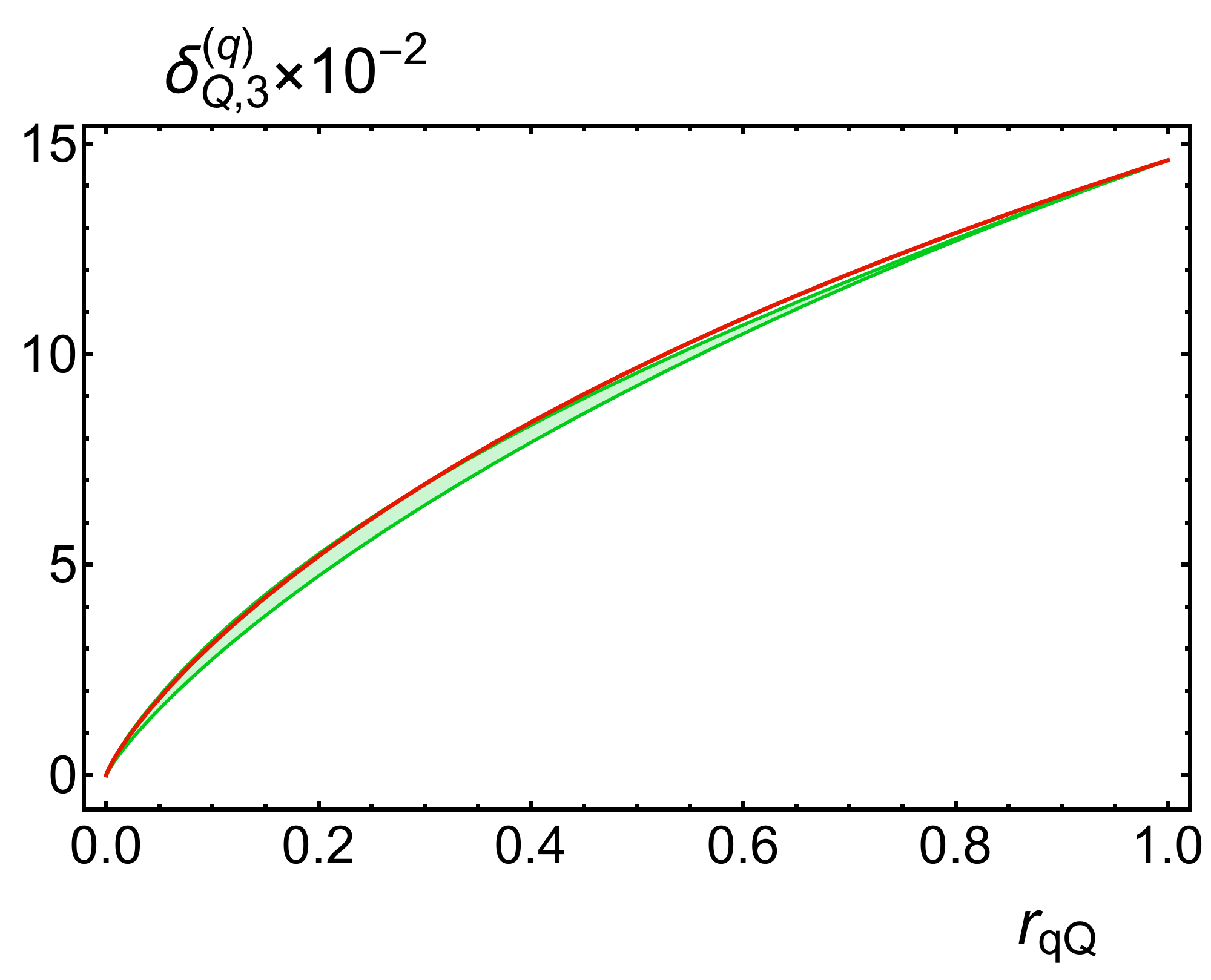}
 \label{fig:deltapredictb}}  
 \caption{\label{fig:deltapredict} 
 (a) Prediction for the ${\cal O}(\alpha_s^4)$ virtual quark mass correction $\delta_{Q,4}^{(q)}(r_{qQ})$ for $\mbar_q\leq\mu\leq\mbar_Q$ (green bands) for $\nq=\nl+1=5$ (lower band) and $\nq=\nl+1=4$ (upper band). The black dashed lines show the prediction for $\mu=\mbar_Q$ which gives the simple approximation formula in Eq.~\eqref{eqn:delta4predict2}.
 (b) The ${\cal O}(\alpha_s^3)$ virtual quark mass correction $\delta_{Q,3}^{(q)}(r_{qQ})$ for $\nq=\nl+1=5$ (red curve). The green band is the prediction for $\delta_{Q,3}^{(q)}(r_{qQ})$ using the method of panel (a) for $\mbar_q\leq\mu\leq\mbar_Q$ showing excellent agreement to the exact result within errors. 
 	}
\end{figure}

In Fig.~\ref{fig:deltapredicta} we show the prediction for $\delta_{Q,4}^{(q)}(r_{qQ})$ for $\mbar_q\leq\mu\leq\mbar_Q$ (green bands) for $\nq=n_q+1=\nl+1=5$ (lower band) and $\nq=n_q+1=\nl+1=4$ (upper band). The prediction satisfies exactly the required boundary condition $\delta_{Q,4}^{(q)}(0) = 0$ and Eq.~\eqref{eqn:Delta2an} for $r_{qQ}=1$ and provides an interpolation for $0<r_{qQ}<1$ with an uncertainty of $\pm3\%$ (for $r_{qQ}\lesssim0.1$) or smaller (for $r_{qQ}>0.1$). To judge the quality of the prediction we apply the same method at ${\cal O}(\alpha_s^3)$  to ``predict'' $\delta_{Q,3}^{(q)}(r_{qQ})$ which gives
\begin{equation}\label{eqn:delta3predict}
 \delta_{Q,3}^{(q)}(r_{qQ}) \approx r_{qQ}\left[\,\delta_{q,3}^{(q)}(1) + 4\,\betazero\delta_2(1)\ln\left(\frac{\mu}{\mbar_q}\right)\right] - 4\betazero\delta_2 (r_{qQ})\ln\left(\frac{\mu}{\mbar_Q}\right) \,.
\end{equation}
The result for the prediction of $\delta_{Q,3}^{(q)}(r_{qQ})$ is shown in Fig.~\ref{fig:deltapredictb} for $\nq=n_q+1=\nl+1=5$. The green band illustrates again the range of predictions for $\mu$-variations  $\mbar_q\leq\mu\leq\mbar_Q$, and represents an uncertainty of $\pm10\%$ (for $r_{qQ}\lesssim0.1$) or smaller (for $r_{qQ}>0.1$). Compared to the ${\cal O}(\alpha_s^4)$ result, the larger $\mu$ variation we observe at ${\cal O}(\alpha_s^3)$ is expected because the infrared sensitivity is weaker and the large order asymptotic behavior is less dominating at the lower order. The red curve is the exact result for $\delta_{Q,3}^{(q)}(r_{qQ})$ obtained from the results in Ref.~\cite{Bekavac:2007tk}, see also Eq.~\eqref{eqn:d3qq}. We see that the prediction is fully compatible with the exact result and that the uncertainty estimate based on the $\mu$-variation is reliable. The prediction for $\delta_{Q,3}^{(q)}(r_{qQ})$ for $\nq=n_q+1=\nl+1=4$ has the same good properties but is not displayed since it is numerically very close to the prediction for $\nq=n_q+1=\nl+1=5$.

Overall, the examination shows that the prediction and the uncertainty estimate for $\delta_{Q,4}^{(q)}(r_{qQ})$ can be considered reliable. We can also provide a very simple closed analytic expression by evaluating Eq.~\eqref{eqn:delta4predict} for $\mu=\mbar_Q$, which gives 
\begin{align}\label{eqn:delta4predict2}
 &\delta_{Q,4}^{(q)}(r_{qQ}) \\ &\approx r_{qQ}\bigg[\,\delta_{q,4}^{(q)}(1) - \left(6\,\betazero\delta_{q,3}^{(q)}(1) + 4\,\betaone\delta_2(1) \right)\ln\left(r_{qQ}\right) + 12\,\delta_2(1)\left(\betazero\ln(r_{qQ})\right)^2\bigg] \nonumber\\
  &= r_{qQ}\bigg[\,(203915.\pm32.) - 22962.\,\nq + 525.2\,\nq^2 + (-130946. +13831.\,\nq -328.5\,\nq^2)\,\ln(r_{qQ}) \nonumber\\ &\hspace{1.5cm}
  + (26599.1 - 3224.1\,\nq + 97.70\,\nq^2 )\,\ln(r_{qQ})^2\,\bigg] \,.\nonumber
\end{align}
The expression depends via the boundary condition of Eq.~\eqref{eqn:Delta2an} entirely on the coefficients $a_n(n_q,n_h)$ of Eq.~\eqref{eqn:coeffanmsbar}, which for this case describe the corrections to the heavy quark $q$ self energy for the case that all lighter quarks are massless, and the coefficients of the $\beta$-function. The expression is shown as the black dashed lines in Fig.~\ref{fig:deltapredicta} for $\nq=\nl+1=5$ (lower line) and $\nq=\nl+1=4$ (upper line).
This approximation for $\delta_{Q,4}^{(q)}(r_{qQ})$  has a simple overall linear behavior on the mass ratio $r_{qQ}=\mbar_q/\mbar_Q$. The behavior is just a manifestation of $\delta_{Q,4}^{(q)}(r_{qQ})$ being dominated by the large order asymptotic behavior due to its ${\cal O}(\Lambda_{\rm QCD})$ renormalon ambiguity which is related to linear sensitivity to small scales. The overall linear dependence of $\delta_{Q,4}^{(q)}(r_{qQ})$ on $\mbar_q$ arises since the mass of quark $q$ represents an infrared cut and thus represents the characteristic physical scale that governs $\delta_{Q,4}^{(q)}(r_{qQ})$. 
This also explains the origin of the logarithms shown in Eq.~\eqref{eqn:delta4predict2}: They arise because all virtual quark mass corrections in  Eqs.~\eqref{eqn:mpoleMSbar}, \eqref{eqn:mpoleMSbarv2} and \eqref{eqn:mpoleMSR} are defined in an expansion in $\alpha_s(\mbar_Q)$.
We note that for the ${\cal O}(\alpha_s^3)$ virtual massive quark correction $\delta_{Q,3}^{(q)}(r_{qQ})$ these aspects were already discussed in Ref.~\cite{Hoang:2000fm} and later in Ref.~\cite{Ayala:2014yxa}, where a direct comparison to the explicit calculations from Ref.~\cite{Bekavac:2007tk} could be carried out.
These analyses were, however, using generic considerations and were not carried out within a systematic RG framework.

The expression of Eq.~\eqref{eqn:delta4predict2} is a special case of the general statement that the asymptotic large order behavior of the coefficients $\delta_{Q,n}^{(q)}(r_{qQ})$ can be obtained from the relation
\begin{align}
\label{deltaqallorderasy}
\delta_2(r_{qQ})&\left(\frac{\alpha_s^{(\nq)}(\mbar_Q)}{4\pi}\right)^2 + \sum_{n=3}^\infty\,\delta_{Q,n}^{(q)}(r_{qQ})\left(\frac{\alpha_s^{(\nq)}(\mbar_Q)}{4\pi}\right)^n
\\
&\approx  r_{qQ}\,\overline{\delta}_{q}^{(q)}(1) = 
r_{qQ}\left [ \delta_2(1)\,\left( \frac{\alpha_s^{(\nq)}(\mbar_q)}{4\pi} \right)^2 + \sum_{n=3}^\infty\delta_{q,n}^{(q)}(1)\left(\frac{\alpha_s^{(\nq)}(\mbar_q)}{4\pi} \right)^n \,\right ] \, ,\nonumber
\end{align}
where on the RHS of the approximate equality $\alpha_s^{(\nq)}(\mbar_q)$ has to be expanded in powers of $\alpha_s^{(\nq)}(\mbar_Q)$, and we have $\delta_2(1)=18.3189$, $\delta_{q,3}^{(q)}(1)=1870.79 - 82.1208\, \nq$ and  \mbox{$\delta_{q,4}^{(q)}(1)=(203915.\pm32.) - 22961.6\, \nq + 525.216\, \nq^2$}. The terms $\delta_{q,n}^{(q)}(1)$ for $n>4$ can be obtained from 
using Eqs.~\eqref{eqn:Deltadef} and \eqref{eqn:Delta2an} together with the large order asymptotic form of the coefficients $a_n$ shown in Eq.~\eqref{eqn:asy1}, giving
\begin{align}
\delta_{q,n>4}^{(q)}(1) &\approx 
a_n^\mathrm{asy}(n_q) - a_n^\mathrm{asy}(n_q+1) =   
a_n^\mathrm{asy}(n_Q-1) - a_n^\mathrm{asy}(n_Q)\label{eqn:deltaqn5} \,,
\end{align}
where we would like to remind the reader that for the case we consider here we have $n_Q = n_q+1 = n_\ell+1$.
Our examination at ${\cal O}(\alpha_s^3)$ and ${\cal O}(\alpha_s^4)$ above showed that this relation provides an approximation for $\delta_{Q,4}^{(q)}$ within a few percent. For the higher-order terms $\delta_{Q,n}^{(q)}$ with $n>4$ it should be even more precise, and we therefore believe that it should be sufficient for essentially all future applications in the context of studies of the pole mass scheme. 

To conclude we note that it is straightforward to extend Eq.~\eqref{eqn:delta4predict} from the case of having only one massive quark $q$ being lighter than heavy quark $Q$, i.e.\ $\nq=n_q+1=\nl+1$, to the case of having a larger number of lighter massive quarks. For example for the case that there are two massive quarks lighter than quark $Q$ (let's say $q$ and $q^\prime$, in order of decreasing mass) with $\nq=n_q+1=n_{q^\prime}+2=\nl+2$, the generalization of the approximation formula~\eqref{eqn:delta4predict} reads
\begin{align}
 &\delta_{Q,4}^{(q,q^\prime)}(r_{qQ},r_{q^\prime Q}) + \delta_{Q,4}^{(q^\prime)}(r_{q^\prime Q}) \approx r_{qQ}\,\Bigg\{ \,\delta_{q,4}^{(q,q^\prime)}(1,r_{q^\prime q}) + \delta_{q,4}^{(q^\prime)}(r_{q^\prime q}) \nonumber\\ &\hspace{1cm}
 + \left[6\,\betazero\left(\delta_{q,3}^{(q,q^\prime)}(1,r_{q^\prime q}) + \delta_{q,3}^{(q^\prime)}(r_{q^\prime q})\right) + 4\,\betaone \left( \delta_2(1) + \delta_2(r_{q^\prime q}) \right)\right] \ln\left(\frac{\mu}{\mbar_q}\right) \nonumber\\
 &\hspace{1cm} + 12\,\left(\delta_2(1) + \delta_2(r_{q^\prime q})\right)\,\left(\betazero\ln\left(\frac{\mu}{\mbar_q}\right)\right)^2\Bigg\} \\ &\hspace{1cm}
 - \left[6\,\betazero\left(\delta_{Q,3}^{(q,q^\prime)}(r_{qQ},r_{q^\prime Q}) + \delta_{Q,3}^{(q^\prime)}(r_{q^\prime Q})\right) + 4\,\betaone \left( \delta_2(r_{qQ}) + \delta_2(r_{q^\prime Q})\right)\right] \ln\left(\frac{\mu}{\mbar_Q}\right) \nonumber\\
 &\hspace{1cm} - 12\,\left(\delta_2(r_{qQ}) + \delta_2(r_{q^\prime Q})\right)\,\left(\betazero\ln\left(\frac{\mu}{\mbar_Q}\right)\right)^2\,.\nonumber
\end{align}
\mbox{}

\subsection{Pole Mass Differences}
\label{sec:polediff}

Using the MSR mass we have set up a conceptual framework to systematically quantify the contributions to the pole mass of a heavy quark coming from the different momentum regions contained in the on-shell self energy diagrams. The pole mass of a heavy quark $Q$ contains the contributions from all momenta, while the $\MSb$ mass $\mbar_Q(\mu)$ and the MSR mass $\msr_Q(R)$ contain the contributions from above the scales $\mu$ and $R$, respectively (see Fig.~\ref{fig:massschemes}). The MSR mass is the natural extension of the $\MSb$ mass, which is applied for scales $\mu>m_Q$, to scales $R<m_Q$, and obeys a RG-evolution equation that is linear in $R$, called R-evolution~\cite{Hoang:2008yj,Hoang:2017suc}. The R-evolution equation quantifies in a way free of the ${\cal O}(\Lambda_{\rm QCD})$ renormalon the change in the MSR mass when contributions from lower momenta are included into the mass when $R$ is decreased, as long as $R>\LQCD$.

In Sec.~\ref{sec:Qintout} we discussed the matching corrections $\Delta m_Q^{(\nq+1\rightarrow\nq)}(\mbar_Q)$ that arise when the virtual loop contributions of quark $Q$ are integrated out by switching from $\mbar_Q$ to $\msr_Q(\mbar_Q)$. In Sec.~\ref{sec:MSRmass} we discussed the MSR mass difference $\Delta m^{(n_Q)}(R,R^\prime) = \msr_Q(R^\prime) - \msr_Q(R)$, which is determined from solving the R-evolution equation of the MSR mass and which systematically sums logarithms of $R/R^\prime$. In Sec.~\ref{sec:MSRMSbmatch} we examined the matching between the QCD corrections to the MSR mass of the heavy quark $Q$ and the $\MSb$ mass of the next lighter massive quark $q$, $\delta m_{q,q^\prime,\dots}^{(Q\rightarrow q)}(\mbar_q,\mbar_{q^\prime},\dots)$ accounting for the mass effects of the quarks $q,q^\prime,\dots\;$. This matching is based on heavy quark symmetry and the small numerical size of $\delta m_{q,q^\prime,\dots}^{(Q\rightarrow q)}(\mbar_q,\mbar_{q^\prime},\dots)$ reflects that the symmetry breaking effects due to the finite quark masses are quite small. These two types of matching corrections and the R-evolution of the MSR mass each are free of ${\cal O}(\LQCD)$ renormalon ambiguities and show excellent convergence properties in QCD perturbation theory.

An interesting application is the determination of the difference of the pole masses of two massive quarks. Due to heavy quark symmetry, the differences of two heavy quark pole masses are also free of ${\cal O}(\LQCD)$ renormalon ambiguities and can therefore be determined to high precision. The matching corrections discussed above and the R-evolution of the MSR mass allow us to systematically sum logarithms of the mass ratios that would remain unsummed in a fixed-order calculation, and to achieve more precise perturbative predictions~\cite{Hoang:2017suc}. Taking the example of the top and bottom mass one can then write the difference of the top quark pole-$\MSb$ mass relation and the bottom quark pole-$\MSb$ mass relation in the form
\begin{equation}\label{eqn:tbrelation}
 \left[ \mpole_t - \mbar_t \right] - \left[ \mpole_b - \mbar_b \right] = \Delta m_t^{(6\rightarrow 5)}(\mbar_t) + \Delta m^{(5)}(\mbar_t,\mbar_b) + \delta m_{b,c}^{(t\rightarrow b)}(\mbar_b,\mbar_c) \,.
\end{equation}
The analogous relation for the bottom and charm quarks reads
\begin{equation}\label{eqn:bcrelation}
 \left[ \mpole_b - \mbar_b \right] - \left[ \mpole_c - \mbar_c \right] = \Delta m_b^{(5\rightarrow 4)}(\mbar_b) + \Delta m^{(4)}(\mbar_b,\mbar_c) + \delta m_{c}^{(b\rightarrow c)}(\mbar_c) \,.
\end{equation}
Each of the mass differences is the sum of universal matching and evolution building blocks which each can be computed to high precision, as shown in Tabs.~\ref{tab:DeltaMRRprime}, \ref{tab:MSRMSbmatch}, \ref{tab:HQbreaking}.

The resulting relations between the top, bottom and charm quark pole masses read
\begin{align}
 \mpole_t - \mpole_b &= \left[ \mbar_t - \mbar_b \right] + \Delta m_t^{(6\rightarrow 5)}(\mbar_t) + \Delta m^{(5)}(\mbar_t,\mbar_b) + \delta m_{b,c}^{(t\rightarrow b)}(\mbar_b,\mbar_c) \label{eqn:tbpole} \,,\\
 \mpole_b - \mpole_c &= \left[ \mbar_b - \mbar_c \right] + \Delta m_b^{(5\rightarrow 4)}(\mbar_b) + \Delta m^{(4)}(\mbar_b,\mbar_c) + \delta m_{c}^{(b\rightarrow c)}(\mbar_c) \label{eqn:bcpole} \,,\\
 \mpole_t - \mpole_c &= \left[ \mbar_t - \mbar_c \right] + \Delta m_t^{(6\rightarrow 5)}(\mbar_t) + \Delta m^{(5)}(\mbar_t,\mbar_b) + \delta m_{b,c}^{(t\rightarrow b)}(\mbar_b,\mbar_c) \nonumber\\ &\hspace{2.13cm}+ \Delta m_b^{(5\rightarrow 4)}(\mbar_b) + \Delta m^{(4)}(\mbar_b,\mbar_c) + \delta m_{c}^{(b\rightarrow c)}(\mbar_c) \label{eqn:tcpole} \,,
\end{align}
and can be readily evaluated from the highest order results given in Tabs.~\ref{tab:DeltaMRRprime}, \ref{tab:MSRMSbmatch}, \ref{tab:HQbreaking} for the case ($\mbar_t,\mbar_b,\mbar_c$) = ($163,4.2,1.3$)~GeV:
\begin{align}
 \mpole_t - \mpole_b &= 158.800 + (0.032\pm0.001) + (9.331\pm0.016) + (0.006\pm0.001) \;{\rm GeV}  \nonumber \\ 
 &= 168.169\pm0.016 \;{\rm GeV} \,, \label{eqn:tbpole2} \\
 \mpole_b - \mpole_c &= 2.9 + (0.004\pm0.001) + (0.423\pm0.017) + (0.004\pm0.001) \;{\rm GeV}   \nonumber \\
 &= 3.331\pm0.017 \;{\rm GeV} \,,  \label{eqn:bcpole2}  \\
 \mpole_t - \mpole_c &= 171.500 \pm 0.024 \;{\rm GeV}  \,, \label{eqn:tcpole2}
\end{align}
where we have added all uncertainties quadratically. 
We can compare our results for the bottom-charm pole mass difference $\mpole_b - \mpole_c$ to the result obtained in Ref.~\cite{Hoang:2005zw} using a fixed-order expansion at ${\cal O}(\alpha_s^3)$ for the mass difference. Their result was based on a linear approximation for the virtual charm quark mass effects derived in Ref.~\cite{Hoang:2000fm} which is similar to Eq.~\eqref{eqn:delta3predict}, but used a numerical calculation of the coefficient linear in $r_{qQ}$ from Ref.~\cite{Melles:1998dj}. In this analysis the pole mass difference was used to eliminate the charm quark mass as a primary parameter in the predictions. They determined 
$\mpole_b - \mpole_c=3.401\pm 0.013$~GeV and obtained $\mbar_c=1.22\pm 0.06$~GeV from the fits using $\mbar_b=4.16\pm 0.05$~GeV as input.  
Their result for $\mpole_b - \mpole_c$ is consistent with ours, but one should keep in mind that 
logarithms of $\mbar_c/\mbar_b$ were not systematically summed and that 
their result also included nontrivial QCD corrections to semileptonic B-meson decay spectra for $B\to X_c\ell\nu$ and $B\to X_s\gamma$ which were only known to ${\cal O}(\alpha_s^2)$. The mutual agreement is reassuring (also for the theoretical approximations made in the context of the B meson analyses) and in particular shows that the summation of logarithms of $\mbar_c/\mbar_b$ is not essential for bottom and charm masses, which is expected, and that the ${\cal O}(\alpha_s^4)$ corrections are tiny, which can also be seen explicitly in our results. The larger error we obtain in our computation of $\mpole_b - \mpole_c$ arises from the renormalization scale scale variation in $\Delta m^{(4)}(\mbar_b,\mbar_c)$ which includes scales as low as $0.6 \,\mbar_c$ while in their analysis the lowest renormalization scale was $\mbar_c$. 
Similar determinations of bottom and charm quark masses from B-meson decay spectra were carried out in Ref.~\cite{Buchmuller:2005zv, Gambino:2016jkc}, and they are also consistent with our result for $\mpole_b - \mpole_c$.	
 	
For the case ($\mbar_t,\mbar_b,\mbar_c$) = ($163,4.2,0$)~GeV, the difference between the top and bottom pole masses reads
\begin{align}\label{eqn:tbpole3}
	\mpole_t - \mpole_b &= 158.800 + (0.032\pm0.001) + (9.331\pm0.016) + (0.005\pm0.001) \;{\rm GeV} \nonumber\\
	&= 168.168\pm0.016 \;{\rm GeV} \,.
\end{align}
This result differs from Eq.~\eqref{eqn:tbpole2} by only $1$~MeV
showing that the effects of the finite charm quark mass are tiny in the difference of the top and bottom pole masses. 
The uncertainties in the pole mass differences are between $16$ and $24$~MeV and should be considered as conservative estimates of the theoretical uncertainties due to missing higher order corrections.

\subsection{Lighter Massive Flavor Decoupling}
\label{sec:decoupling}

Another very instructive application of the RG framework to quantify and separate the contributions to the pole mass of a heavy quark coming from the different physical momentum regions is to examine the effective massive flavor decoupling at large orders. It was observed in Ref.~\cite{Ayala:2014yxa} that the sum of the known ${\cal O}(\alpha_s^2)$ and ${\cal O}(\alpha_s^3)$ charm quark mass effects in the bottom quark pole-$\MSb$ mass series expressed in four flavor coupling $\alpha_s^{(4)}(\mbar_b)$ (where they amount to about $35$~MeV) are essentially fully captured simply by expressing the series in the three flavor coupling $\alpha_s^{(3)}(\mbar_b)$ (where they amount to only $-2$~MeV). This observation entails that one can simply neglect the charm quark mass corrections by computing the bottom quark pole-$\MSb$ mass relation right from start in the three flavor theory without any charm quark (which corresponds to an infinitely heavy charm quark). This effective decoupling of lighter massive quarks is obvious and truly happening at asymptotic large orders. The importance of the observation made in Ref.~\cite{Ayala:2014yxa} was that the finite charm quark mass corrections in the decoupled calculation at ${\cal O}(\alpha_s^2)$ and ${\cal O}(\alpha_s^3)$ were so tiny that there was no need to compute them explicitly in the first place. If this decoupling property would be true in general (i.e.\ the remaining light quark mass correction become negligible) it would represent a great simplification because it may make an explicit calculation of the lighter massive quark corrections and also the summation of the associated logarithms irrelevant. 

Using the RG framework for the lighter massive flavor dependence of the pole mass we can examine systematically in which way this effective lighter massive quark decoupling property is realized. In the following we analyze this issue for $(\mbar_t,\mbar_b,\mbar_c)=(163,4.2,1.3)$~GeV.
We start with the effects of the charm quark mass in the bottom pole-$\MSb$ mass relation examined in Ref.~\cite{Ayala:2014yxa}. Applying the same considerations as for the pole mass differences in Sec.~\ref{sec:polediff} for this case we can write down the relation
\begin{align}\label{eqn:mbmcdecoup}
\mpole_b- & \bigg[\, \mbar_b+\mbar_b\sum_{n=1}^\infty a_n(\nl=3,0)\left(\frac{\alpha_s^{(3)}(\mbar_b)}{4\pi}\right)^n\,\bigg] \\
= & \,\,\Delta m_b^{(5\to 4)}(\mbar_b)+\Delta m^{(4)}(\mbar_b,\mbar_c)+\delta m_{c}^{(b\to c)}(\mbar_c)+\Delta m_c^{(4\to 3)}(\mbar_c) \nonumber \\
&-\Delta m^{(3)}(\mbar_b,\mbar_c) \,\nonumber \\
= &\,\,  (0.004\pm0.001) + (0.423\pm0.017) + (0.004\pm0.001) + (0.005\pm0.002)\nonumber \\
& \,\, - (0.434\pm0.020)  \;{\rm GeV} \nonumber\\
= &\,\, 0.002\pm0.026 \;{\rm GeV}\,.
\end{align}
The RHS represents a computation of the charm quark mass corrections that remain within a calculation where the charm mass effects are approximated by making the charm infinitely heavy (i.e.\ $\nl=3$).  
The individual numerical results have been taken from the highest order results in Tabs.~\ref{tab:DeltaMRRprime}, \ref{tab:MSRMSbmatch} and \ref{tab:HQbreaking}, and for the final numerical result we have conservatively added all uncertainties quadratically.
We see that these remaining corrections are essentially zero, fully confirming the observation of Ref.~\cite{Ayala:2014yxa}. This is not surprising since the bottom and charm quark masses are similar in size and the ratio $\mbar_c/\mbar_b$ does not lead to large logarithms. So the summation of these logarithms which is contained in our computation does not make an improvement, and the agreement with Ref.~\cite{Ayala:2014yxa} simply represents a computational cross check of both calculations. The scale uncertainty is larger than the one shown in Ref.~\cite{Ayala:2014yxa} because we considered variations of the renormalization scale down to $\mu=0.6\,\mbar_c$, which were not considered by them, and because we do not attempt to eliminate the strong correlation in scale-dependence between $\Delta m^{(4)}(\mbar_b,\mbar_c)$ and $\Delta m^{(3)}(\mbar_b,\mbar_c)$ from these low scales here.

Let us now investigate the case of the bottom quark mass corrections in the top quark pole-$\MSb$ mass relation assuming a massless charm quark. We can simply adapt Eq.~\eqref{eqn:mbmcdecoup} through trivial modifications and obtain the relation
\begin{align}\label{eqn:mtmbdecoup}
\mpole_t-& \bigg[\, \mbar_t+\mbar_t\sum_{n=1}^\infty a_n(\nl=4,0)\left(\frac{\alpha_s^{(4)}(\mbar_t)}{4\pi}\right)^n\,\bigg] \\
= & \,\,\Delta m_t^{(6\to 5)}(\mbar_t)+\Delta m^{(5)}(\mbar_t,\mbar_b)+\delta m_{b,c}^{(t\to b)}(\mbar_b,0)+\Delta m_b^{(5\to 4)}(\mbar_b)\nonumber\\
&-\Delta m^{(4)}(\mbar_t,\mbar_b)\nonumber \\
= &\,\, (0.032\pm0.001) + (9.331\pm0.016) + (0.005\pm0.001)  + (0.004\pm0.001) \nonumber \\
& \,\, - (9.114\pm0.014)  \;{\rm GeV} \nonumber\\
= &\,\, 0.258\pm0.021 \;{\rm GeV}\,. \nonumber
\end{align}
We see that using the approximation of an infinitely heavy bottom quark for a calculation of the bottom mass effects in the top quark pole-$\MSb$ mass relation gives a result that is about $260$~MeV too small.

We can now go one step further and also consider the case where the masses of both the bottom and charm quark are accounted for. Generalizing the previous two calculations to this case is straightforward and we obtain
\begin{align}\label{eqn:mtmbmcdecoup}
\mpole_t-& \bigg[\, \mbar_t+\mbar_t\sum_{n=1}^\infty a_n(\nl=3,0)\left(\frac{\alpha_s^{(3)}(\mbar_t)}{4\pi}\right)^n\,\bigg] \\
= & \,\,\Delta m_t^{(6\to 5)}(\mbar_t)+\Delta m^{(5)}(\mbar_t,\mbar_b)+\delta m_{b,c}^{(t\to b)}(\mbar_b,\mbar_c)\nonumber\\
&+\Delta m_b^{(5\to 4)}(\mbar_b)+\Delta m^{(4)}(\mbar_b,\mbar_c)+\delta m_{c}^{(b\to c)}(\mbar_c)+\Delta m_c^{(4\to 3)}(\mbar_c) \nonumber \\
&-\Delta m^{(3)}(\mbar_t,\mbar_c) \,.\nonumber \\
= &\,\, (0.032\pm0.001) + (9.331\pm0.016) + (0.006\pm0.001) \nonumber \\
& \,\, + (0.004\pm0.001) + (0.423\pm0.017) + (0.004\pm0.001) + (0.005\pm0.002)\nonumber \\
& \,\, - (9.111\pm0.032)  \;{\rm GeV} \nonumber\\
= &\,\, 0.694\pm0.040 \;{\rm GeV}\,. \nonumber
\end{align}
In this case using the approximation of infinitely heavy bottom and charm quarks for a calculation of the bottom and charm mass effects in the top quark pole-$\MSb$ mass relation gives a result that is almost $700$~MeV too small.

Our results show that the approximation of computing the  lighter heavy flavor mass corrections in a theory where these heavy flavors are decoupled is an excellent approximation for the charm mass corrections in the bottom quark pole mass, but it is considerably worse for the top quark, where the discrepancy even reaches the $1$~GeV level. The reason is that the decoupling limit can in general not capture the true size of the lighter quark mass effects if the hierarchy of scales is large. One should therefore not use this approximation to determine bottom or charm quark mass effects for the top quark.

\section{The Top Quark Pole Mass Ambiguity}
\label{sec:topmass}

\begin{figure}
\center
 \includegraphics[width=.6\textwidth]{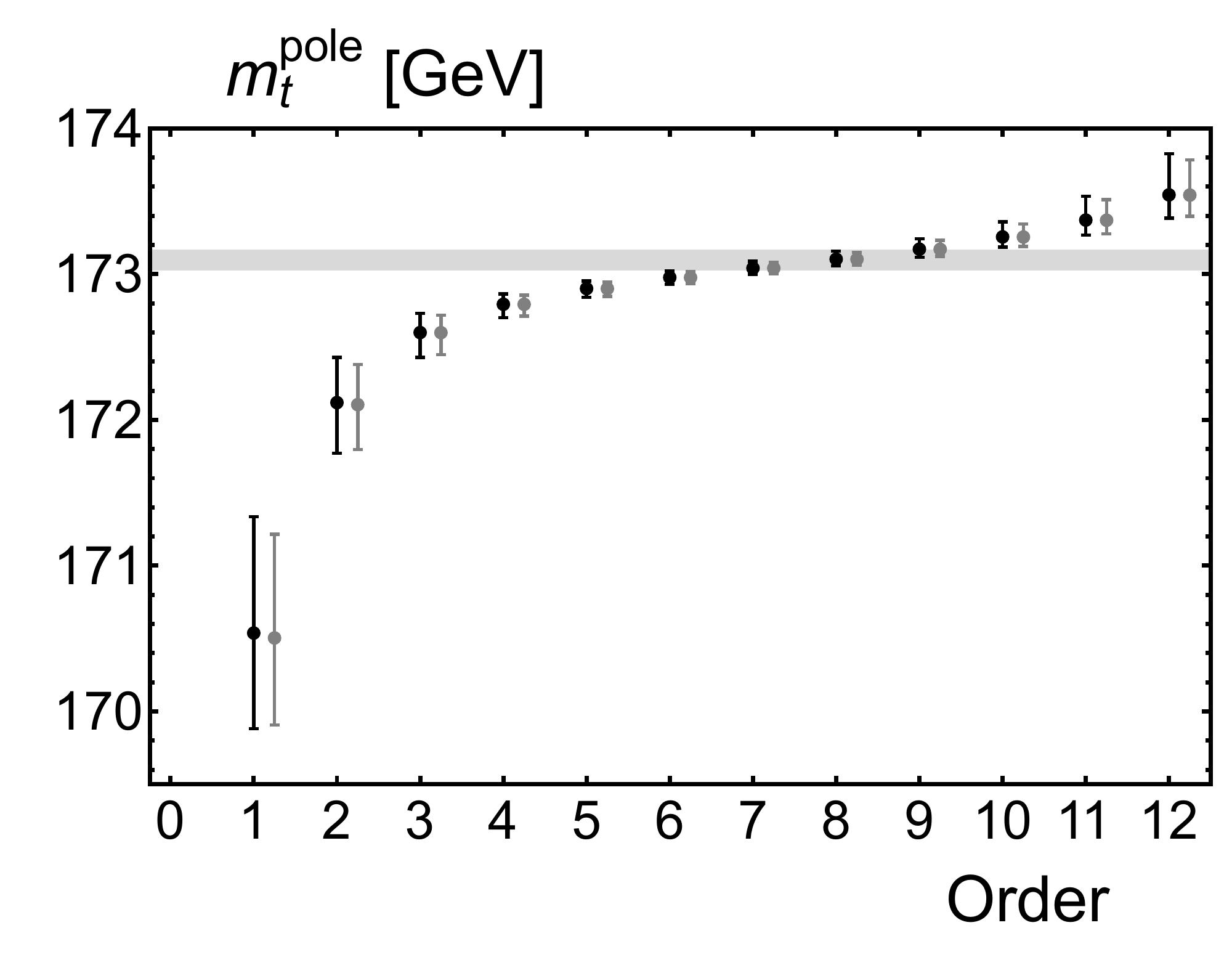}
 \caption{\label{fig:mtpole1} 
 	Top quark pole mass as a function of order obtained from the MSR mass $m_t^{\rm MSR}(\mbar_t)$ (black) and $\mbar_t=\mbar_t(\mbar_t)=163$~GeV (gray) for massless bottom and charm quarks. The central dots refer to the renormalization scale $\mu=\mbar_t$ for the strong coupling. The error bars arise from renormalization scale variation $\mbar_t/2 \le \mu \le 2\,\mbar_t$. The gray horizontal band represents the region $m_t^{\rm pole}=173.10\pm 0.07$, which indicates   	
 	the top quark pole mass and its scale uncertainty obtained from $m_t^{\rm MSR}(\mbar_t)$ at the 8th order. 
 	}
\end{figure}

\subsection{General Comments and Estimation Method}
\label{sec:method}

In this section we address the question of the best possible approximation and the ambiguity of the top quark pole mass $m_t^{\rm pole}$ using the RG formalism for the top mass described in the earlier sections. As a reminder and for illustration we show in Fig.~\ref{fig:mtpole1} $\mpole_t$ as a function of the order obtained from the series for $\mpole_t - \msr_t(\mbar_t)$ in powers of $\alpha_s^{(5)}$ given in Eq.~\eqref{eqn:mpoleMSR} for massless bottom and charm quarks, where the central dots are obtained for the default choice of renormalization scale $\mu=\mbar_t$ in the strong coupling and the error bars represent the scale variation $\mbar_t/2\leq\mu\leq2\,\mbar_t$. The corresponding results from the series for $\mpole_t-\mbar_t$ given in Eq.~\eqref{eqn:mpoleMSbar} in powers of $\alpha_s^{(6)}$, also for massless bottom and charm quarks, are shown in gray. We have used the asymptotic form of the perturbative coefficients  
shown in Tab.~\ref{tab:aasy} for the series coefficients beyond $\mathcal{O}(\alpha_s^4)$\footnote{The uncertainties of the normalization factors $N_{1/2}^{(\nq)}$ are about an order of magnitude smaller than the renormalization scale variation of the series beyond ${\cal O}(\alpha_s^4)$ and therefore not significant for our analysis.}. 
We note that focusing on the approximation of massless bottom and charm quarks by itself is phenomenologically valuable because it is employed for most current predictions in the context of top quark physics, and since the analytic expressions are most transparent for this case.

The graphics illustrates visually the problematic features associated to the top quark ${\cal O}(\LQCD)$ pole mass renormalon, and in particular the specific properties of the series for $\mu\sim m_t$ already mentioned in Sec.~\ref{sec:intro}: The minimal term of the series is obtained at order $n_{\rm min} = 8$, which according to the theory of asymptotic series is the order that provides the best possible approximation for the top quark pole mass. Furthermore, the corrections are numerically close to the eighth order correction for the orders in the range 6 to 10, i.e.\ $\Delta n \approx 5$, for which the partially summed series increases linearly with the order. According to the theory of asymptotic series it is this region of orders that is relevant for the size of the principle uncertainty of this best approximation.  We also see two very important practical issues appearing already at lower orders which can make dealing with the pole mass in mass determinations difficult: First, the higher order corrections are much larger than indicated by usual renormalization scale variations of the lower order prediction and, second, the common renormalization scale variation at any given truncation order is not an appropriate tool to estimate the perturbative uncertainty. In this context it is easy to understand that specifying a concrete numerical value for the principle uncertainty of the top quark pole mass is non-trivial even if the series is known precisely to all orders. 	
So to obtain a top quark pole mass determination with uncertainties close to the principle uncertainty within a phenomenological analysis based on a usual truncated finite order calculation may be quite difficult.
As a comparison let us recall the much better perturbative behavior of a series that is free of an ${\cal O}(\LQCD)$ renormalon ambiguity such as the MSR mass differences $\Delta m^{(n_Q)}(R,R^\prime)$ of Eq.~\eqref{eqn:rrge} with numerical evaluations given in Tab.~\ref{tab:DeltaMRRprime}.

Prior to this work the issue of the best possible estimate and the ambiguity of the top quark pole mass were already studied in
Ref.~\cite{Beneke:2016cbu}. They examined the pole-$\MSb$ mass relation of Eq.~\eqref{eqn:mpoleMSbar} for massless bottom and charm quarks (i.e.\ $\nq=n_t=\nl=5$) and their analysis addressed the numerical uncertainty of the top quark pole mass accounting for all series terms displayed in Fig.~\ref{fig:mtpole1} for $\mu=\mbar_t$. They adopted a prescription given in Ref.~\cite{Beneke:1998ui}, which {\it defined} the top quark pole mass uncertainty as the imaginary part of the inverse Borel integral of Eq.~\eqref{eqn:borelambiguity}, $\Delta m_{\rm Borel}^{(\nl=5)}$, divided by $\pi$, which gives about $65$~MeV. Since this agrees in size with the minimal series term\footnote{
In Ref.~\cite{Beneke:1998ui} the order of the minimal series term $n_{\rm min}$ and the size of the minimal term $\Delta(n_{\rm min})$ were not chosen from the set of the actual series terms but computed from the minimum of a quadratic fit to the series terms in the vicinity of the minimum, so that their $n_{\rm min}$ was a non-integer value and their  $\Delta(n_{\rm min})$ value is slightly smaller than the minimal term in the series. 
There are neither practical nor conceptual advantages of this procedure, and the numerical results are unchanged within their errors if $\Delta(n_{\rm min})$ is taken as the minimal terms in the series. 
}, 
which arises at order $\alpha_s^8$, 
they argued that $\Delta m_{\rm Borel}^{(\nl=5)}/\pi$ (or the size of the minimal term) is a reliable quantification of
the top quark pole mass ambiguity, which they finally specified as $70$~MeV. Interpreting the specification like a numerical uncertainty, this gives $m_t^{\rm pole}=173.10\pm 0.07$, which is shown in Fig.~\ref{fig:mtpole1} as the thin gray horizontal band. The uncertainty band is about the same size as the renormalization scale variation of the series truncated at the eighth order.

We believe that quoting $70$~MeV for the top quark pole mass ambiguity for massless bottom and charm quarks is too optimistic. 
Given (i) the overall bad behavior of the series, 
(ii) that there is a sizable range of orders where the corrections have very similar size and (iii) that the partially summed series increases linearly with the order in the range $6$ to $10$ ($\Delta n \approx 5$), we see no compelling reason to truncate precisely at the order $n_{\rm min}=8$ and to quote a number at the level of the scale variation of the truncated series or the size of the correction at this order as the principle uncertainty.
Our view is also supported by heavy quark symmetry (HQS)~\cite{Isgur:1989vq} which states that the pole mass ambiguity is independent of the mass of the heavy quark up to power corrections of ${\cal O}(\Lambda_{\rm QCD}^2/m_Q)$. This is the first aspect following from HQS we discussed in Sec.~\ref{sec:intro}. HQS requires that the criteria and the outcome of the method used to determine the top quark pole mass ambiguity are independent of the top mass value (as long as it is sufficiently bigger than $\Lambda_{\rm QCD}$). So it is straightforward to carry out a test concerning HQS by changing the value of $\mbar_t$ while keeping $\mu/\mbar_t=1$ and checking whether the approach to estimate the ambiguity provides stable results.

Concerning Ref.~\cite{Beneke:2016cbu} this check is best carried out in the five-flavor scheme for the strong coupling, and we therefore evaluate the size of the minimal term in the series for $m_t^{\rm pole}-m_t^{\rm MSR}(\mbar_t)$. Adopting the values $163$, $20$, $4.2$, $2$ and $1.3$~GeV for $\mbar_t$ we obtain $62$, $75$, $91$, $113$ and $131$~MeV for the minimal term $\Delta(n_{\rm min})$. This behavior is roughly described by the approximate formula $\Delta(n_{\rm min})\approx(4\pi\alpha_s^{(\nl=5)}(\mu)/\beta_0^{(\nl=5)})^{1/2}\Lambda_{\rm QCD}^{(\nl=5)}$,
already mentioned in Sec.~\ref{sec:intro} and shows that the basic dependence on $\mu$ is logarithmic. We can even render the minimal term arbitrarily small if we adopt for $\mbar_t$ values much larger than $163$~GeV. 
We see that  $\Delta m_{\rm Borel}^{(\nl=5)}/\pi$, which is independent of the top mass value and therefore proportional to the ambiguity, agrees with the size of the minimal term only for $\mu\sim 163$~GeV, but disagrees for other choices. So the 
line of reasoning used for the analysis of the top quark pole mass ambiguity in  Ref.~\cite{Beneke:2016cbu} is not independent of the top quark mass value, and one has to conclude that the ambiguity must be larger than $\Delta m_{\rm Borel}^{(\nl=5)}/\pi$ and certainly larger than $130$~MeV, which is the size of the minimal term for a very small value of $\mbar_t$. Concerning the quoted numbers, we emphasize that we still discuss the case of massless bottom and charm quarks. From the relation 
$\Delta n\times\Delta(n_{\rm min})\propto \pi^2 \Lambda_{\rm QCD}^{(\nl)}/\beta_0\propto \Delta m_{\rm Borel}$
we see in particular that a reliable method consistent with HQS has to explicitly account for the range $n_{\rm min}\pm\Delta n/2$ in orders 
for which the terms in the series have values close to $\Delta(n_{\rm min})$. We stress that the latter issue is not at all new and has been known since the work of Refs.~\cite{Bigi:1994em,Beneke:1994sw}. It was also argued  
in~\cite{Beneke:2016cbu} that their approach to estimate the size of the top quark pole mass ambiguity is consistent concerning that issue. However, their approach did not account for the actual size of $\Delta n$, which is about $5$ for the case discussed in~\cite{Beneke:2016cbu} and also shown in Fig.~\ref{fig:mtpole1}.

In the following subsections we apply a method to determine the best possible estimate and the ambiguity of the top quark pole mass
which explicitly accounts for the range $n_{\rm min}\pm\Delta n/2$ in orders where the $\Delta(n)$ are very close to $\Delta(n_{\rm min})$. It also accounts for the practical problems in an order-by-order determination of the pole mass from a series containing the ${\cal O}(\Lambda_{\rm QCD})$ renormalon which we discussed above in the context of Fig.~\ref{fig:mtpole1}.
To describe the method we define, for a given series to calculate the top quark pole mass,
\begin{equation}
 \Delta(n) \equiv \mpole_t(n) - \mpole_t(n-1) \,,
\end{equation}
where $\mpole_t(n)$ is the partial sum  at ${\cal O}(\alpha_s^n)$ of the series for the top quark pole mass that contains the ${\cal O}(\LQCD)$ pole mass renormalon, and thus $\Delta(n)$ is the $n$-th order correction. The method we use is as follows:
\begin{enumerate}
 \item We determine the minimal term $\Delta(n_{\rm min})$ and the set of orders $\{n\}_f\equiv\{n:\,\Delta(n)\leq f\,\Delta(n_{\rm min})\}$ in the series for a default renormalization scale, where $f$ is a number larger but close to unity.
 \item We use half of the range of values covered by $\mpole_t(n)$ with $n\in\{n\}_f$ evaluated for this setup and include renormalization scale variation in a given range as an estimate for the ambiguity of the top quark mass. We use the midpoint of the covered range as the central value.
\end{enumerate}
While $n_{\rm min}$, $\Delta(n_{\rm min})$ and $\Delta n$ each can vary substantially depending on which setup one uses to determine $m_t^{\rm pole}$, the method provides results that are setup-independent and is therefore consistent with HQS.
Through the RG formalism we developed in the previous sections we can explicitly implement the other important requirement of HQS, namely that the ambiguities of the pole masses of all heavy quarks agree. To do this
we apply our method for three different scenarios which differ on whether the bottom and charm quarks are treated as massive or massless and we furthermore study the pole-MSR mass difference for different values of $R$.

\subsection{Massless Bottom and Charm Quarks}
\label{sec:mbmczero}

For the case that the bottom and charm quarks are treated as massless we can calculate the top quark pole mass from the top MSR mass $\msr_t(R)$ at different scales $R\leq\mbar_t$. Using the $\MSb$-MSR mass matching contribution $\Delta m_t^{(6\rightarrow 5)}(\mbar_t)$ of Eq.~\eqref{eqn:MSRMSbmatch2} and R-evolution from the scale $\mbar_t$ to $R$ of Eq.~\eqref{eqn:rrge} with $n_t=5$ active dynamical flavors one can write the top quark pole mass as
\begin{equation}\label{eqn:mtpolembmc0}
 \mpole_t = \mbar_t + \Delta m_t^{(6\rightarrow 5)}(\mbar_t) + \Delta m^{(5)}(\mbar_t,R) + R\,\sum_{n=1}^{\infty} a_n(n_\ell=5,0)\left(\frac{\alpha_s^{(5)}(R)}{4\pi}\right)^n \,,
\end{equation}
where the sum of the second and third term on the RHS is just $\msr_t(R)-\mbar_t$. The terms $\Delta m_t^{(6\rightarrow 5)}(\mbar_t)$ and $\Delta m^{(5)}(\mbar_t,R)$ are free of an ${\cal O}(\LQCD)$ renormalon ambiguity and can be evaluated to the highest order given in Tabs.~\ref{tab:DeltaMRRprime} and \ref{tab:MSRMSbmatch}. We can then determine the best estimate of the top quark pole mass and its ${\cal O}(\LQCD)$ renormalon ambiguity from the $R$-dependent series which is just equal to $\mpole_t-\msr_t(R)$. The outcome of the analysis using the method described in Sec.~\ref{sec:method} for $\mbar_t=163$~GeV and $R=163,20,4.2$ and $1.3$~GeV and $f=5/4$ is shown in the upper section of Tab.~\ref{tab:mtpole}.

\begin{table}
 \renewcommand{\arraystretch}{1.2}
 \newcolumntype{A}{>{\centering\arraybackslash} m{.4cm} }
 \newcolumntype{B}{>{\centering\arraybackslash} m{2.7cm} }
 \newcolumntype{C}{>{\centering\arraybackslash} m{.6cm} }
 \newcolumntype{E}{>{\centering\arraybackslash} m{1.5cm} }
 \newcolumntype{H}{>{\centering\arraybackslash} m{1.25cm} }
 \newcolumntype{I}{>{\centering\arraybackslash} m{1.65cm} }
 \newcolumntype{J}{>{\centering\arraybackslash} m{1.7cm} }
 \newcolumntype{K}{>{\centering\arraybackslash} m{1.8cm} }
 \small
 \centering
 \begin{tabular}{|A|B|C|H|I|E|J|K|}
   \hline
   \multicolumn{8}{|c|}{$\mbar_t = 163\,{\rm GeV},\quad\mbar_b = \mbar_c = 0\,{\rm GeV},\quad \nl = n_t =  5$}\\\hline
   $R$ & $\msr_t(R)-\mbar_t$ & $n_{\rm min}$ & $\Delta(n_{\rm min})$ & $\sum_{n=5}^{n_{\rm min}} \Delta(n)$ & $\{n\}_{5/4}$ & \vspace{.1cm}\shortstack{$\mpole_t\qquad\quad$\\ $-\;\msr_t(R)$}\vspace{.1cm} & $\mpole_t$ \\\hline
   $163$ & $0.032(1)\hphantom{0}$ & $8$ & $0.062(3)$ & $0.310(17)$ & $\{6,7,8,9\}$ & $10.054(157)$ & $173.086(157)$ \\
   $20$ & $8.038(9)\hphantom{0}$ & $6$  & $0.075(4)$ & $0.150(8)\hphantom{0}$ & $\{5,6,7\}$ & $\hphantom{0}2.140(166)$ & $173.178(166)$ \\
   $4.2$ & $9.363(16)$ & $4$ & $0.091\hphantom{(0)}$  & $0$ & $\{3,4,5\}$ & $\hphantom{0}0.832(217)$ & $173.195(218)$ \\
   $1.3$ & $9.748(23)$ & $3$ & $0.098\hphantom{(0)}$ & $0$ & $\{2,3,4\}$ & $\hphantom{0}0.394(186)$ & $173.142(187)$ \\\hline
\multicolumn{8}{|c|}{$\mbar_t = 163\,{\rm GeV},\quad\mbar_b = 4.2\,{\rm GeV},\quad\mbar_c = 0\,{\rm GeV},\quad \nl= n_t - 1 = 4$}\\\hline
   $R$ & \vspace{.1cm}\shortstack{$\mpole_t - \mpole_b\hspace{.7cm}$\\$ +\; \msr_b(R)-\mbar_t$}\vspace{.1cm} & $n_{\rm min}$ & $\Delta(n_{\rm min})$ & $\sum_{n=5}^{n_{\rm min}} \Delta(n)$ &  $\{n\}_{5/4}$ & \vspace{.1cm}\shortstack{$\mpole_b\qquad\quad$\\ $-\;\msr_b(R)$}\vspace{.1cm} & $\mpole_t$ \\\hline
   $163$ & $0.258(21)$ & $7$ & $0.087(3)$ & $0.324(11)$ & $\{6,7,8,9\}$ & $9.904(227)$ & $173.162(228)$ \\
   $20 $ & $8.035(17)$ & $5$ & $0.104(3)$ & $0.104(3)$ & $\{4,5,6\}$ & $2.120(211)$ & $173.155(212)$ \\
   $4.2$ & $9.372(16)$ & $4$ & $0.135$ & $0$ & $\{3,4\}$ & $0.855(211)$ & $173.227(212)$ \\
   $1.3$ & $9.795(23)$ & $2$ & $0.124$ & $0$ & $\{1,2,3\}$ & $0.331(214)$ & $173.126(215)$ \\\hline
   \multicolumn{8}{|c|}{$\mbar_t = 163\,{\rm GeV},\quad\mbar_b = 4.2\,{\rm GeV},\quad\mbar_c = 1.3\,{\rm GeV},\quad \nl= n_t -2 = 3$}\\\hline
   $R$ & \vspace{.1cm}\shortstack{$\mpole_t - \mpole_c\hspace{.7cm}$\\$ +\; \msr_c(R)-\mbar_t$}\vspace{.1cm} & $n_{\rm min}$ & $\Delta(n_{\rm min})$ & $\sum_{n=5}^{n_{\rm min}} \Delta(n)$ & $\{n\}_{5/4}$ & \vspace{.1cm}\shortstack{$\mpole_c\qquad\quad$\\ $-\;\msr_c(R)$}\vspace{.1cm} & $\mpole_t$ \\\hline
   $163$ & $0.694(40)$ & $7$ & $0.098(2)$ & $0.355(8)$ & $\{6,7,8,9\}$ & $9.471(260)$ & $173.165(263)$ \\
   $20$  & $8.076(33)$ & $5$ & $0.116(3)$ & $0.116(3)$ & $\{4,5,6\}$ & $2.085(243)$ & $173.161(245)$ \\
   $4.2$ & $9.371(31)$ & $3$ & $0.154$ & $0$ & $\{3,4\}$ & $0.888(257)$ & $173.259(259)$ \\
   $1.3$ & $9.805(24)$ & $2$ & $0.128$ & $0$ & $\{1,2,3\}$ & $0.354(243)$ & $173.159(244)$ \\\hline
 \end{tabular}
\caption{
 Details of the numerical results of our method to determine $m_t^{\rm pole}$ for the cases of massless bottom and charm quarks (upper section),  massless charm quarks (middle section) and  finite bottom and charm quarks (lower section) and exploring different setups to determine $m_t^{\rm pole}$. The final respective results for $m_t^{\rm pole}$ are shown in the last column. See the text for details.
 All numbers for masses and mass differences are in units of GeV. Errors are quoted in parentheses.
 \label{tab:mtpole}}
\end{table}

\begin{figure}
\center
 \subfigure[]{\includegraphics[width=.48\textwidth]{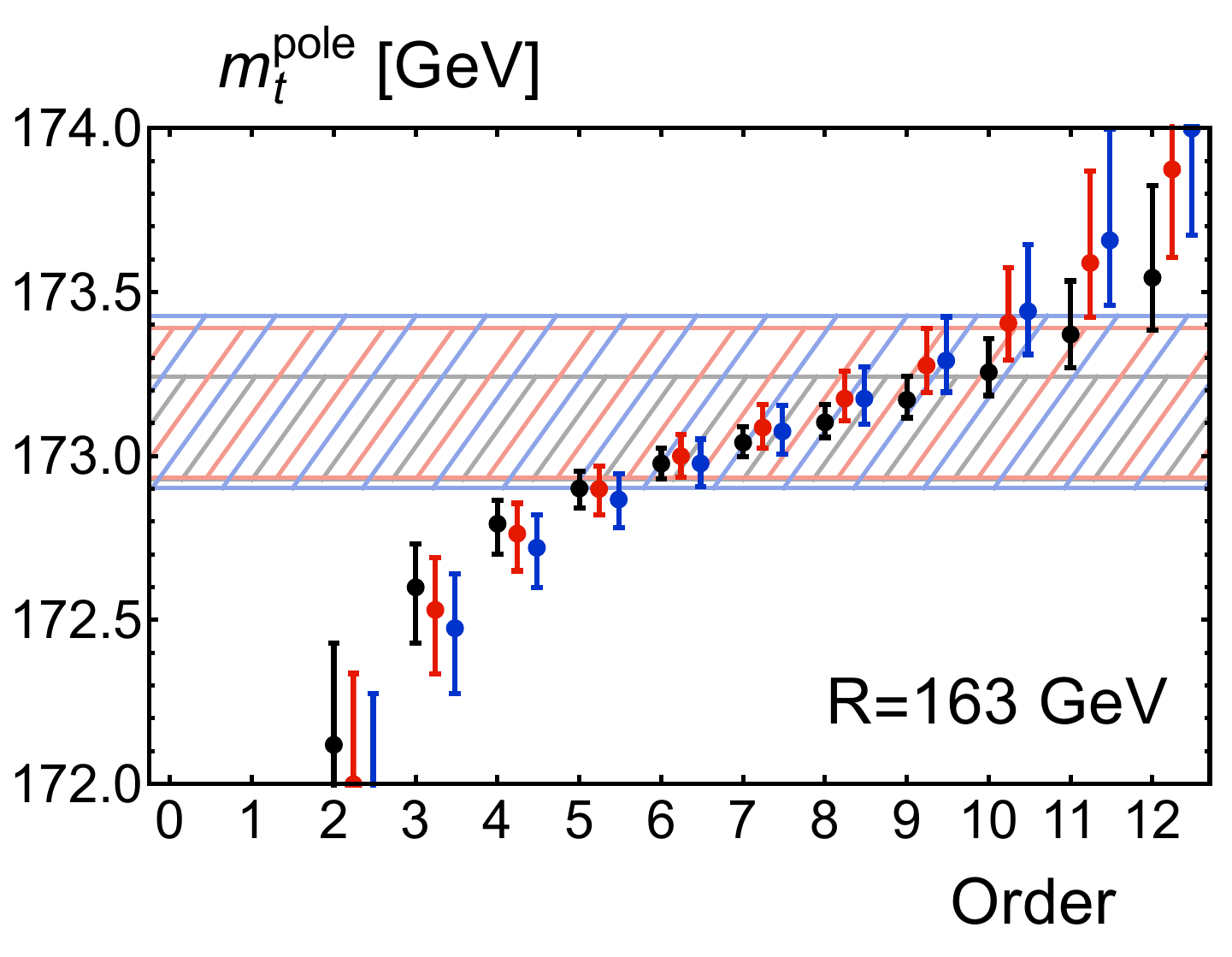}
 \label{fig:mtpolea}}  
 \subfigure[]{\includegraphics[width=.48\textwidth]{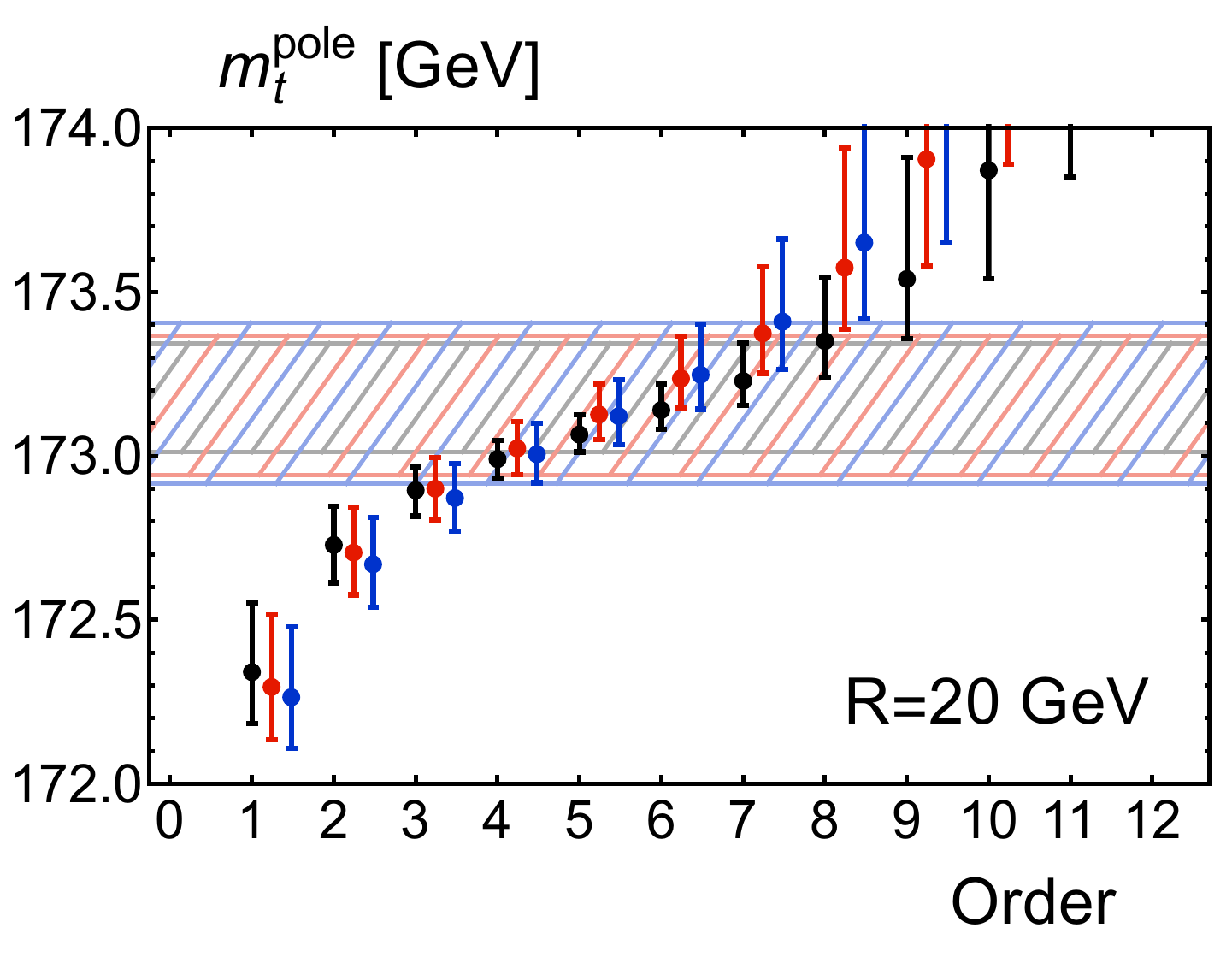}
 \label{fig:mtpoleb}}\\  
 \subfigure[]{\includegraphics[width=.48\textwidth]{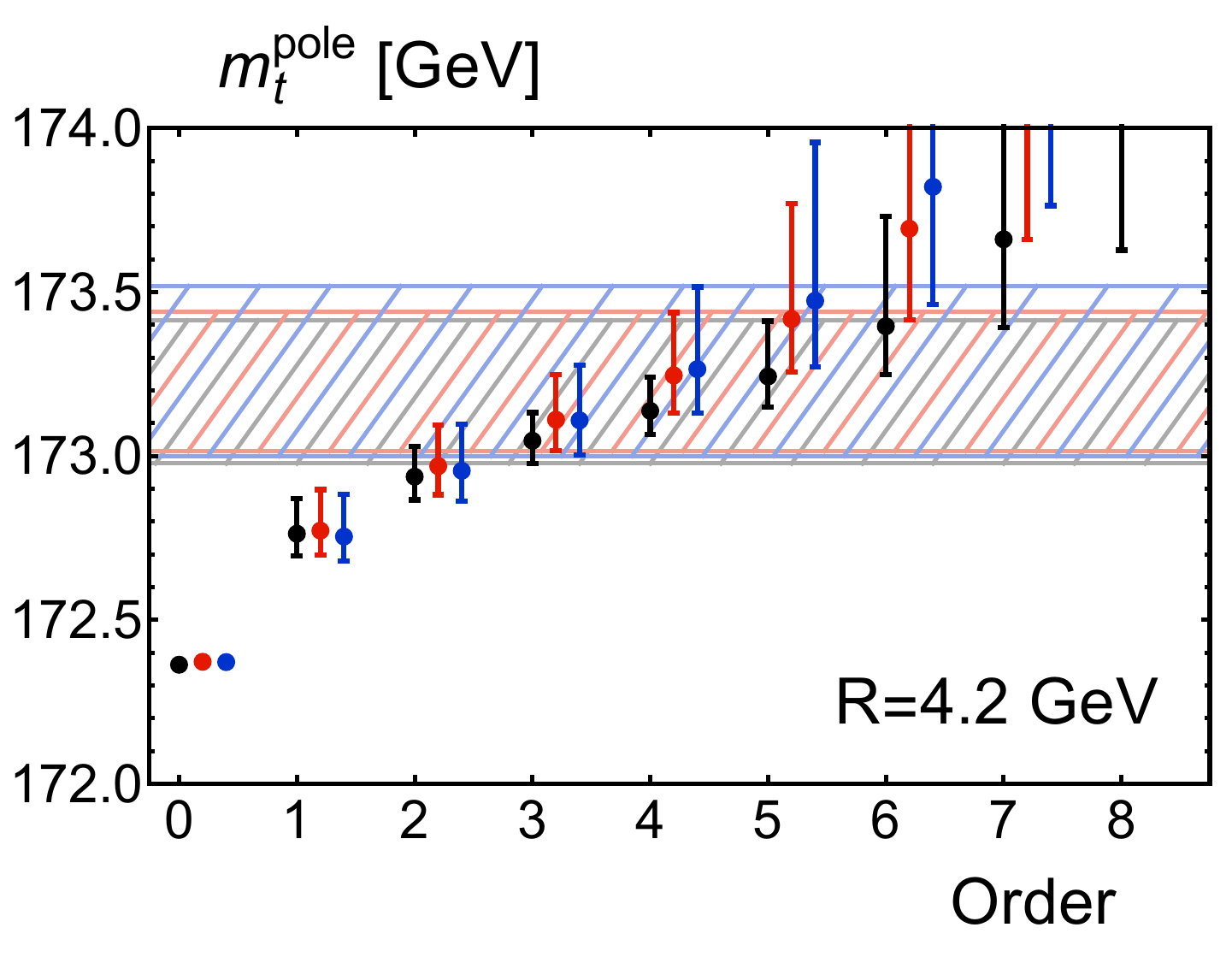}
 \label{fig:mtpolec}}  
 \subfigure[]{\includegraphics[width=.48\textwidth]{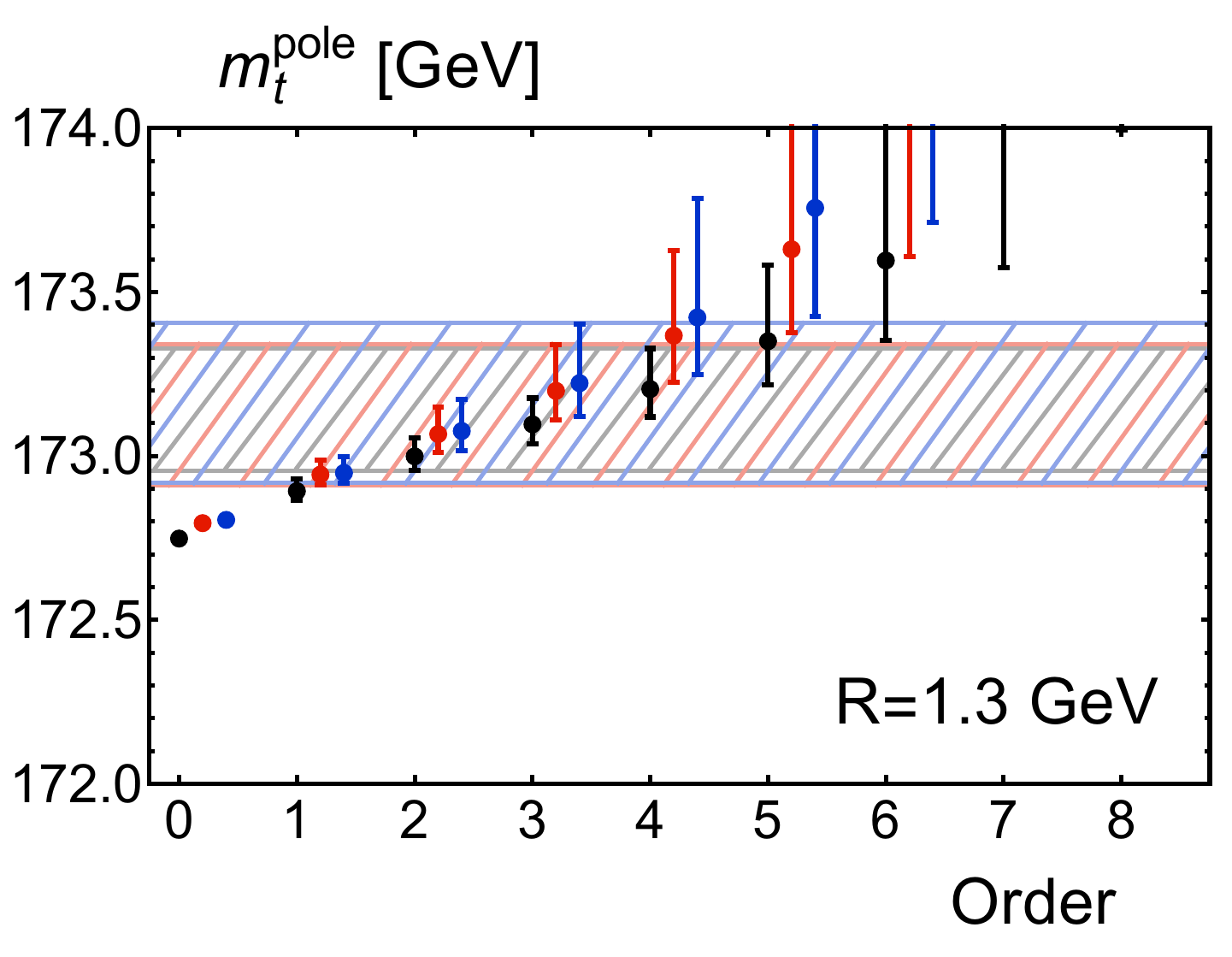}
 \label{fig:mtpoled}}  
 \caption{\label{fig:mtpole}
 Top quark pole mass $m_t^{\rm pole}$ as a function of order taking $\mbar_t=\mbar_t(\mbar_t)=163$~GeV as input and using different methods to obtain the best possible estimate and the ambiguity. The central dots are obtained for the default renormalization scales for the strong coupling and the error bands represent the scale variation as explained in the text. The light colored hatched horizontal bands bounded by equal colored lines show the best possible estimate for the respective method also given in the last column in Tab.~\ref{tab:mtpole}.
 All results obtained for massless bottom and charm quarks are in black, all results for 
 $(\mbar_b,\mbar_c)=(4.2,0)$~GeV are in red, all results for 
 $(\mbar_b,\mbar_c)=(4.2,1.3)$~GeV are in blue. 
 Panel (a) shows results for $R=163$~GeV, 
 panel (b) for $R=20$~GeV, panel (c) for $R=4.2$~GeV and panel (d) for $R=1.3$~GeV. 
 	}
\end{figure}

The entries are as follows: The second column shows $\msr_t(R)-\mbar_t = \Delta m_t^{(6\rightarrow 5)}(\mbar_t) + \Delta m^{(5)}(\mbar_t,R)$ at the highest order. The third and fourth column show the order $n_{\rm min}$ and $\Delta(n_{\rm min})$ for the default renormalization scale $\mu=R$ for the cases $R=163,20$ and $4.2$~GeV and $\mu=2\mbar_c$ for $R=1.3$~GeV. The values for $\Delta(n_{\rm min})$ for $R=163$ and $20$~GeV have an uncertainty because for these cases $n_{\rm min}>4$ and the values for $\Delta(n>4)$ are determined from the asymptotic large order values given in Tab.~\ref{tab:aasy} which have a numerical uncertainty from the normalization factor $N_{1/2}^{(5)}$ in Eqs.~\eqref{eqn:N12msr}.
The fifth column shows the sum of the perturbative corrections beyond the explicitly calculated ${\cal O}(\alpha_s^4)$ terms up to order $n_{\rm min}$ showing the amount of extrapolation needed to obtain the best possible top quark mass based on the asymptotic approximation.
The sixth column shows the set of orders $\{n\}_{f=5/4}$ for which $\Delta(n)\leq f\,\Delta(n_{\rm min})$ and which are used for determining the best estimate and the uncertainty of the top quark pole mass. The seventh column then contains the best estimate and the ambiguity of the series for $\mpole_t - \msr_t(R)$ using the method from Sec.~\ref{sec:method}. To obtain the uncertainties we used renormalization scale variation for $\alpha_s^{(5)}(\mu)$ in the range $R/2\leq\mu\leq 2\,R$ for the cases $R=163,20,4.2$~GeV and in the range $1.5\,{\rm GeV} \leq \mu\leq 5$~GeV for $R = 1.3$~GeV. 
For $R=1.3$~GeV we always use renormalization scales $\mu$ of the strong coupling that are larger than $1.5$~GeV because the dependence on the renormalization scale grows rapidly for smaller scales. The last column contains the final result for $\mpole_t$ combining the results for $\msr_t(R) - \mbar_t$ and $\mpole_t - \msr_t(R)$ where the uncertainties of both are added quadratically to give the final number for the ambiguity of $m_t^{\rm pole}$. These results are also displayed graphically in Figs.~\ref{fig:mtpolea}-\ref{fig:mtpoled} as the gray hatched horizontal bands.

In Figs.~\ref{fig:mtpole} we have also shown in black the results for $\mpole_t(n)$ over the order $n$ for the different setups where the dots are the results for the default renormalization scales that are used to determine $n_{\rm min}$, $\Delta(n_{\rm min})$ and $\{n\}_f$. The error bars represent the range of values at each order of the truncated series coming from the variations of the renormalization scale of the strong coupling. The black dot at $n=0$ visible in Figs.~\ref{fig:mtpolec}, \ref{fig:mtpoled} shows the highest order result for $\msr_t(R)$. 

We see that the results for the top quark pole mass $m_t^{\rm pole}$ for the different $R$ values are fully compatible to each other. In particular, the ambiguity estimates based on our method agree within $\pm 15\%$ and average to $182$~MeV. Furthermore, the central values for the best estimates vary by at most $110$~MeV and average to $173.150$~GeV. It is reassuring that the spread of the central values is smaller than the size of the ambiguity. We emphasize that the consistency of our results for the different $R$ values to each other cannot be interpreted in any way statistically since the analyses for different $R$ values are not theoretically independent. The agreement just shows that our method is consistent since the best estimate (and also the ambiguity) of the top quark pole mass is independent of $R$.  Interestingly our estimate for the ambiguity of the top quark pole mass agrees quite well with $\LQCD^{(\nl=5)} = 166$~MeV given in Eq.~\eqref{eqn:lambda}.

As already pointed out in Sec.~\ref{sec:method}, the minimal correction $\Delta(n_{\rm min})$ increases from around $60$~MeV for $R=163$~GeV to about 
$100$~MeV\footnote{
This number is obtained for the default renormalization scale $\mu=2\,\mbar_c=2.6$~GeV.
In the short analysis of Sec.~\ref{sec:method} we quoted $131$~MeV for the size of the minimal term for $R=1.3$~GeV, which was obtained for $\mu=1.3$~GeV.} 
for $R=1.3$~GeV.
At the same time, the order $n_{\rm min}$ where the minimal correction $\Delta(n_{\rm min})$ arises decreases from $n_{\rm min}=8$ at $R=\mbar_t$ down to $n_{\rm min}=4$ and $3$ for $R=4.2$ and $1.3$~GeV. Moreover,
the contribution in the best estimate for $m_t^{\rm pole}$ from orders beyond $n=4$ until order $n_{\rm min}$ decreases from about $310$~MeV at $R=\mbar_t$ to about $150$~MeV at $R=20$~GeV.
For $R$ scales around the bottom quark mass and below, where $n_{\rm min}\le 4$, there is no need any more to extrapolate beyond the explicitly calculated four orders to get the best value for $m_t^{\rm pole}$. This information is not just of academic importance but it is also relevant for phenomenology: The MSR mass $m_t^{\rm MSR}(R)$ for some low scale $R$ can serve as a low-scale short-distance mass for a physical application where the characteristic physical scale is $R$. Typical examples include the top pair inclusive cross section at the production threshold where $R \sim m_t\alpha_s \sim 25$~GeV~\cite{Hoang:2000yr}, or the reconstructed invariant top quark mass distribution where $R$ is in the range of $5$ to $10$~GeV~\cite{Fleming:2007qr,Butenschoen:2016lpz,Hoang:2017kmk}. The behavior of the series for $m_t^{\rm pole}-m_t^{\rm MSR}(R)$ thus reflects the typical behavior of the QCD corrections to the mass for the respective physical applications. 
The observations we make for the $R$-dependence of the behavior of the series show that the best possible determination of the top quark mass from an observable characterized by a low characteristic physical scale can in general be achieved at a lower order and also involves smaller perturbative corrections compared to an observable characterized by high characteristic physical scales  
(such as inclusive top pair cross sections at high energies or virtual top quark effects). This general property is also reflected visually in the graphical illustrations shown in Fig.~\ref{fig:mtpole}.

We note that our numerical analysis has a rather weak overall dependence on the choice of $f$ and that the results change by construction in a non-continuous way. Using $f=4/3$ only the outcome for $R=20$~GeV is modified to $\mpole_t-\msr_t(R) = 2.100\pm0.206$. Using $f=6/5$ only the outcome for $R=163$~GeV is modified to $\mpole_t-\msr_t(R) = 10.088\pm0.123$. This leaves the overall conclusion about the ambiguity of the top quark pole mass unchanged and we therefore consider $f=5/4$ as a reasonable default choice.

Comparing our results to those of Ref.~\cite{Beneke:2016cbu}, we find that our estimate of the top quark pole mass ambiguity of $180$~MeV exceeds theirs of $70$~MeV by a factor of $2.5$. The discrepancy arises since their result was only related to the size of the minimal term $\Delta(n_{\rm min})$ for an $R$ value close to $163$~GeV and did not account for the number of orders $\Delta n$ for which the $\Delta(n)$ are close to the minimal term $\Delta(n_{\rm min})$. For $R = 163$~GeV we have $\Delta n = 4$ for $f=5/4$ and we see the discrepancy is roughly compatible with $\Delta n/2$. 
Since for other choices of $R$ the values of $\Delta(n_{\rm min})$ and $\Delta n$ vary individually substantially (while their product is stable) we believe that a specification of the top quark pole mass ambiguity of $70$~MeV is not consistent with heavy quark symmetry.

\subsection{Massless Charm Quark}
\label{sec:mczero}

For the case of a massive bottom quark and treating the charm quark as massless we can calculate the top quark pole mass from the bottom MSR mass $m_b^\MSR(R\leq\mbar_b)$ using the top-bottom mass matching contribution $\delta m_{b,c}^{(t\to b)}(\mbar_b,0)$ of Eq.~\eqref{eqn:tbmatch2} for $\mbar_c=0$ in combination with the top and bottom $\MSb$-MSR mass matching contributions, $\Delta m_t^{(6\to 5)}(\mbar_t)$ and $\Delta m_b^{(5\to 4)}(\mbar_b)$ of Eq.~\eqref{eqn:MSRMSbmatch2} and R-evolution, see Eq.~\eqref{eqn:rrge}, with $n_t=5$ active dynamical  flavors from $\mbar_t$ to $\mbar_b$ and with $n_b=4$ active dynamical flavors from $\mbar_b$ to $R$. The resulting expression for the top quark pole mass systematically sums all logarithms $\log(\mbar_b/\mbar_t)$ and uses that the bottom quark pole-MSR mass relation, which specifies the bottom quark pole mass ambiguity, fully encodes the top quark pole mass ambiguity due to heavy quark symmetry. The expression for the top quark pole mass we use reads
\begin{align}
\label{eq:mtpolemczero}
 m_t^\pole=\mbar_t &+\Delta m_t^{(6\to 5)}(\mbar_t)+\Delta m^{(5)}(\mbar_t,\mbar_b)+\delta m_{b,c}^{(t\to b)}(\mbar_b,0) +\Delta m_b^{(5\to 4)}(\mbar_b) \nonumber\\
 &+\Delta m^{(4)}(\mbar_b,R) + R\sum_{n=1}^{\infty}a_n(\nl=4,0)\left(\frac{\alpha_s^{(4)}(R)}{4\pi}\right)^n\,,
\end{align}
where the sum of the first four terms on the RHS is just $m_t^\pole-m_b^\pole+\mbar_b$, using Eq.~\eqref{eqn:tbpole}, and the sum of the fifth and sixth term is the difference of the bottom MSR and $\MSb$ masses $m_b^\MSR(R)-\mbar_b$. Both quantities are free of an $\Ord(\LQCD)$ renormalon ambiguity and can be evaluated to the highest order given in Tabs.~\ref{tab:DeltaMRRprime}, \ref{tab:MSRMSbmatch} and \ref{tab:HQbreaking}. We can then study the uncertainty of the top quark pole mass and its $\Ord(\LQCD)$ renormalon ambiguity from the $R$-dependent series which is just equal to $m_b^\pole-m_b^\MSR(R)$.

\begin{table}
 \renewcommand{\arraystretch}{1.2}
 \newcolumntype{A}{>{\centering\arraybackslash} m{1cm} }
 \newcolumntype{B}{>{\centering\arraybackslash} m{4.1cm} }
 \centering
 \begin{tabular}{|A|B|B|B|}
   \hline
\multicolumn{4}{|c|}{$\mbar_t = 163\,{\rm GeV},\quad\mbar_b = 4.2\,{\rm GeV},\quad\mbar_c = 0\,{\rm GeV},\quad \nl= n_t - 1 = 4$}\\\hline
   $R$ & $\mpole_t$ & $\mpole_b$ & $\mpole_c$ \\\hline
   $163$ & $173.162\pm0.228$ & $4.994\pm0.227$ & -- \\
   $20$ & $173.155\pm0.212$ & $4.987\pm0.211$ & -- \\
   $4.2$ & $173.227\pm0.212$ & $5.059\pm0.211$ & -- \\
   $1.3$ & $173.126\pm0.215$ & $4.958\pm0.215$ & -- \\\hline
   \multicolumn{4}{|c|}{$\mbar_t = 163\,{\rm GeV},\quad\mbar_b = 4.2\,{\rm GeV},\quad\mbar_c = 1.3\,{\rm GeV},\quad \nl= n_t -2 = 3$}\\\hline
   $R$ & $\mpole_t$ & $\mpole_b$ & $\mpole_c$ \\\hline
   $163$ & $173.165\pm0.263$ & $4.996\pm0.263$ & $1.665\pm0.262$ \\
   $20$ & $173.161\pm0.245$ & $4.992\pm0.245$ & $1.661\pm0.244$ \\
   $4.2$ & $173.259\pm0.259$ & $5.090\pm0.258$ & $1.759\pm0.258$ \\
   $1.3$ & $173.159\pm0.244$ & $4.990\pm0.244$ & $1.659\pm0.243$ \\\hline
 \end{tabular}
\caption{Upper section: Best estimate for the top and bottom quark pole masses for the case $(\mbar_t,\mbar_b,\mbar_c)=(163,4.2,0)$~GeV for $R=163, 20, 4.2, 1.3$~GeV. Lower section: Best estimate for the top, bottom and charm quark pole masses for the case $(\mbar_t,\mbar_b,\mbar_c)=(163,4.2,1.3)$~GeV for $R=163, 20, 4.2, 1.3$~GeV.
	All numbers are in units of GeV. \label{tab:mQpole}}
\end{table}

The outcome of the analysis using the method described in Sec.~\ref{sec:method} for $(\mbar_t,\mbar_b)=(163,4.2)$~GeV as well as $R=163, 20, 4.2, 1.3$~GeV and $f=5/4$ is shown in the middle section of Tab.~\ref{tab:mtpole}. Except for the second and seventh column the entries are analogous to the analysis for $\mbar_b=\mbar_c=0$ in Sec.~\ref{sec:mbmczero}. Here, the second column shows $m_t^\pole-m_b^\pole+m_b^\MSR(R)-\mbar_t$ and the seventh shows $m_b^\pole-m_b^\MSR(R)$, which contains the $\Ord(\LQCD)$ renormalon ambiguity. The default choices and the ranges of variation for the renormalization scale in the strong coupling in the series for $m_b^\pole-m_b^\MSR(R)$ are the same as for our analysis for $\mbar_b=\mbar_c=0$ in Sec.~\ref{sec:mbmczero} for the corresponding $R$ values.
The last column contains again the final result for $m_t^\pole$ combining the results for $m_t^\pole-m_b^\pole+m_b^\MSR(R)-\mbar_t$ and $m_b^\pole-m_b^\MSR(R)$ where the uncertainties of both are added quadratically.
The results are also displayed graphically in Figs.~\ref{fig:mtpolea}-
\ref{fig:mtpoled} as the light red hatched horizontal bands. 
In the upper section of Tab.~\ref{tab:mQpole} we also show the best estimate for the bottom quark pole mass $m_b^{\rm pole}$ obtained for the respective $R$ values, which can be obtained using Eq.~\eqref{eq:mtpolemczero} and the result for the top-bottom pole mass difference of Eq.~\eqref{eqn:tbpole3}. 

In Figs.~\ref{fig:mtpole} we have shown in red the results for $m_t^\pole(n)$ over the order $n$ for the different setups where the dots are again the results for the default renormalization scales that are used to determine $n_\mathrm{min}$, $\Delta(n_\mathrm{min})$ and $\{n\}_f$. The error bars are the range of values coming from the variations of the renormalization scale of the strong coupling. The red dots at $n=0$ visible in Figs.~\ref{fig:mtpolec} and \ref{fig:mtpoled} show the highest order results for $m_t^\pole-m_b^\pole+m_b^\MSR(R)$.

We again see that the results for the top quark pole mass for the different $R$ values are compatible each other. The ambiguity estimates average to $217$~MeV. Interestingly this estimate for the ambiguity of the top quark pole mass roughly agrees with $\LQCD^{(\nl=4)}=225$~MeV given in Eq.~\eqref{eqn:lambdanum}. This is larger than $\LQCD^{(5)}=166$~MeV since the infrared sensitivity of the top quark pole mass increases when the number of massless quarks is decreased (i.e.\ $\beta_0^{(4)}>\beta_0^{(5)}$). Furthermore, we observe that the central values for the top quark pole mass cover a range that is compatible with case of a massless bottom quark. The central values average to $173.168$~GeV which is about $20$~MeV larger than for a massless bottom quark, which is, however, insignificant given the range of values covered by the central values or even the size of the ambiguity. So the bottom quark mass does essentially not affect the overall value of the top quark pole mass. We also note that the minimal corrections $\Delta(n_\mathrm{min})$ are all larger than the corresponding terms for the case of massless bottom and charm quarks. For $R=4.2$ and $1.3$~GeV they amount to about $130$~MeV.

\subsection{Massive Bottom and Charm Quarks}
\label{sec:finitembmc}

We now, finally, consider the case that both the bottom and the charm quark masses are accounted for. Since this situation involves three scales, it is the most complicated concerning matching and evolution that systematically sums logarithms $\log(\mbar_t/\mbar_b)$ and $\log(\mbar_b/\mbar_c)$. However, the case can be treated in a straightforward way by iterating the top-bottom mass matching procedure of the previous section one more time concerning the bottom-charm mass matching. The resulting formula for the top quark pole mass reads
\begin{align}\label{eqn:mtpolembmc}
 \mpole_t=\mbar_t&+\Delta m_t^{(6\to 5)}(\mbar_t)+\Delta m^{(5)}(\mbar_t,\mbar_b)+\delta m_{b,c}^{(t\to b)}(\mbar_b,\mbar_c)\nonumber\\
 &+\Delta m_b^{(5\to 4)}(\mbar_b)+\Delta m^{(4)}(\mbar_b,\mbar_c)+\delta m_{c}^{(b\to c)}(\mbar_c)+\Delta m_c^{(4\to 3)}(\mbar_c)\\
 &+\Delta m^{(3)}(\mbar_c,R)+ R\sum_{n=1}^\infty a_n(\nl=3,0)\left(\frac{\alpha_s^{(3)}(R)}{4\pi}\right)^n\,.\nonumber
\end{align}
The expression combines the top-bottom and bottom-charm mass matching contributions $\delta m_{b,c}^{(t\to b)}(\mbar_b,\mbar_c)$ and $\delta m_c^{(b\to c)}(\mbar_c)$ from Eqs.~\eqref{eqn:tbmatch2} and \eqref{eqn:bcmatch2}, respectively, and the top, bottom and charm $\MSb$-MSR mass matching contributions $\Delta m_t^{(6\to 5)}(\mbar_t)$, $\Delta m_b^{(5\to 4)}(\mbar_b)$ and $\Delta m_c^{(4\to 3)}(\mbar_c)$ of Eq.~\eqref{eqn:MSRMSbmatch2}. Furthermore it contains contributions from R-evolution with $n_t=5$ active dynamical flavors from $\mbar_t$ to $\mbar_b$, with $n_b=4$ active dynamical flavors from $\mbar_b$ to $\mbar_c$ and with $n_b=3$ active dynamical flavors from $\mbar_c$ to $R$. We do not employ any evolution to scales below $\mbar_c$ due to instabilities of perturbation theory for the charm pole-MSR mass relation at such low scales but we can explore scales above $\mbar_c$ using the R-evolution.

On the RHS of Eq.~\eqref{eqn:mtpolembmc} the sum of the first seven terms is just $m_t^\pole-m_c^\pole+\mbar_c$, using Eq.~\eqref{eqn:tcpole}, and the eighth term is the charm $\MSb$-MSR matching contribution. Both quantities are free from an $\Ord(\LQCD)$ renormalon ambiguity and can be evaluated to the highest order given in Tabs.~\ref{tab:DeltaMRRprime}, \ref{tab:MSRMSbmatch} and \ref{tab:HQbreaking}. We can then study the ambiguity of the top quark pole mass due to the $\Ord(\LQCD)$ renormalon from the $R$-dependent series which is just equal to $m^\pole_c-\msr_c(R)$. 
This relation specifies the charm quark pole mass ambiguity, and it fully encodes the top and bottom quark pole mass ambiguities due to heavy quark symmetry. 

We note that among all the terms shown in Eq.~\eqref{eqn:mtpolembmc}
the contributions from the MSR mass differences $\Delta m^{(5)}(\mbar_t,\mbar_b)$, $\Delta m^{(4)}(\mbar_b,\mbar_c)$ and $\Delta m^{(3)}(\mbar_c,R)$, determined with R-evolution, and the series proportional to $R$, which contains the $\Ord(\LQCD)$ renormalon, constitute the numerically most important terms. They exceed by far the contributions from the matching corrections, which amount to only $50$~MeV and, therefore, fully encode the large order asymptotic behavior of the top quark pole-$\MSb$ mass series $m_t^{\rm pole}-\mbar_t$ as defined in Eq.~\eqref{eqn:mpoleMSbar} in the presence of finite bottom and charm quark masses. The large order asymptotic form of the coefficients in the expansion in powers of $\alpha_s^{(6)}(\mbar_t)$ may then be determined directly from these terms for $R=\mbar_c$ using the analytic solution for the MSR mass differences provided in Eq.~(4.2) of Ref.~\cite{Hoang:2017suc} and expanding in  $\alpha_s^{(6)}(\mbar_t)$. However, the resulting series suffers from the large logarithms involving the ratios of the top, bottom and charm quark masses, and is therefore less reliable for applications than the result shown in Eq.~\eqref{eqn:mtpolembmc}. 

The outcome of the analysis using the method described in Sec.~\ref{sec:method} for ($\mbar_t,\mbar_b,\mbar_c)=(163,4.2,1.3)$~GeV, as well as $R=163, 20, 4.2, 1.3$~GeV and $f=5/4$ is shown in the lower section of Tab.~\ref{tab:mtpole}. Except for the second and seventh column the entries are analogous to the previous two analyses in Secs.~\ref{sec:mbmczero} and \ref{sec:mczero}. Here, the second column shows $m_t^\pole-m_c^\pole+\msr_{c}(R)-\mbar_t$ and the seventh shows $m_c^\pole-\msr_c(R)$, which contains the $\Ord(\LQCD)$ renormalon ambiguity of the top quark pole mass. The default choices and the ranges of variation for the renormalization scale in the strong coupling in the series for $m_c^\pole-m_c^\MSR(R)$ are the same as for the two previous analyses in Secs.~\ref{sec:mbmczero} and \ref{sec:mczero} for the corresponding $R$ values.
The last column contains again the final result for $m_t^\pole$ combining the results for $m_t^\pole-m_c^\pole+\msr_{c}(R)-\mbar_t$ and $m_c^\pole-\msr_c(R)$ where all uncertainties are added quadratically. These results are also displayed graphically in Fig.~\ref{fig:mtpolea}-\ref{fig:mtpoled} as the light blue hatched horizontal bands. In the lower section of Tab.~\ref{tab:mQpole} we also show the best estimate for the charm and bottom quark pole masses $m_c^{\rm pole}$ and $m_b^{\rm pole}$, respectively, for the different $R$ values, which can be obtained using Eq.~\eqref{eqn:mtpolembmc} and the result for the top-bottom and top-charm pole mass difference of Eqs.~\eqref{eqn:tbpole} and \eqref{eqn:tcpole}.

In Fig.~\ref{fig:mtpole} we have also shown in blue the results for $m_t^\pole(n)$ over the order $n$ for the different setups where the dots are again the results for the default renormalization scales that are used to determine $n_\mathrm{min}$, $\Delta(n_\mathrm{min})$ and $\{n\}_f$. The error bars are the range of values coming from the variations of the renormalization scale of the strong coupling. The blue dots visible in Figs.~\ref{fig:mtpolec} and \ref{fig:mtpoled} at $n=0$ shows the highest order result for $m_t^\pole-m_c^\pole+\msr_c(R)$.

We see that the results for the top quark pole mass for the different $R$ values are again fully consistent to each other. The ambiguity estimates average to $253$~MeV, which is more than twice the $110$~MeV ambiguity obtained in Ref.~\cite{Beneke:2016cbu}. The reason for the discrepancy is the same as for the analysis for massless bottom and charm quarks already explained in Secs.~\ref{sec:method} and \ref{sec:mbmczero}, and we therefore do not discuss it here further. 
Concerning the size of the minimal corrections $\Delta(n_\mathrm{min})$, we find that they reach  $116$, $154$ and $128$~MeV for $R=20$, $4.2$ and $1.3$~GeV, respectively, each of which is larger than $110$~MeV.
As in the two previous analyses our result for the ambiguity agrees very well with the corresponding value of $\LQCD$, given in Eq.~\eqref{eqn:lambda}, which in this case is also $\LQCD^{(\nl=3)}=253$~MeV. This is larger than the uncertainties we obtained for the cases discussed in the two previous analyses, where either the bottom and charm quarks were massless or just the charm quark, and thus again follows the pattern that the infrared sensitivity of the top quark pole mass increases when the number of massless quarks decreases (i.e.\ $\beta_0^{(3)}>\beta_0^{(4)}>\beta_0^{(5)}$).

Furthermore, we find that the central values for the top quark pole mass cover a range that is within errors in agreement with the two previous analyses. The range is, however, shifted slightly upwards by about $70$~MeV with respect to the case of massless bottom and charm quarks. For the value of the average we have $173.186$~GeV which is about $40$~MeV higher than the average $173.150$~GeV we obtained for massless bottom and charm quarks. This shift may represent a slight trend, but it is overall insignificant compared to the range of values covered by the central values or the size of the ambiguity.   
This shows that the charm quark mass, like the bottom quark mass, does not affect the value of the top quark pole mass. We can compare to the result of Ref.~\cite{Beneke:2016cbu}, where they found that the finite bottom and charm quark masses increase the top quark pole mass by $80\pm 30$~MeV, where the $30$ is their estimate for the uncertainty in their computation of the bottom and charm mass effects. This is consistent with the dependence on the bottom and charm masses we find in our analysis. Their prescription was based on a successive order-dependent reduction of the effective flavor number in the series motivated by the decoupling property observed in Ref.~\cite{Ayala:2014yxa}.
It incorporated some basic features of the bottom and charm mass corrections beyond the third order but is otherwise heuristic and does not systematically sum logarithms of $\mbar_b/\mbar_t$ and $\mbar_c/\mbar_t$. The consistency shows that concerning the estimate of the top quark pole mass ambiguity and within errors their prescription provides an adequate approximation.

\subsection{Overall Assessment for the Pole Mass Ambiguity}
\label{sec:overall}
The overall outcome of the analyses above concerning the best possible estimates (and the ambiguities) of the top quark pole mass and the pole masses of the bottom and charm quarks is summarized as follows:

\begin{enumerate}
\item Heavy quark symmetry states that the ambiguity of a heavy quark pole mass is independent of the mass of the heavy quark and that the ambiguities of the pole masses of all heavy quarks are equivalent. 
Our method for estimating the ambiguity is insensible to the masses of the heavy quarks and, within any given setup for the heavy quark mass spectrum, obtains the same ambiguities for all heavy quark pole masses. It is therefore fully consistent with heavy quark symmetry.

\item Our examinations for different setups for the spectrum of the masses of the bottom and charm quarks show that the top quark pole mass ambiguity increases when the number $\nl$ of massless quarks is decreased (which arises when the number of lighter massive quarks is increased). The numerical size we find agrees very well with $\LQCD^{(\nl)}$ defined in Eqs.~\eqref{eqn:lambda}. So our studies show that the well-accepted statement that ``heavy quark pole masses have an ambiguity of order $\LQCD$'' can be specified to the more precise statement that ``the ambiguity of the heavy quark pole masses is $\LQCD^{(\nl)}$, where $\nl$ is the number of massless quarks''. 

\item Considering the value of the top quark pole mass (and not its ambiguity) we find essentially no dependence on whether the bottom and charm quarks are treated massive or massless. This also implies that there is no dependence on actual values of the bottom and charm quark masses (which are know to a precision of a few $10$~MeV in the $\MSb$ scheme).
Likewise we also find that the value of the bottom quark pole mass has no dependence on whether the charm quark is treated massive or massless. 
These observations are important because, although the pole mass concept depends, due to the linear sensitivity to small momenta, intrinsically on the spectrum of the lighter massive quarks, they imply that one can give the top and the bottom quark pole masses a unique global meaning irrespective which approximation is used for the bottom and charm masses. In such a global context, however, one has to assign the largest value for $\LQCD$ as the ambiguity of the pole mass. This value is obtained for finite bottom and charm quark masses and amounts to $250$~MeV which we adopt as our final specification of the top quark pole mass ambiguity. 
\end{enumerate}

\section{Conclusions}
\label{sec:conclusions}

In this work we have provided a systematic study of the mass effects of virtual massive quark loops in the relation between the pole mass $\mpole_Q$ and short-distance masses such as the $\MSb$ mass $\mbar_Q(\mu)$ and the MSR mass $\msr_Q(R)$ \cite{Hoang:2008yj,Hoang:2017suc} of a heavy quark $Q$, where we mean virtual loop insertions of quarks $q$ with $\LQCD < m_q < m_Q$. In this context it is well-known that the virtual loops of a massive quark act as an infrared cut-off on the virtuality of the gluon exchange that eliminates the effects of that quark from the large order asymptotic behavior of the series. This effect arises from the ${\cal O}(\Lambda_{\rm QCD})$ renormalon contained in the pole mass which means that the QCD corrections have a linear sensitivity of small momenta that increases with the order in the perturbative expansion. The primary aim of this work was to study this effect in detail at the qualitative and quantitative level. We established a renormalization group formalism that allows to discuss the mass effects coming from virtual quark loops in the on-shell self energy diagrams of heavy quarks in a coherent and systematic fashion. We in particular examined (i) how the logarithms of mass ratios that arise in this multi-scale problem can be systematically summed to all orders, (ii) the large order asymptotic behavior and structure of the mass corrections themselves and (iii) the consequences of heavy quark symmetry (HQS). 

The basis of our formalism is that the difference of the pole mass and a short-distance mass contains the QCD corrections from all momentum scales between zero and the scale at which the short-distance mass is defined, which is $\mu$ for the $\MSb$ mass $\mbar_Q(\mu)$ or $R$ for the MSR mass $m_Q^{\rm MSR}(R)$. The MSR mass $\msr_Q(R)$, which is derived from self energy diagrams like the $\MSb$ mass, is particularly suited to describe the scale-dependence for momentum scales $R<m_Q$ since its renormalization group (RG) evolution is linear in $R$, called R-evolution~\cite{Hoang:2008yj,Hoang:2017suc}. When the finite masses of lighter heavy quarks are accounted for, the MSR mass concept allows to establish a RG evolution and matching procedure where the number of active dynamical flavors governing the evolution changes when the evolution crosses a mass threshold and where threshold corrections arise when a massive flavor is integrated out. This follows entirely the common approach of logarithmic RG equations as known from the $n_f$ flavor dependent $\mu$-evolution of the strong coupling $\alpha_s^{(n_f)}(\mu)$ and reflects the properties of HQS.

Due to heavy quark symmetry, the procedure allows for example to relate the QCD corrections in the top quark pole-$\MSb$ mass difference $\mpole_t-\mbar_t(\mbar_t)$ that are coming from scales smaller than the bottom mass, to the bottom quark pole-$\MSb$ mass difference $\mpole_b-\mbar_b(\mbar_b)$. This relation can be used to generically study and determine the large order asymptotic behavior and the structure of the lighter virtual quark mass corrections in the pole-$\MSb$ mass difference of a heavy quark $Q$. Within the RG framework we have proposed, we find that the bulk of the lighter virtual quark mass corrections is determined by their large order asymptotic behavior already at ${\mathcal O(\alpha_s^3)}$ (very much like the QCD corrections for massless virtual quarks), which confirms
earlier observations made in Refs.~\cite{Hoang:1999us,Hoang:2000fm} and \cite{Ayala:2014yxa}.
Using our RG framework and heavy quark symmetry we used this property
to predict the previously unknown ${\mathcal O(\alpha_s^4)}$ lighter virtual quark mass corrections to within a few percent from the available information on the ${\mathcal O(\alpha_s^4)}$ corrections for massless lighter quarks without an additional loop computation, see Eq.~\eqref{eqn:delta4predict2}. Furthermore we calculated the differences of the top, bottom and charm quark pole masses with a precision of around $20$~MeV, and we analyzed in detail the quality of the coupling approximation of Ref.~\cite{Ayala:2014yxa}, which works in an excellent way for the charm mass effects in the bottom quark pole mass, where in the context of the top quark, it fails.  

The second aim of the paper was to use the formalism to determine a concrete numerical specification of the ambiguities of the heavy quark pole masses and in particular of the top quark pole mass. This is of interest because the top quark pole mass is still the most frequently used mass scheme in higher order theoretical predictions for the LHC top physics analyses. The ambiguity of the pole mass is the precision with which the pole mass can be determined {\it in principle} given that the complete series is known. This ambiguity is universal (i.e.\ it exists in equivalent size in any context and cannot be circumvented) and its size can therefore be quantified from the relation of the pole mass and any short-distance mass alone for which all terms in the series can be determined to high precision. With the renormalization group formalism we have proposed we carried out an analysis accounting explicitly for the constraints coming from HQS. HQS states (i) that the ambiguity of a heavy quark is independent of its mass, and (ii) that the QCD effects in the heavy quark masses coming from momenta below the lightest massive quark are all equivalent, which implies that the ambiguities of all heavy quarks are equal.  

With our formalism both aspects were incorporated and validated in detail at the qualitative and quantitative level.
We considered different scenarios for the treatment of the bottom and charm quark masses and employed a method to estimate the ambiguity that does not depend on the mass of the heavy quark in a way that is consistent with heavy quark symmetry.
For the case  of massless bottom and charm quarks we found that the ambiguity of the top quark pole mass is $180$~MeV, 
when the charm quark is  massless we found $215$~MeV and when the finite masses of both the bottom and charm quarks are accounted for we obtained $250$~MeV. Numerically, the ambiguity turns out be essentially equal to the hadronization scale $\Lambda_{\rm QCD}^{(n_\ell)}$, defined in Eq.~\eqref{eqn:lambda}, where $\nl$ is the number of massless quarks. 
Thus, our analysis allows to specify the well-known qualitative statement ``the heavy quark pole masses have an ambiguity of order $\LQCD$'' to the more specific statement ``the ambiguity of heavy quark pole masses is $\LQCD^{(\nl)}$, where $\nl$ is the number of massless quarks''.
This dependence of the top quark pole mass ambiguity on the number of massless flavors is fully consistent with the behavior expected from the pole mass renormalon.
Furthermore, we have found that there is no significant dependence of the central value of the top quark pole mass on whether the bottom and charm quarks are treated as massive or massless. 

Our results for the ambiguities differ considerably from those of Ref.~\cite{Beneke:2016cbu}. They estimated the top quark pole mass ambiguity as $70$~MeV for the case that bottom and charm masses are neglected and as $110$~MeV when the bottom and charm masses are accounted for. We have shown in detail in which ways these values are incompatible with heavy quark symmetry and why our ambiguity estimates should be considered more reliable.

If one considers the top quark pole mass as a globally defined mass scheme valid for all choices of approximations for the bottom and charm quark masses, one should assign it an intrinsic principle ambiguity due to the ${\cal O}(\LQCD)$ renormalon of $250$~MeV. 
We stress, that this intrinsic uncertainty refers to the best possible precision with which one can in principle theoretically determine the top quark pole mass, and does not account in any way for issues unrelated to the pole mass renormalon in applications for actual phenomenological quantities, which typically involve NLO, NNLO or even NNNLO corrections from perturbative QCD. Furthermore, in order to achieve this theoretical precision it is required to have access to orders where the corrections (in the relation involving the pole mass) become minimal. The order where this happens in an actual phenomenological analysis also depends on the typical physical scale (i.e.\ the value of $R$) governing the examined quantity. If the top quark mass is determined from a quantity which has a low characteristic physical scale (e.g.\ top pair production close to threshold, kinematic endpoints, reconstructed top invariant mass distributions) then the minimal term is reached at very low orders, which may well be within the orders that can be calculated explicitly. If the top quark mass is determined from a quantity which has a high characteristic scale of the order or the top quark mass (e.g.\ total inclusive cross sections at high energies, virtual top quark effects) then the minimal term is reached only at high orders, which are not accessible to full perturbative computations. This also explains why top mass sensitive observables involving low characteristic physical scales are more sensitive for top quark mass determinations than observables involving high characteristic physical scales. So reaching the uncertainties in top quark pole mass determinations that come close to the ambiguity limit is in general much harder for observables governed by high physical scales.  

Currently, the most precise measurements of the top quark mass from the D0 and CDF experiments at the Tevatron~\cite{Tevatron:2014cka,Abazov:2017ktz} and the ATLAS and CMS collaborations at the LHC~\cite{Aaboud:2016igd,Khachatryan:2015hba} use the top reconstruction method and already reach the level of $500$ to $700$~MeV. Projections for LHC Run-2 further indicate that this uncertainty can be reduced significantly in the future and may reach the level of $200$~MeV for the high-luminosity LHC run~\cite{HL-LHC}. 
The outcome of our analysis disfavors the top quark pole mass as a practically adequate mass parameter in the theoretical interpretation of these measurements.

As a final comment we would like to remind the reader that all tricky issues concerning the convergence of the perturbative series and 
the way how to properly estimate the ambiguity of top quark pole mass become irrelevant if one employs an adequate short-distance mass definition. This may of course not mean in general that switching to a short-distance mass scheme will automatically lead to smaller uncertainties simply because other unresolved issues may then dominate.
The outcome of our analysis, however, implies that even reaching a $250$~MeV uncertainty for the top quark pole mass in a reliable way within a practical application is difficult. This is because the ${\cal O}(\Lambda_{\rm QCD})$ renormalon prevents using common ways such as scale variation for the truncated series to estimate theoretical uncertainties, and can affect the behavior of the series already at low orders where the corrections still decrease.  It is therefore advantageous to abandon the pole mass scheme in favor of an adequately chosen short-distance mass at latest when the available QCD corrections for a mass sensitive quantity yield perturbative uncertainties in the pole mass that become of the order of its ambiguity, which we believe is when they approach $0.5$~GeV.

\section*{Acknowledgments}
We  acknowledge partial support by the FWF Austrian Science Fund under the Doctoral Program No.\ W1252-N27 and the Project No.\ P28535-N27 and the U.S.\ Department of Energy under the Grant No.\ DE-SC0011090. We also thank the Erwin-Schr\"odinger International Institute for Mathematics and Physics for  partial support.
\vspace*{0.3cm}

\begin{appendix}
\section{Virtual Quark Mass Corrections up to \texorpdfstring{${\cal O}(\alpha_s^3)$}{3-Loop Order}}
\label{app:Delta}
 
The virtual quark mass corrections of ${\cal O}(\alpha_s^2)$ were determined in Ref.~\cite{Gray:1990yh} and read
\begin{align}\label{eqn:d21}
 \delta_2(1) & =	8\left(\frac{\pi^2}{3}-1\right)\, = \, 18.3189 \\
	\label{eqn:d2r}
 \delta_2(r) &= \frac{8}{9}\pi^2 +\frac{16}{3}\ln^2 r - \frac{16}{3} r^2\left(\frac{3}{2} + \ln r\right) \nonumber\\
 &\hspace{1cm}+ \frac{16}{3}(1+r)(1+r^3)\left(\frac{\pi^2}{6} - \frac{1}{2}\ln^2 r + \ln r \,\ln(1+r) + {\rm Li}_2(-r)\right)\\
 &\hspace{1cm}+ \frac{16}{3}(1-r)(1-r^3)\left(-\frac{\pi^2}{3} - \frac{1}{2}\ln^2 r + \ln r \,\ln(1-r) + {\rm Li}_2(r)\right) \,.\nonumber
\end{align}
The expansion of $\delta_2$ for small $r$ has the form $\delta_2(r)=(8\pi^2/3) r -16 r^2+(8\pi^2/3) r^3+\ldots$.
At ${\cal O}(\alpha_s^3)$ the virtual quark mass corrections were determined semi-analytically in Ref.~\cite{Bekavac:2007tk} for the case of one more massive quark $q$ in the heavy quark $Q$ self-energy. The corrections from the insertions of virtual loops of two different massive quarks $q$ and $q^\prime$ were not provided and are given in Eq.~\eqref{eqn:dttr1r2}. In the following we provide the results for the full set of 
${\cal O}(\alpha_s^3)$ virtual quark mass corrections using the results
from Ref.~\cite{Bekavac:2007tk} in the expansion for $\mbar_q/\mbar_Q\ll1$ adapted to our notation. The expressions for general $\mbar_q/\mbar_Q$, which are extensive, can be downloaded at \url{https://backend.univie.ac.at/fileadmin/user_upload/i_particle_physics/publications/hpw.m}. 

We consider the ${\cal O}(\alpha_s^3)$ virtual quark mass corrections to the pole-$\MSb$ mass relation of the heavy quark $Q$ coming from $n$ lighter massive quarks $q_1$, $q_2$, \ldots $q_n$ in the {\it order of decreasing mass} and $n_\ell$ additional quarks lighter than $\Lambda_{\rm QCD}$, which we treat as massless. So, the number $n_Q$ of quark flavors lighter than quark $Q$ is $n_Q=n+\nl$.
The expressions for the functions $\delta_{Q,3}$ defined in Eqs.~\eqref{eqn:Deltadef} and \eqref{eqn:Deltadef2} can be written in the form
\begin{align}
 \delta_{Q,3}^{(Q,q_1,q_2,\dots,q_n)}(1,r_{q_1 Q},\ldots,r_{q_n Q}) &= h(1) + (\nq+1)\,p(1) + \sum_{i=1}^n w(1,r_{q_i Q}) \label{eqn:d31qq} \,,\\
 \delta_{Q,3}^{(q_1,q_2,\dots,q_n)}(r_{q_1 Q},r_{q_2 Q},\ldots,r_{q_n Q}) &= h(r_{q_1 Q}) + \nq\,p(r_{q_1 Q}) + \sum_{i=2}^n w(r_{q_1 Q},r_{q_i Q}) \label{eqn:d3qq} \,,\\
 \delta_{Q,3}^{(q_m,q_{m+1},\dots,q_n)}(r_{q_m Q},r_{q_{m+1} Q},\ldots,r_{q_n Q}) &= h(r_{q_m Q}) + (\nq-m+1)\,p(r_{q_m Q}) \nonumber\\ &\hspace{3.3cm}+ \sum_{i=m+1}^n w(r_{q_m Q},r_{q_i Q}) \,.
\end{align}
All three formulae follow the same general scheme, where the number multiplying the function $p(r)$ is just the number of massive quarks in the superscript plus the number of massless quarks, $\nl$. We have displayed them nevertheless for clarity. The explicit form of the functions $h$, $p$ and $w$ is
\begin{align}
 h(1) &= 1870.7877 \,, \\
 h(r) &= r\,(1486.55 - 1158.03 \ln r) \nonumber\\
 &\hspace{.5cm} + r^2\,(-884.044-683.967\ln r 
 ) + r^3\,(906.021 - 1126.84 \ln r) \nonumber\\
 &\hspace{.5cm}+ r^4\,(225.158 + 11.4991\ln r - 80.3086 \ln^2 r + 21.3333 \ln^3 r) \\
 &\hspace{.5cm} + r^5\,(126.996 -182.478\ln r) + r^6\,(-22.8899 + 38.3536\ln r - 54.5284 \ln^2 r) \nonumber\\
 &\hspace{.5cm} + r^7\,(15.3830 - 34.8914\ln r) + r^8\,(2.52528 - 3.82270\ln r - 20.4593 \ln^2 r) 
 + {\cal O}(r^9) \,,\nonumber
\end{align}
and
\begin{align}
 p(1) &= -82.1208 \,, \\
 p(r) &= \frac{32}{27}\int_0^\infty\dd z\left[\frac{z}{2} + \left(1-\frac{z}{2}\right)\sqrt{1+\frac{4}{z}}\,\right] {\rm P}\left(\frac{r^2}{z}\right)\left(\ln z-\frac{5}{3}\right)\\
 &= r\,(-66.4668 + 70.1839 \ln r) + r^2\, 14.2222 + r^3\,(15.4143 + 70.1839 \ln r) \nonumber\\
 &\hspace{.5cm} + r^4\,(-23.1242 + 18.0613\ln r + 15.4074\ln^2 r - 4.74074\ln^3 r) - 31.5827\,r^5 \nonumber \\
 &\hspace{.5cm} + r^6\,(11.9886 - 1.70667\ln r) - 4.17761\,r^7 + r^8\,(2.40987 - 0.161088\ln r)
 + {\cal O}(r^9)\,, \nonumber
\end{align}
as well as 
\begin{align}
 w(1,1) &= 6.77871 \,, \\
 w(1,r) &= r^2\,14.2222 
  - 18.7157\,r^3 + r^4\,(7.36885 - 11.1477\ln r) \nonumber\\
 &\hspace{.5cm} + r^6\,(3.92059 - 3.60296\ln r + 1.89630\ln^2 r) \nonumber\\
 &\hspace{.5cm} + r^8\,(0.0837382 - 0.0772789 \ln r + 0.457144 \ln^2 r) 
 + {\cal O}(r^9)\,,
 \label{eqn:dtt1r}
\\
 w(r_1,r_2) &= p(r_2) + \frac{32}{27}\int_0^\infty\dd z\left[\frac{z}{2} + \left(1-\frac{z}{2}\right)\sqrt{1+\frac{4}{z}}\,\right] {\rm P}\left(\frac{r_1^2}{z}\right){\rm P}\left(\frac{r_2^2}{z}\right)\,, \label{eqn:dttr1r2}
\end{align}
where
\begin{align} 
 \Pi(x) &= \frac{1}{3} - (1-2\,x)\left[2-\sqrt{1+4\,x}\ln\left(\frac{\sqrt{1+4\,x}+1}{\sqrt{1+4\,x}-1}\right)\right] \,, \\
 {\rm P}(x) &= \Pi(x) + \ln x + \frac{5}{3} \,.
\end{align}
 
\end{appendix}

\bibliography{./sources}
\bibliographystyle{JHEP}

\end{document}